\documentclass[12pt]{article}
\usepackage{draft}

\begin{document}

\unitlength = .8mm

\begin{titlepage}

\begin{center}

\hfill \\
\hfill \\
\vskip 1cm

\title{A Low Temperature Expansion for \\ Matrix Quantum Mechanics}

\author{Ying-Hsuan Lin$^a$, Shu-Heng Shao$^b$, Yifan Wang$^b$, Xi Yin$^{a}$}

\address{$^a$Jefferson Physical Laboratory, Harvard University, \\
Cambridge, MA 02138 USA
\\
$^b$Center for Theoretical Physics, Massachusetts Institute of Technology, \\
Cambridge, MA 02139 USA}

\email{yhlin@physics.harvard.edu, shshao@mit.edu, \\ yifanw@mit.edu,
xiyin@fas.harvard.edu}

\end{center}

\abstract{ We analyze solutions to loop-truncated Schwinger-Dyson equations in massless ${\cal N}=2$ and ${\cal N}=4$ Wess-Zumino matrix quantum mechanics at finite temperature, where conventional perturbation theory breaks down due to IR divergences. We find a rather intricate low temperature expansion that involves fractional power scaling in the temperature, based on a consistent ``soft collinear" approximation. We conjecture that at least in the ${\cal N}=4$ matrix quantum mechanics, such scaling behavior holds to all perturbative orders in the $1/N$ expansion. We discuss some preliminary results in analyzing the gauged supersymmetric quantum mechanics using Schwinger-Dyson equations, and comment on the connection to metastable microstates of black holes in the holographic dual of BFSS matrix quantum mechanics. }

\vfill

\end{titlepage}

\eject

\tableofcontents


\section{Introduction}

The sixteen supercharge $SU(N)$ gauged matrix quantum mechanics was famously proposed by Banks, Fischler, Shenker and Susskind to give a non-perturbative formulation of M-theory \cite{Banks:1996vh}.  It was later recognized \cite{Balasubramanian:1997kd,Itzhaki:1998dd,Susskind:1998vk,Polchinski:1999br} that the precise holographic dual of the matrix quantum mechanics is an asymptotically null compactification of M-theory, along the general lines of gauge/gravity duality \cite{Maldacena:1997re,Gubser:1998bc,Witten:1998qj}.  Despite intense efforts over a number of years (see \cite{Taylor:2001vb} and references therein), most of the successful tests of the duality were realized to be consequences of supersymmetric non-renormalization theorems \cite{Paban:1998ea,Paban:1998qy,Hyun:1999hf,Sethi:1999qv,Nicolai:2000ht}, while unsuccessful attempts \cite{KeskiVakkuri:1997wr,Becker:1997cp,Helling:1999js} in testing the duality involved comparisons of the gravity and gauge theory results in regimes that do not overlap. It had become clear that in order to explore the semi-classical gravity (either type IIA or 11-dimensional supergravity) regime in the bulk, one must work in the genuinely strong coupling regime of the matrix quantum mechanics.

A strong coupling test of the duality within the 't Hooft scaling regime was performed by numerically computing the free energy of the matrix quantum mechanics at finite temperature using Monte Carlo method \cite{Anagnostopoulos:2007fw, Hanada:2008gy, Hanada:2008ez, Hanada:2009ne, Hanada:2011fq}. The result was shown to be consistent with the expected free energy of black holes in the gravity dual. In order for the black hole horizon to lie in the semi-classical gravity regime, one needs to take the low temperature and large $N$ limits of the matrix quantum mechanics. However, these are precisely the limits where the Monte Carlo computation becomes costly.

An alternative approach to the thermal free energy of the matrix quantum mechanics was pioneered by Kabat, Lifschytz and Lowe, using the one-loop truncated Schwinger-Dyson equations \cite{Kabat:1999hp,Kabat:2000zv,Kabat:2001ve}. This is the approach we will follow, and refine, in this paper. In place of the numerical solutions to the Schwinger-Dyson equations investigated in \cite{Kabat:1999hp,Kabat:2000zv,Kabat:2001ve}, we will analyze the solutions to these equations analytically. We will see that in the low temperature limit, intricate cancelation occurs due to an approximate restoration of supersymmetry, which allows for nontrivial scaling behavior with temperature. Despite the fact that the naive 't Hooft coupling goes to infinity in the low temperature limit, the effective coupling parameter that controls the loop expansion of Schwinger-Dyson equation could be finite, and the scaling behavior of solutions to the one-loop Schwinger-Dyson equation could survive to all order in the $1/N$ expansion, when there are sufficiently many supersymmetries. 


Before describing our results, let us briefly review the connection between black holes in the bulk geometry and the thermal free energy of the dual quantum mechanics. The gravity dual of the $SU(N)$ BFSS matrix quantum mechanics at finite temperature is given by type IIA string theory in the near horizon limit of a near extremal black hole carrying $N$ units of D0-brane charge \cite{Balasubramanian:1997kd,Itzhaki:1998dd,Susskind:1998vk,Polchinski:1999br}
\ie
ds^2_{IIA} &= -f^{-1/2} A dt^2 + f^{1/2} A^{-1} (dr^2 + r^2 d\Omega_8^2), \\
C_1 &= - f^{-1} \left( {1 + A \over 2}\right)dt, \quad e^\phi = f^{3/4}, \\
f &\equiv {c_0 g_s N l_s^7 \over r^7}, \quad c_0 = 60 \pi^3, \quad A = 1 - {r_0^7 \over r^7},
\label{IIAbkgd}
\fe
where $d\Omega_8^2$ is the metric on a unit eight-sphere, and the black hole horizon is at $r=r_0$. The Hawking temperature of the black hole is 
\ie
T_H = {7 \over (2\pi)^{7/2} \sqrt{30}} (g^2_{YM} N) ^{-1/2} \left( {r_0 \over l_s^2} \right)^{5/2},
\label{hawkingT}
\fe
where $g_{YM}^2 = g_s / 4\pi^2 l_s^3$. The free energy of the black hole is \cite{Klebanov:1996un}
\ie
\B F = - \left( {2^{21} 3^2 5^7 \pi^{14} \over 7^{19} } \right)^{1/5} N^2 \left(  {T \over (g_{YM}^2 N)^{1/3} } \right)^{9/5}.
\label{sugraF}
\fe
The black hole horizon lies in the type IIA supergravity regime when $g_{YM}^{2/3} N^{1/7} \ll T \ll (g_{YM}^2 N)^{1/3}$. Working in the 't Hooft scaling limit on the gauged quantum mechanics side, one takes the large $N$ limit while keeping the dimensionless 't Hooft coupling $g_{YM}^2N/T^3$ finite. The $T^{9/5}$ scaling of the free energy or the entropy of the black hole is expected to hold in the matrix quantum mechanics in the limit of large 't Hooft coupling, or equivalently, the low temperature limit.

One must be cautious about the meaning of the free energy of the matrix quantum mechanics at large $N$. Since BFSS quantum mechanics has exactly flat directions, at finite $N$, the free energy is infinite, reflecting the continuum of scattering states. There are only $N$ flat directions, however, whereas the entropy of the black hole in the gravity dual scales like $N^2$ times a function of the dimensionless 't Hooft coupling. If we regularize the volume divergence by an IR cutoff at distances much greater than the horizon size of the black hole but much smaller than $e^N$, we expect the black hole states to dominate the contribution to the free energy. In particular, if we take the infinite $N$ limit first, and then take the volume to infinity while maintaining that the volume grows no faster than exponential in $N$, the $N^2$ coefficient of the free energy should remain a finite function of the 't Hooft coupling. This is the planar free energy of interest. The corresponding, finite, entropy of order $N^2$ generally counts metastable states rather than exact energy eigenstates in the quantum mechanics. Nonetheless, the lifetime of these metastable are expected to go to infinity (exponentially) in the infinity $N$ limit.

In the gravity dual, the metastable states are the microstates of the D0 black hole, which decays by emitting D0-branes via Hawking radiation. Note that the D0-branes and their BPS bound states are the only particles that can escape to infinity in the type IIA 0-brane decoupling geometry. To see this, consider the Born-Infeld effective action for a probe D0-brane in the background (\ref{IIAbkgd})
\ie
S_{D0} &= T_{D0}\left[ \int dt ~ f^{-1} A^{1/2} \sqrt{1 + f A^{-2} \dot r^2 } - \int dt ~ f^{-1} {1 + A \over 2} \right].
\fe
While this effective action a priori holds only in the supergravity regime, there is strong evidence that it in fact continues to hold at arbitrary large radial distance where the curvature is at string scale (while the string coupling goes to zero). This is presumably due to the supersymmetry preserved by the asymptotic geometry. Note that in the large $r$ limit, the mass of the D0-brane cancels the potential energy, and the action reduces to that of a non-relativistic particle. The situation is in contrast to the Hawking radiation of black holes in global AdS where all emitted particles bounce back in finite time, making it possible for the black hole to be in equilibrium with a thermal bath.  We conclude that the black hole in the 0-brane decoupling geometry is unstable via the emission of D0-branes. The Hawking decay rate of emitting a D0-brane is computed in Appendix~\ref{app:decay}. One finds
\ie
\Gamma \sim { R_{10} \over N^{43 \over 14} l_P^2 } \left( r_0 \over l_P \right)^{359 \over 14} e^{-{2\pi \over 7} \sqrt{ 2 r_0^9 \over 15 N l_P^9} },
\fe
where $l_P \equiv (2\pi g_s)^{1/3} l_s$ is the 11D Planck length and $R_{10} \equiv g_s l_s$ is the radius of the M-theory circle.  The exponential factor can be understood in terms of the chemical potential for the D0-brane charge. This formula is valid in the type IIA supergravity as well as in the M-theory regime where the lift of the black hole solution to 11 dimensions is thermodynamically stable. The latter is expected when $r_0 \gg N^{1/9} l_P$. This is also the regime where the decay rate is exponentially suppressed. 


The metastable microstates of the D0 black hole should be counted, to leading order in the $1/N$ expansion, by the planar free energy of the matrix quantum mechanics. Even in the high temperature regime, where the black hole horizon spills into the stringy part of the bulk geometry, and where one naively expects the matrix quantum mechanics to be weakly coupled, one encounters infrared divergences in conventional perturbation theory. Similar IR divergences were previously encountered in the two-loop computation of scattering amplitudes on the Coulomb branch of the theory, and have been essentially ignored \cite{Becker:1997wh,Becker:1997xw}. The IR divergence can be cured (non-perturbatively) if one solves for the exact propagators using Schwinger-Dyson equations. This is the approach of \cite{Kabat:1999hp,Kabat:2000zv,Kabat:2001ve}, where the authors studied the self-energies of various fields, as well as the free energy of the theory at finite temperature, using the one-loop truncated Schwinger-Dyson equations, which amounts to a mean-field approximation. In \cite{Kabat:1999hp,Kabat:2000zv,Kabat:2001ve}, the solutions of the self-energies were found numerically. While in a certain temperature range the result seemed to be consistent with the expectation from the gravity side, the numerical solution appears to break down below a certain temperature. From the gravity side, it is clear that (\ref{sugraF}) should be valid for $T/(g_{YM}^2N)^{1/3}\gg N^{-4/21}$, and thus the temperature can be taken to be arbitrarily small in the 't Hooft limit.  It would be highly desirable to have an analytic understanding of the $T^{9/5}$ scaling of the low temperature free energy/entropy in the gauged quantum mechanics.

Though broken at finite temperature, supersymmetry plays an important role in the low temperature limit of the solutions to the Schwinger-Dyson equations. As pointed out in \cite{Kabat:1999hp,Kabat:2000zv,Kabat:2001ve}, in working with a truncated set of Schwinger-Dyson equations, either to a certain loop order or by including a certain finite subset of renormalized vertices, the equations must be manifestly supersymmetric in the zero temperature limit (though the solutions may be singular in the zero temperature limit) in order to have any chance of capturing the correct low temperature limiting behavior. In particular, the self-energies of the auxiliary fields in a supermultiplet must be included in the S-D equations. In a gauge theory, the solution to a truncated set of S-D equations will depend on the choice of gauge. While the solution to the exact S-D equations should clearly be independent of the gauge-fixing condition, in working with the truncated S-D equations, a manifestly (off-shell) supersymmetric gauge-fixing is necessary. The familiar Wess-Zumino gauge breaks all supersymmetries, and cannot be applied for our purpose, namely to extract the low temperature physics from the solutions to the truncated S-D equations. The authors of \cite{Kabat:1999hp,Kabat:2000zv,Kabat:2001ve} considered a gauge fixing condition that preserves manifest ${\cal N}=2$ supersymmetries, out of the ${\cal N}=16$ supersymmetries of BFSS matrix quantum mechanics. This leads to rather unconventional kinetic terms and fermion coupling. Alternatively, one may choose to work with gauge fixing conditions that manifest ${\cal N}=4$ or ${\cal N}=8$ supersymmetries. As a matter of fact, the ${\cal N}=4$ gauge fixing results in a rather complicated looking, fully nonlinear, action, and the ${\cal N}=8$ gauge fixing based on harmonic superspace requires the inclusion of infinitely many auxiliary fields.

As a first step towards understanding the low temperature scaling behavior in BFSS matrix quantum mechanics, we study its truncation to the matter sector in the cases of ${\cal N}=2$ and ${\cal N}=4$ gauge fixing. These may also be thought of as supersymmetric deformations of the BFSS matrix quantum mechanics, by turning off the coupling of the ${\cal N}=2$ or ${\cal N}=4$ gauge multiplet. For instance, the truncation to ${\cal N}=4$ matter multiplets results in the ${\cal N}=4$ quantum mechanics with three matrix matter multiplets $\Phi^a$ and the cubic superpotential $W = - { i\kappa \over 3\sqrt{2} } \epsilon^{abc} \Tr( \Phi^a \Phi^b \Phi^c)$. We refer to such a theory as an ${\cal N}=4$ Wess-Zumino (matrix) quantum mechanics. The Schwinger-Dyson equations of the ${\cal N}=4$ Wess-Zumino quantum mechanics have the particularly nice property that, going beyond the one-loop truncation, there are in fact no two-loop contributions, and there are no three-loop {\it planar} contributions (and the first planar correction to the one-loop S-D equations shows up at four-loop order). It is conceivable that the solution to the one-loop S-D equations in fact captures the correct scaling behavior of the planar free energy.

In this paper, we will find an analytic low temperature expansion of the solution to the one-loop truncated Schwinger-Dyson equations for the ${\cal N}=4$ Wess-Zumino quantum mechanics. The key observation will be that loops containing the zero mode of the bosonic field $\phi^a$ dominate the contribution to the self-energies of the nonzero modes and the auxiliary zero mode. This allows for the solution of the nonzero mode and auxiliary zero mode self-energies in terms of the boson zero mode self-energy. The S-D equation for the boson zero mode self-energy, on the other hand, is nontrivial only if we work to the next-to-next-to-leading order contributions in the low temperature expansion. In the end we find a nontrivial scaling behavior of the boson zero mode self-energy, which controls the scaling of the self-energies of all other modes.

The matter multiplet of ${\cal N}=4$ Wess-Zumino quantum mechanics contains the following component fields, schematically: the boson $\phi$, the fermion $\psi$, and the auxiliary field $f$. At finite temperature, i.e. in the Euclidean theory where the Euclidean time is compactified with periodicity $\beta = 1/T$, let the self-energies for the momentum modes of $\phi$, $f$ and $\psi$ on the Euclidean time circle be $\sigma_n$, $\eta_n$, and $h_r$, respectively. Here $n$ is an integer, whereas $r$ is a half integer, reflecting the anti-periodic thermal boundary condition for the fermionic field. The self-energies for the nonzero modes are solved in terms of the boson zero mode $\sigma_0$, with the following results:
\ie
\sigma_{n\neq0} &= { 2\pi |n| \over \B} \sqrt{2 \over \B\sigma_0} \left[ 
1 + ({\B\sigma_0 \over 2})^{3/2} {1 \over 2\pi} \left( \text{sign}(n) C_n + {3 \over |n|} \right) \right] + \cO(\B^{-2}), \\
\eta_{n\neq0} &= {\B \over 2\pi |n|} \sqrt{2 \over \B\sigma_0} \left[ 1 + ({\B\sigma_0 \over 2})^{3/2} {1 \over 2\pi} \left( \text{sign}(n) C_n - {3\over |n|} \right) \right] + \cO(\B^{0}), \\
h_r &= \text{sign}(r) \sqrt{2 \over \B\sigma_0} \left[ 1 + ({\B\sigma_0 \over 2})^{3/2} {1 \over 2\pi} \text{sign}(r) C_r \right] + \cO(\B^{-1}),
\label{finalsol}
\fe
where $C_n = \sum_{ k \neq 0, n } { \text{sign}(k) \text{sign}(n-k) /k}$. Here we expressed the results in units where the dimensionful 't Hooft coupling $\kappa^2 N$ is set to 1. For the zero mode self-energies $\sigma_0$ and $\eta_0$, of the boson $\phi$ and auxiliary field $f$, we will find
\ie
\eta_0 &= {1 \over \B\sigma_0^2} + {\B^2\sigma_0 \over 24} + \cO(\B^0), \\
\sigma_0 &= 2({\pi \over 3})^{2/5} \B^{-7/5} +\cO(\B^{-2}).
\fe
Somewhat surprisingly, the low temperature expansion parameter is a fractional power of the temperature, namely $\B^{-3/5}$. The planar free energy can then be computed in the mean-field approximation \cite{Kabat:1999hp}. The result is 
\ie
\B F = \text{const} - {5 \over 2} ({\pi \over 3})^{6/5} N^2 N_f \B^{-6/5} + {\cal O}(\B^{-9/5}).
\fe
where $N_f$ is the number of chiral superfields. Curiously, this result differs from the expected scaling of BFSS matrix quantum mechanics (\ref{sugraF}) by one power of the expansion parameter $\beta^{-3/5}$.

While we have not evaluated explicitly the four-loop correction to the planar Schwinger-Dyson equations and to the free energy, it appears that such higher-loop corrections could contribute at the {\it same} order as the one-loop contributions to the self-energies, in the low temperature limit, despite the fact that the naive dimensionless 't Hooft coupling is infinite in this limit. It is conceivable that the $T^{6/5}$ scaling is exact for the large $N$ ${\cal N}=4$ Wess-Zumino matrix quantum mechanics with cubic superpotential, and that as we continuously deform the BFSS matrix quantum mechanics to the ${\cal N}=4$ Wess-Zumino model by turning off the gauge coupling, the low temperature scaling behavior of the planar free energy interpolates between $T^{9/5}$ and $T^{6/5}$.

One might be puzzled by the following. Consider Model I, the ${\cal N}=4$ Wess-Zumino matrix quantum mechanics with a single matter multiplet $\Phi$ and superpotential $W={\rm Tr}\,\Phi^3$, and Model II, the ${\cal N}=4$ Wess-Zumino matrix quantum mechanics with three matter multiplets $X,Y,Z$ and superpotential $W={\rm Tr}(XYZ)$. In the former case the spectrum is gapped, whereas in the latter case there is a continuum of scattering states due to flat directions. At finite $N$, the free energy of the former should be exponentially suppressed in the low temperature limit, whereas that of the latter diverges due to the continuous spectrum. How are these consistent with our claimed scaling as follows from the Schwinger-Dyson equations? Our result suggests that in Model I, even though the spectrum is gapped at finite $N$, in the large $N$ limit the gap in the spectrum becomes very small, and if we take $N$ to infinity first and then take the low temperature (or equivalently, strong 't Hooft coupling) limit, the free energy exhibits power scaling in the temperature. In Model II, on the other hand, there are {\it different} solutions to the Schwinger-Dyson equation. The scaling solution we described above, in particular, treats $X,Y,Z$ on equal footing. There are other, singular, solutions that sets the self-energy of one of $X,Y$ or $Z$ to zero and giving infinite self-energies to the remaining two fields. These singular solutions describe the phase of the theory where one of the three fields acquires a large expectation value, while the other two fields are very massive, in contrast to the ``unbroken phase" described by the scaling solution. We conjecture that these distinct phases exist in the infinite $N$ limit, and the tunneling between different phases are exponentially suppressed in the large $N$ limit.

When vector multiplets, along with ghosts in the supersymmetric gauge fixing, are included in the one-loop truncated Schwinger-Dyson equations, it appears that our low temperature expansion scheme is spoiled. It is not clear to us whether the one-loop S-D equation captures the correct low temperature physics in this case. It is likely that loop corrections to the quartic and possibly higher vertices of the vector multiplet, which are not taken into account by the one-loop S-D equation, are needed to obtain a nontrivial low temperature scaling. We will nonetheless discuss preliminary results on the low temperature effective action of the vector multiplet by integrating out matter multiplets using the one-loop truncated S-D equations, as well as the high temperature expansion. The hope is that an improved S-D equation for vector multiplets will produce the $T^{9/5}$ scaling of BFSS matrix quantum mechanics. Perhaps more auxiliary fields need to be included in the S-D equations, or a gauge fixing that preserves more manifest supersymmetries is needed. This is left for the future.

The rest of the paper is organized as follows.
In Section~\ref{sec:N=2}, we introduce the $\cN = 2$ Wess-Zumino matrix quantum mechanics and derive its one-loop truncated Schwinger-Dyson equations.  In Section~\ref{sec:lowT}, we solve the Schwinger-Dyson equations in the low temperature limit using a ``soft collinear'' approximation scheme, and compute the mean-field free energy.  We then investigate the corrections from higher-loop diagrams, and argue that although the temperature scalings appear to be spoiled by these corrections in the $\cN = 2$ Wess-Zumino matrix quantum mechanics, they should remain valid in the $\cN = 4$ version of the theory in the planar limit.  In Section~\ref{sec:N=4}, we introduce the $\cN = 4$ Wess-Zumino matrix quantum mechanics, and repeat the low temperature analysis.  In Section~\ref{sec:BFSS}, we discuss how our results for the Wess-Zumino quantum mechanics can be extended to the full BFSS matrix theory by coupling it to a vector multiplet; in particular, we present the supersymmetric gauge-fixing conditions and write down the corresponding Schwinger-Dyson equations.  We also discuss the various phases of solutions.  In Section~\ref{sec:highT}, we explore the high temperature limit of BFSS.  In Section~\ref{sec:discuss}, we discuss future prospects of this program, including ways to write down the BFSS action that preserves more manifest supersymmetries, and applications of our methods to supersymmetric quantum field theories in other dimensions such as the two-dimensional $\cN = (2, 2)$ Landau-Ginzburg model.  The Hawking decay rate of the black hole in the 0-brane decoupling geometry is derived in Appendix~\ref{app:decay}.  Details on the convention of ${\cal N}=2$ and ${\cal N}=4$ superspace, the one-loop Schwinger-Dyson equations, and the low temperature expansion of the solutions are given in Appendices~\ref{app:N=2}-\ref{app:free}.  Finally, the high temperature expansion of the Schwinger-Dyson equations for BFSS matrix theory is analyzed in Appendix~\ref{app:highT}.


\section{$\cN = 2$ Wess-Zumino Matrix Quantum Mechanics}
\label{sec:N=2}

In the 1D $\cN = 2$ language, the BFSS matrix quantum mechanics consists of one vector multiplet and seven matter multiplets.  Its truncation to the matter sector is an $\cN = 2$ Wess-Zumino matrix quantum mechanics with the 7 matter multiplets interacting through a $G_2$-invariant cubic superpotential.
The one-loop truncated Schwinger-Dyson equations for this system have been studied numerically in \cite{Kabat:1999hp}.
In this section we recall the form of the Schwinger-Dyson equations for the $\cN = 2$ Wess-Zumino model at finite temperature, and set up the notations for the analytic results in subsequent sections.

\subsection{The Action}
\label{subsec:N=2action}

Consider $\cN = 2$ Wess-Zumino matrix quantum mechanics with flavor symmetry $G$. Let the matter multiplets be in some $N_f$-dimensional representation of $G$, labeled by index $a$. We will assume that $G$ has a rank-3 totally antisymmetric invariant tensor $\epsilon^{abc}$ normalized by
\ie
\epsilon^{abc} \epsilon^{abd} = c \delta^{cd}
\label{epsnorm}
\fe
in this $N_f$-dimensional representation.
If $G$ is the $\bf 7$ of $G_2$ as for the matter sector of BFSS, then $N_f = 7$ and $c = 3/2$.  The 1D $\cN = 2$ superspace is introduced in Appendix~\ref{app:N=2}.  In this language, the $\cN = 2$ Wess-Zumino quantum mechanics in Euclidean signature contains $N_f$ real superfields\footnote{ The $i$ in front of $f$ gives the right sign for the kinetic term of $f$. }
\ie
\Phi^a = \phi^a + i \psi^a_\A \theta_\A + i f^a \theta^2,
\fe
interacting through a cubic superpotential
\ie
W = - {i \kappa \over 6 \sqrt c} \epsilon^{abc} \Tr \left( \Phi^a [\Phi^b, \Phi^c] \right).
\fe
Here $\Phi^a$ are matrix superfields in the adjoint representation of $SU(N)$\footnote{ This $SU(N)$ becomes the color symmetry once we embed this theory into BFSS. } and in flavor symmetry $G$.  The coefficient for the superpotential is chosen so that the form of the Schwinger-Dyson equations will not depend on the choice of $G$ (but the free energy will).

After integrating out the fermionic coordiantes, the action is\footnote{ We let $\{ T^A \}$ be a basis for $SU(N)$ such that $\Tr (T^A T^B) = \delta^{AB}$, and write $\phi^a = \phi^{aA} T^A$. }
\ie
S &= \int d\tau ~ \Tr \bigg\{ {1 \over 2} \dot \phi^a \dot \phi^a + {1 \over 2} \psi^a_\A \dot \psi^a_\A + {1 \over 2} f^a f^a \\
& \hspace{1.5in} + {\kappa \over 2 \sqrt c} \epsilon^{abc} f^a [\phi^b, \phi^c] + {\kappa \over 2 \sqrt c} \epsilon_{\A\B} \epsilon^{abc} \phi^a [\psi^b_\A, \psi^c_\B] \bigg\},
\label{S}
\fe
where dot stands for Euclidean time derivative.  The SUSY transformations are
\ie
~[Q_\A, \phi] &= - i \psi_\A, \\
\{Q_\A, \psi_\B\} &= i \epsilon_{\A\B} f + i \delta_{\A\B} \dot \phi, \\
[Q_\A, f] &= -i \epsilon_{\A\B} \dot\psi_\B.
\fe

\subsection{Schwinger-Dyson Equations}
\label{subsec:SD}

The perturbative field theoretic approach to a (classically) massless theory in one dimension suffers from infrared divergences. In many examples, such infrared divergences are cured non-perturbatively. A framework that improves the ordinary perturbation theory and naturally resolves the IR divergences is the Schwinger-Dyson equations. While the usual formulation of Schwinger-Dyson equations are a set of recursive integral equations that express exact correlation functions or 1PI vertices in terms of higher point vertices, such equations are often hard to solve due to the general momentum dependence in the exact vertices. In this paper, we will work with the equations that express the exact self-energies in terms of the integrals of exact propagators and tree-level vertices. We will make a truncation on the loop order of these equations, study the solutions to the truncated equations, and then discuss the validity of such truncations.
 
\begin{figure}[htb]
\begin{center}
\includegraphics[width=0.8\textwidth]{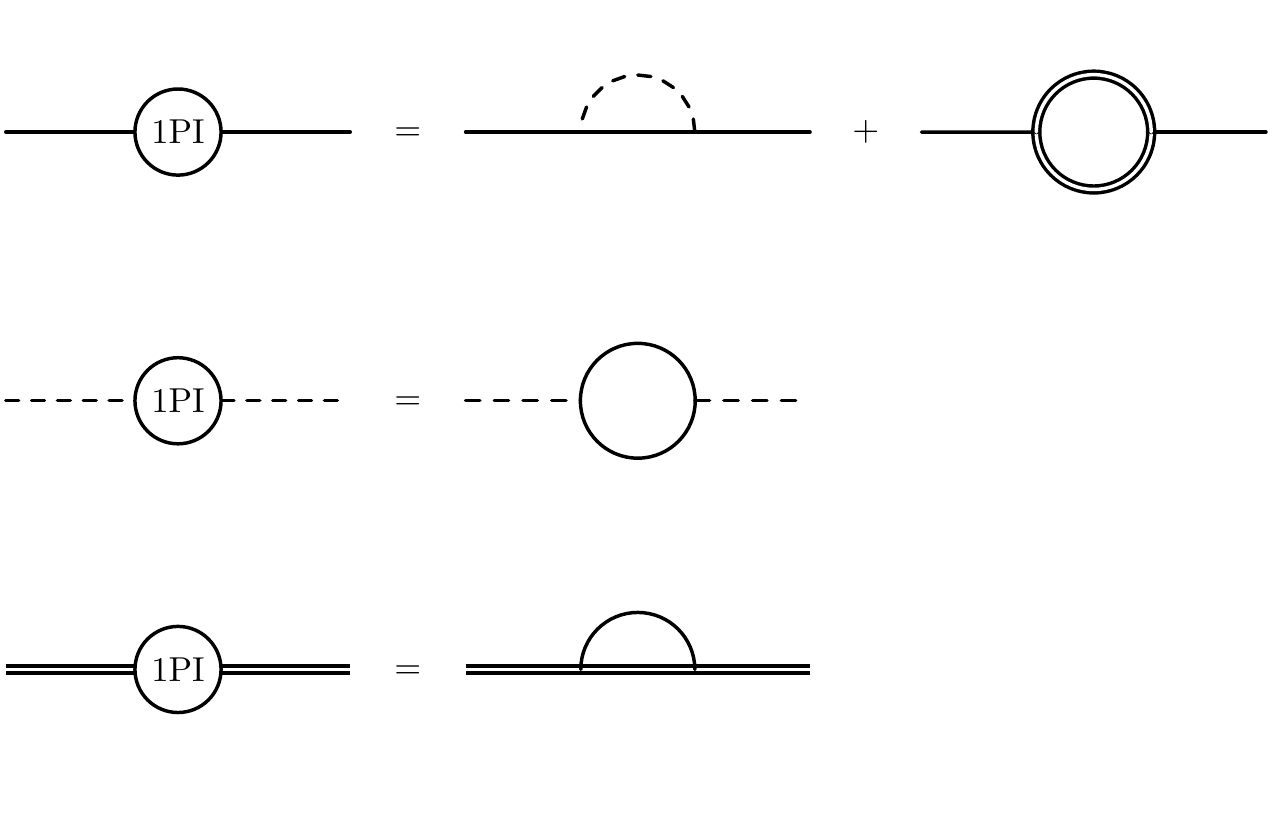}
\caption{ The one-loop truncated Schwinger-Dyson equations with bare cubic coupling. The solid, dashed, and double lines represent the boson propagator $\Delta_n$, the auxiliary field propagator $\epsilon_n$, and the fermion propagator $g_r$, respectively. }
\label{fig:sd}
\end{center}
\end{figure}

Starting from the action (\ref{S}), the Schwinger-Dyson equations at finite temperature are formulated as follows.  We compactify Euclidean time on a circle of circumference $\B$, and expand the fields in their Kaluza-Klein modes along the Euclidean time circle\footnote{ Unless otherwise noted, throughout this paper $n, k, \ell$ are integral and $r, s$ are half-integral. }
\ie
\phi^a &= {1 \over \sqrt \B} \sum_{n \in \mathbb{Z}} \phi^a_n e^{2\pi i n \tau / \B}, \\
\psi^a_\A  &=  {1 \over \sqrt \B} \sum_ {r \in \mathbb{Z} + {1 \over 2}} \psi^a_{\A, r} e^{2\pi i r \tau / \B} , \\
f^a &= {1 \over \sqrt \B} \sum_{n \in \mathbb{Z}} f^a_n e^{2\pi i n \tau / \B}.
\fe
The action (\ref{S}) written in terms of these Kaluza-Klein modes becomes
\ie
S &= \Tr \bigg\{ \sum_n {1 \over 2} ({2\pi n \over\B})^2 \phi^a_{-n} \phi^a_n
+ {1 \over 2} \sum_r {2\pi i r \over\B} \psi^a_{\A, -r} \psi^a_{\A, r}
+ {1 \over 2} \sum_n f_{-n}^a f^a_n \\
& \hspace{1in} + {i\kappa \over 2\sqrt{c\B}} \sum_{n,k} \epsilon^{abc} f^a_{-n-k} \phi^b_n \phi^c_k
+ {i\kappa \over 2\sqrt{c\B}} \sum_{r,s} \epsilon^{abc} \epsilon^{\A\B} \phi^a_{-r-s} \psi^b_{\A,r}\psi^c_{\B,s} \bigg\}.
\fe
Let us denote the exact propagators by
\ie
\la \phi^a_n \phi^b_m \ra &\equiv \Delta_n \delta^{ab} \delta_{n, -m}, \\
\la \psi^a_{\A, r} \psi^b_{\B, s} \ra &\equiv -i  g_r \delta^{ab} \delta_{\A\B} \delta_{r, -s}, \\
\la f^a_n f^b_m \ra &\equiv \epsilon_n \delta^{ab} \delta_{n,-m}.
\label{exactprops}
\fe
Note that $\la \phi f \ra$ is prohibited by the following $\bZ_2$-symmetry (R-parity) of the action (\ref{S}):
\ie
\phi \to -\phi, \quad \psi_1 \to -\psi_1.
\fe
Working in 't Hooft units where the dimensionful 't Hooft coupling $\kappa^2N$ is set to 1, i.e., $(\kappa^2 N)^{1/3} \B \to \B$, the ``one-loop truncated'' Schwinger-Dyson equations are\footnote{ The sign for the fermion loop is compensated by the factor of $(-i)^2$ from $\la \psi^a_{\A, r} \psi^b_{\B, s} \ra \equiv -i  g_r \delta^{ab} \delta_{\A\B} \delta_{r, -s}$. } (see Figure~\ref{fig:sd})
\ie
{1 \over \Delta_n}  &= ({2\pi n\over \B})^2 +  {2 \over \B} \sum_k \Delta_k \epsilon_{n-k} + {2 \over \B} \sum_r g_r g_{n-r}, \\
{1 \over \epsilon_n} &= 1+ {1 \over \B} \sum_k \Delta_k \Delta_{n-k},\\
{1\over g_r} &= {2\pi r\over \B} + {2 \over \B} \sum_k \Delta_k g_{r-k}.  \label{SDprops}
\fe
The terminology here requires some explanation.  The equations (\ref{SDprops}) are the Schwinger-Dyson equations for the exact two-point functions, where the three- and higher-point functions are approximated by their bare values.  We refer to (\ref{SDprops}) as the ``one-loop truncated'' Schwinger-Dyson equations.
The higher loop corrections will be discussed in Section~\ref{subsec:N=2higherloops}.  This approach is somewhere in between a fully non-perturbative treatment and perturbation theory, but the results clearly go beyond perturbation theory since the nontrivial solution for the self-energies cuts off the IR divergence that would invalidate the conventional perturbation theory.

Our convention for the self-energies $\sigma_n$, $h_r$, and $\eta_n$, for the boson $\phi$, fermion $\psi$, and auxiliary field $f$ respectively, is such that they are related to the exact propagators by
\ie
\Delta_n &\equiv {1 \over ({2\pi n \over \B})^2 + \sigma_n}, \\
g_r &\equiv {1 \over {2\pi r \over \B} + h_r}, \\
\epsilon_n &\equiv {1 \over 1 + \eta_n},
\label{prop2self}
\fe
where $\sigma_{-n} = \sigma_n$, $h_{-r} = -h_r$, and $\eta_{-n} = \eta_n$.  The Schwinger-Dyson equations (\ref{SDprops}), written in terms of the self-energies, are
\ie
\sigma_n &= {2 \over \B} \sum_k {1 \over \left[ ({2\pi k\over \B})^2+ \sigma_k \right] \left[ 1+\eta_{n-k} \right] }
+ {2 \over \B} \sum_r {1 \over \left[ {2\pi r \over \B} + h_r \right] \left[ {2\pi(n-r) \over \B} + h_{n-r} \right] }, \\
\eta_n &= {1 \over \B} \sum_k  {1 \over \left[ ({2\pi k\over \B})^2+\sigma_k \right] \left[ ({2\pi (n-k) \over \B})^2 + \sigma_{n-k} \right] }, \\h_r &= {2 \over \B} \sum_k {1 \over \left[ ({2\pi k \over \B})^2 +\sigma_k \right] \left[ {2\pi (r-k) \over \B} + h_{r-k} \right] }.
\label{SDself}
\fe

\subsection{$\cN=2$ SUSY Ward Identities}

Although supersymmetry is broken at finite temperature, one expects supersymmetry to be ``approximately" restored in the low temperature limit. This is subtle due to IR divergences, as the self-energies in the strict zero temperature limit are singular. Nonetheless, as shown later, the Ward identities we derive here assuming supersymmetry will hold approximately for the self-energies of the nonzero frequency modes of the fields in the low temperature limit. Such relations are useful in organizing the low temperature expansion of the solutions to the Schwinger-Dyson equations (\ref{SDself}).

Naively, at strictly zero temperature, where the momentum (in Euclidean time) is continuous, we write the exact two-point functions (\ref{exactprops}) as
\ie
\la \phi(p) \phi(-p) \ra &= {1 \over p^2 + \sigma(p)}, \\
\la \psi_\A(p) \psi_\B(-p) \ra &= {-i \delta_{\A\B} \over p + h(p)}, \\
\la f(p ) f(-p) \ra &= {1 \over 1 + \eta(p)}.
\fe
If SUSY is not spontaneously broken, then
\ie
0 = \la \{ Q_\A , \phi(\tau) \psi_\B(\tau') \} \ra
= -i \la \psi_\A(\tau) \psi_\B(\tau')\ra + i \delta_{\A\B} \la \phi(\tau) \dot \phi(\tau') \ra,
\label{ward1}
\fe
where we used $\la \phi f \ra = 0$.  Similarly we have
\ie
0= \la \{ Q_\A, f(\tau) \psi_\B(\tau') \} \ra
= -i\epsilon_{\A\gamma}\la \dot \psi_\gamma(\tau) \psi_\B (\tau')\ra
+ i \epsilon_{\A\B} \la f(\tau)  f(\tau') \ra.
\label{ward2}
\fe
In momentum space, (\ref{ward1}) and (\ref{ward2}) read
\ie
\sigma(p) = p h(p) = p^2 \eta(p).  \label{ward}
\fe

At finite but low temperatures, if supersymmetry is only slightly broken, one may anticipate that there exists a continuous function $s(p)$ such that
\ie
\sigma_n &= s({2\pi n \over \B}) \left[ {2\pi n \over \B}  + \cO(\B^{<-1}) \right], \\
h_r &= s({2\pi r \over \B}) \left[ 1 + \cO(\B^{<0}) \right], \\
\eta_n &= s({2\pi n \over \B}) \left[  { 2\pi n \over\beta} + \cO(\B^{<-1}) \right]^{-1} .
\label{approxward}
\fe
In particular, if we assume $s(0)$ is finite, then one expects that $\sigma_0$ should be much smaller than $\sigma_{n\neq0}$, and $\eta_0$ should be much larger than $\eta_{n\neq0}$. This will indeed be the case.  The separation of zero modes and nonzero modes is the key to the scaling ansatz (\ref{soft})  that will allow us to solve the Schwinger-Dyson equations at low temperatures.


\section{The Low Temperature Limit}
\label{sec:lowT}

\subsection{A ``Soft-Collinear'' Approximation}

In the low temperature, or large $\beta$ limit, we will demonstrate that the Schwinger-Dyson equations (\ref{SDself}) admit solutions of the following leading order scaling in $\B$ (here the mode numbers $n, r$ are assumed to be $\cO(1)$)
\ie
\begin{array}{lll}
\sigma_0 \sim \B^{-7/5}, & \quad \eta_0 \sim \B^{9/5} & \\
\sigma_{n \neq 0} \sim \B^{-4/5}, & \quad \eta_{n \neq 0} \sim \B^{6/5}, & \quad h_{r} \sim \B^{1/5}.
\label{soft}
\end{array}
\fe
Note that the scaling of the zero modes $\sigma_0$ and $\eta_0$ are order $\B^{-3/5}$ lower and higher, respectively, than their nonzero mode counterparts.  This $\B^{-3/5}$ turns out to be the appropriate expansion parameter for the Wess-Zumino quantum mechanics at low temperatures.

One key property of the ansatz (\ref{soft}) is that at large $\B$ the boson zero mode propagator $\Delta_0 \sim \beta^{7/5}$ dominates the loop sums on the right-hand-side of the Schwinger-Dyson equations (\ref{SDself}).  Hence, to leading order in $1/\B$, the S-D equations reduce to simple algebraic expressions for the self-energies of the nonzero modes and for $\eta_0$ in terms of $\sigma_0$.  Once we have the leading order expressions, we can compute corrections in $1/\B$ order by order. Finally, substituting these expressions into the S-D equation for $\sigma_0$, we find an equation involving $\sigma_0$ only.  If a solution for $\sigma_0$ with the assumed scaling exists, then we know that the scaling ansatz (\ref{soft}) is indeed consistent. The consistency of the ansatz is highly nontrivial, however, and as we will see require delicate cancelations.  By a slight abuse of terminology, we refer to our ansatz as a ``soft collinear'' approximation, where we regard the nonzero modes as ``hard" and zero modes as ``soft".

\subsection{Nonzero Modes $\sigma_n$, $\eta_n$, $h_r$ and the Auxiliary Zero Mode $\eta_0$}

In the following we implicitly take $n \neq 0$.  Let us reorganize the Schwinger-Dyson equations (\ref{SDself}) for the nonzero modes $\sigma_n$, $\eta_n$, and $h_r$ and the auxiliary zero mode $\eta_0$ in the form
\ie
\sigma_n &= {2 \over \B\sigma_0} {1 \over 1 + \eta_n} + {2\pi n \over \B} A_n, \\
\eta_n &= {2\over \B\sigma_0} {1\over ({2\pi n \over \B})^2 + \sigma_n} + {\B \over 2\pi n} B_n, \\
h_r &= {2 \over \B\sigma_0} {1 \over {2\pi r \over \B} + h_r} + C_r, \\
\eta_0 &= {1 \over \B\sigma_0^2} + D,
\label{SD}
\fe
where
\ie
A_n &\equiv {1 \over \pi n} \left( \sum_{k \neq 0} \Delta_k \epsilon_{n-k} + \sum_r g_r g_{n-r} \right), \\
B_n &\equiv {2\pi n\over \B^2} \sum_{k \neq 0, n} \Delta_k \Delta_{n-k}, \\
C_r &\equiv {2 \over \B} \sum_{k \neq 0} \Delta_k g_{r-k}, \\
D &\equiv {1 \over \B} \sum_{k \neq 0} \Delta_k \Delta_{-k}
\fe
According to the scaling ansatz (\ref{soft}), 
\ie
\begin{array}{*2{>{\displaystyle}l}}
{2 \over \B\sigma_0} {1 \over 1 + \eta_n} \sim \B^{-4/5},
& \quad {2\pi n \over\B} A_n \sim \B^{-7/5}, \\
{2 \over \B\sigma_0} {1 \over ({2\pi n \over \B})^2 + \sigma_n} \sim \B^{6/5}, & \quad {\B\over 2\pi n} B_n \sim \B^{3/5}, \\ {2 \over \B\sigma_0} {1 \over {2 \pi r\over\B} + h_r} \sim \B^{1/5},
&\quad C_r \sim \B^{-2/5}, \\
{1 \over \B\sigma_0^2} \sim \B^{9/5},
& \quad D \sim \B^{3/5}.
\end{array}
\fe
Since the terms on the RHS involving $A_n$, $B_n$, $C_r$, and $D$ are subleading in $1/\B$, the leading order Schwinger-Dyson equations (\ref{SD}) are simple algebraic equations as advertised earlier.  Notice that because we expanded in $1/\B$ while taking $n, r \sim \cO(1)$, the loop sums where $n$ and $r$ range over $-\infty$ to $\infty$ become divergent and will require regularization.

The fermion self-energy $h_r$ and the product $\sigma_n \eta_n$ can be directly read off
\bea
h_r &=& \text{sign}(r) \sqrt{2 \over \B\sigma_0} + \cO(\B^{-2/5}), \label{hr} \\
\sigma_n \eta_n &=& {2 \over \B\sigma_0} + \cO(\B^{-1/5}),
\eea
where sign$(r)$ comes from the odd parity of $h_r$.  To find $\sigma_n$ and $\eta_n$ individually, let
\ie
\sigma_n &= {2\pi n\over\B} s_n (1+x_n), \\
\eta_n &= {\B\over 2\pi n} s_n (1+x_n)^{-1},
\label{ansatz}
\fe
where
\ie
s_n \equiv \text{sign}(n) \sqrt{2 \over \B\sigma_0} + \cO (\B^{-2/5}).  \label{sn}
\fe
If the approximate SUSY Ward identities (\ref{ward}) are to hold in the low temperature limit, then we expect $x_n$ to scale with a negative power in $\B$.  In Appendix~\ref{appx} we solve for $x_n$ directly from the Schwinger-Dyson equations, and find
\ie
x_n = ({\B\sigma_0 \over 2})^{3 \over 2} {3 \over 2\pi |n|} + \cO(\B^{-6/5}) \sim \B^{-3/5},
\label{xn}
\fe
in accordance with (\ref{ward}).

\subsection{Boson Zero Mode $\sigma_0$}
\label{subsec:bosonzm}

Let us examine the Schwinger-Dyson equation for the boson zero mode $\sigma_0$
\ie
\sigma_0 = {2 \over \B\sigma_0 (1 + \eta_0)} + {2 \over \B} \left( \sum_{n \neq 0} {1 \over s_n^2} - \sum_r {1 \over h_r^2} \right) + {2\pi \over \B^2} \left( {\B\sigma_0 \over 2} \right)^{3 \over 2} + \cO(\B^{-16/5}).
\fe
If one substitutes the leading order expressions for the other self-energies into the above equation, then each term becomes order $\B^{-7/5}$, and one finds that the two sides exactly cancel.  It turns out that the next order $\B^{-2}$ terms also cancel, and therefore the solution to $\sigma_0$ is determined at order $\B^{-13/5}$.  This requires expanding $s_n$ and $h_r$ to order $\B^{-1}$ and $\eta_0$ to order $\B^{3/5}$ (each two orders of $\B^{-3/5}$ down from the leading piece).

For the auxiliary zero mode $\eta_0$, the relevant correction is given straightforwardly by $D$ in (\ref{SD}), and we find
\ie
\eta_0 = {1 \over \beta\sigma_0^2} + {1 \over \beta} \sum_{n \neq 0} {\beta^2\over 4\pi^2 n^2} {1\over s_n^2(1+x_n)^2} + {\cal O}(\beta^{-3/5})
= {1\over \beta\sigma_0^2} + {\beta^2\sigma_0\over 24} + {\cal O}(\beta^{0}).
\label{eta0}
\fe
For $s_n$ and $h_r$, let us write
\ie
s_n &= \text{sign}(n) \sqrt{2 \over \B\sigma_0} + s_n^{(1)} + s_n^{(2)} + \cO(\B^{-8/5}), \\
h_r &= \text{sign}(r) \sqrt{2 \over \B\sigma_0} + h_r^{(1)} + h_r^{(2)} + \cO(\B^{-8/5}),
\label{s0}
\fe
where $s_n^{(1)}$, $h_r^{(1)}$ are of order $\B^{-2/5}$ and $s_n^{(2)}$, $h_r^{(2)}$ are of order $\B^{-1}$.  Next, define
\ie
g \equiv \sqrt{\B\sigma_0 \over 2}, \quad C_k \equiv \sum_{\ell \neq 0, k} {\text{sign}(\ell) \text{sign}(k - \ell) \over \ell},
\fe
where $k$ can take both integral and half-integral values.  The computation for $s_n^{(1)}$, $s_n^{(2)}$, $h_r^{(1)}$, and $h_r^{(2)}$ can be found in Appendix~\ref{app:shcorr}.  The results are
\ie
s_n^{(1)} &= {\B\sigma_0 \over 4\pi} C_n + \mathcal{O}(\B^{-1}), \\
h_r^{(1)} &=  {\B\sigma_0 \over 4\pi} C_r + \mathcal{O}(\B^{-1}), \\
s_n^{(2)} &=
- {\pi n \over \B} + \text{sign}(n) {g^5 \over 8\pi^2} C_n^2 
-{g^5 \over 4\pi^2} \sum_{k \neq 0, n}
{ \text{sign}(n-k) C_k +\text{sign}(k) C_{n-k} \over k }\\
& \hspace{.5in} + {g^5 \over 4\pi^2 n} \sum_{k \neq 0, 2n} (-)^k \text{sign}(n - {k\over2}) C_{k/2}
- {g^2 \over 2\pi} \sum_{k \neq 0, n} 
{ \text{sign}(k) \text{sign}(n-k)  x_k \over k } \\
& \hspace{.5in} + {g^5\over2\pi^2n^2} + \cO(\B^{-8/5}),\\
h_r^{(2)} &=
- {\pi r \over \B} + \text{sign}( r) {g^5 \over 8\pi^2} C_r^2 
- {g^5\over 4\pi^2} \sum_{k \neq 0}
{ \text{sign}(r-k) C_k + \text{sign}(k) C_{r-k} \over k  }\\
& \hspace{.5in} - {g^2\over 2\pi} \sum_{k \neq 0} 
{ \text{sign}(k) \text{sign}(r-k)  x_k \over k } + \cO(\B^{-8/5}).
\label{shcorr}
\fe
The relevant contribution to the Schwinger-Dyson equation for $\sigma_0$ is
\ie
\sum_{n \neq 0} s_n^{-2} - \sum_r h_r^{-2} = -g^2 \left( 1 + {\pi g \over 2\B} + {5g^6 \over 6} \right) + \cO(\B^{-11/5}).
\label{s2-h2}
\fe
Using (\ref{eta0}) and (\ref{s2-h2}), the equation for $\sigma_0$ becomes
\ie
\sigma_0 = \sigma_0 + {\pi \over 2\sqrt2} \sigma_0^{3/2} \B^{-1/2} - {3 \over 16} \B^3 \sigma_0^4 + \cO(\B^{-16/5}),  \label{sigma0SD}
\fe
from which we obtain
\ie
\sigma_0 = 2\left( {\pi \over 3} \right)^{2 \over 5} \B^{-{7 \over 5}} + \cO(\B^{-2}).  \label{sigma0}
\fe

To summarize, we solved the Schwinger-Dyson equations and obtained the self-energies at low temperatures.  Consistency of the ansatz (\ref{soft}) is thus verified.  Note that although we solved $s_n$ to second order in (\ref{shcorr}), $x_n$ is only computed to zeroth order in (\ref{xn}), so according to (\ref{ansatz}) we only have first order expressions for $\sigma_n$ and $\eta_n$.  Fortunately, for the purpose of solving $\sigma_0$ and later on computing the free energy, we only need $s_n$ to second order but not $\sigma_n$ and $\eta_n$ separately.

\subsection{Continuum Limit}

Before computing the free energy, let us note the following property of our solution of self-energies.  In terms of the low temperature expansion parameter $g$, the boson self-energies are
\ie
\sigma_0 &= {2g^2 \over \B},  \\
\sigma_n &= {2\pi |n| \over g\B} \left( 1 + {3g^3 \over 2\pi |n|} \right) \left( 1 + {g^3 |C_n| \over 2\pi} \right) + \cO(\B^{-2}),
\fe
where the first equation is the definition of $g$.  If we write $p=2\pi n/\B$, then the self-enengy for the nonzero modes can be written as
\ie
\sigma(p) &= {|p| \over g} \left( 1 + {3g^3 \over \B|p|} \right) \left[ 1 + {g^3 \over \B} \left( 2 \sum_{k=1}^{|n|} {\B \over 2\pi k} - {1 \over |p|} \right) \right] + \cO(\B^{-2}) \\
&= {|p| \over g} \left[ 1 + {2g^3 \over \B|p|} + {g^3 \over \pi} HN({\B|p| \over 2\pi}) + \cO(g^6) \right],
\fe
where $HN(x)$ is the analytic continuation of the harmonic function.  Since $HN(x)$ is regular at $x=0$, taking $p \to 0$ gives
\ie
\sigma(0) = {2g^2 \over \B} + \cO(g^5).
\fe
Curiously, $\sigma(0)$ coincides with $\sigma_0$.  In other words, the zero mode self-energy $\sigma_0$ could be regarded as the $p\to 0$ limit of the self-energy function $\sigma(p)$ of a continuous momentum $p$. This fact may be useful in applying the Schwinger-Dyson equation to Wess-Zumino quantum mechanics coupled to gauge fields, where a continuous distribution of the eigenvalues of the holonomy matrix along the thermal circle is involved.

\subsection{Free Energy}

Having obtained the propagators from the one-loop truncated Schwinger-Dyson equation, we can now compute the free energy in the mean-field approximation \cite{Kabat:1999hp}.  Let us define a quadratic Fourier space ``trial action" $S_0$, whose propagators are solutions to the one-loop truncated Schwinger-Dyson equations:
\ie
S_0 &= \Tr \bigg\{ \sum_n {\phi^a_{n} \phi^a_{-n} \over 2\Delta_n} +i \sum_r {\psi^a_{\A, r} \psi^a_{\A, -r} \over 2g_r} + \sum_n {f^a_{n} f^a_{-n} \over 2\epsilon_n} \bigg\}.
\label{trialS}
\fe
Next we write the true action as $S = S_2 + S_3$, where $S_2$ and $S_3$ consist of the quadratic and cubic terms, respectively.  Then an approximate free energy is given by
\ie
\B F = \B F_0 + \la S_2 -S_0 \ra_0 - {1 \over 2} \la S^2_3\ra_0.
\label{BFdef}
\fe
The subscript 0 here indicates that the expectation value is taken with respect the trial action, i.e.,
\ie
\la \cdots\ra_0 \equiv \int D\phi D\psi Df ~ e^{-S_0} \cdots.
\fe

In the planar limit, and dropping an overall factor of $N^2 N_f $, we have
\ie
\B F_0 &= - {1 \over 2} \sum_n \ln \Delta_n - {1 \over 2} \sum_n \ln \epsilon_n + {1 \over 2} \sum_r \ln g_r^2, \\
\la S_2 - S_0 \ra_0 &= {1 \over 2} \sum_n \left[ \left( {2\pi n \over \B} \right)^2 \Delta_n - 1 \right] + {1 \over 2} \sum_n (\epsilon_n - 1) 
- \sum_r \left( {2\pi r \over \B} g_r - 1 \right), \\
- {1 \over 2}\la S^2_3 \ra_0 &= {1 \over 2\B} \sum_{m, k} \Delta_m \Delta_k \epsilon_{m+k}
+ {1 \over \B} \sum_{r, s} g_r g_s \Delta_{r+s}.
\fe
In Appendix~\ref{app:free}, we compute these contributions to the free energy using solutions computed in Section~\ref{subsec:bosonzm}, but without plugging in the explicit value of $\sigma_0$ in (\ref{sigma0}).  The results are
\ie
\B F_0 &= \text{const} + g \left( \sum_{n \neq 0} s^{(2)}_{|n|} - \sum_r h^{(2)}_{|r|} \right) \\
& \hspace{1in} + {4\pi g \over \B} \left( \sum_{n>0} n - \sum_{r>0} r \right) + {5 \over 192} (\B\sigma_0)^3 + \cO(\B^{-9/5}), \\
\la S_2 - S_0\ra_0 &= 1 - {\pi \over 2\sqrt2} ({\sigma_0 \over \B})^{1/2} + \cO(\B^{-9/5}), \\
- {1 \over 2} \la S^2_3 \ra_0 &= -g \left( \sum_{n \neq 0} s^{(2)}_{|n|} -\sum_r h^{(2)}_{|r|} \right) \\
& \hspace{1in} - {4\pi g \over \B} \left( \sum_{n>0} n - \sum_{r>0} r \right) + {1 \over 192} (\B\sigma_0)^3 + \cO(\B^{-9/5}).
\fe
We see that the potentially divergent sums cancel between $\B F_0$ and $-{1 \over 2} \la S^2_3 \ra_0$, and thus
\ie
\B F = \text{const} + {1 \over 32} (\B\sigma_0)^3 - {\pi \over 2\sqrt2} ({\sigma_0 \over \B})^{1/2} + \cO(\B^{-9/5}).
\label{BFsigma0}
\fe

In~\cite{Kabat:1999hp}, the one-loop truncated Schwinger-Dyson equations are obtained by extremizing the mean-field free energy (\ref{BFdef}) with respect to the propagators in the trial action (\ref{trialS}) (which is why it was called the trial action in the first place).  As a consistency check, we should find that our solution for $\sigma_0$ (\ref{sigma0}) minimizes the free energy (\ref{BFsigma0}).  This is indeed the case as one can easily verify.  Replacing $\sigma_0$ by its explicit solution (\ref{sigma0}), we get
\ie
\B F = \text{const} - {5 \over 4} ({\pi \over 3})^{6/5} N^2 N_f \B^{-6/5} + \cO(\B^{-9/5}),  \label{BF}
\fe
where we have restored the overall $N^2 N_f$ factor.

To conclude, we solved the one-loop truncated Schwinger-Dyson equations in the large $\B$ limit, and found that the non-constant part of $\B F$ has $\B^{-6/5}$ scaling. We have also confirmed it numerically, including the coefficient in front of $\beta^{-6/5}$. For comparison, the numerical analysis of \cite{Kabat:1999hp} by fitting up to $\beta\simeq 28$ gave the scaling $\sim \beta^{-1.09}$. In fact, a careful inspection of various terms in the $\sigma_0$ S-D equation shows that the correct $\beta^{-1.2}$ low temperature scaling is visible only when $\beta$ is greater than $\sim 100$ (in the normalization convention of \cite{Kabat:1999hp}).

\subsection{Higher Loop Corrections}
\label{subsec:N=2higherloops}

A priori, there seems to be no reason to expect the one-loop truncation of the Schwinger-Dyson equations to capture the correct low temperature physics, since the dimensionless 't Hooft coupling is strong in the low temperature limit. Naively, one may expect two and higher loop contributions to the S-D equations to overwhelm the one-loop contribution and completely destroy the scaling behavior we found.
However, due to the peculiar scaling behavior of the propagators, the effective expansion parameter in the low temperature limit is not the 't Hooft coupling. If we proceed with our ansatz for the one-loop self-energies, the question is whether the correction to the S-D equation for $\sigma_0$ respects the scaling ansatz.

\begin{figure}[htb]
\begin{center}
\includegraphics[width=0.8\textwidth]{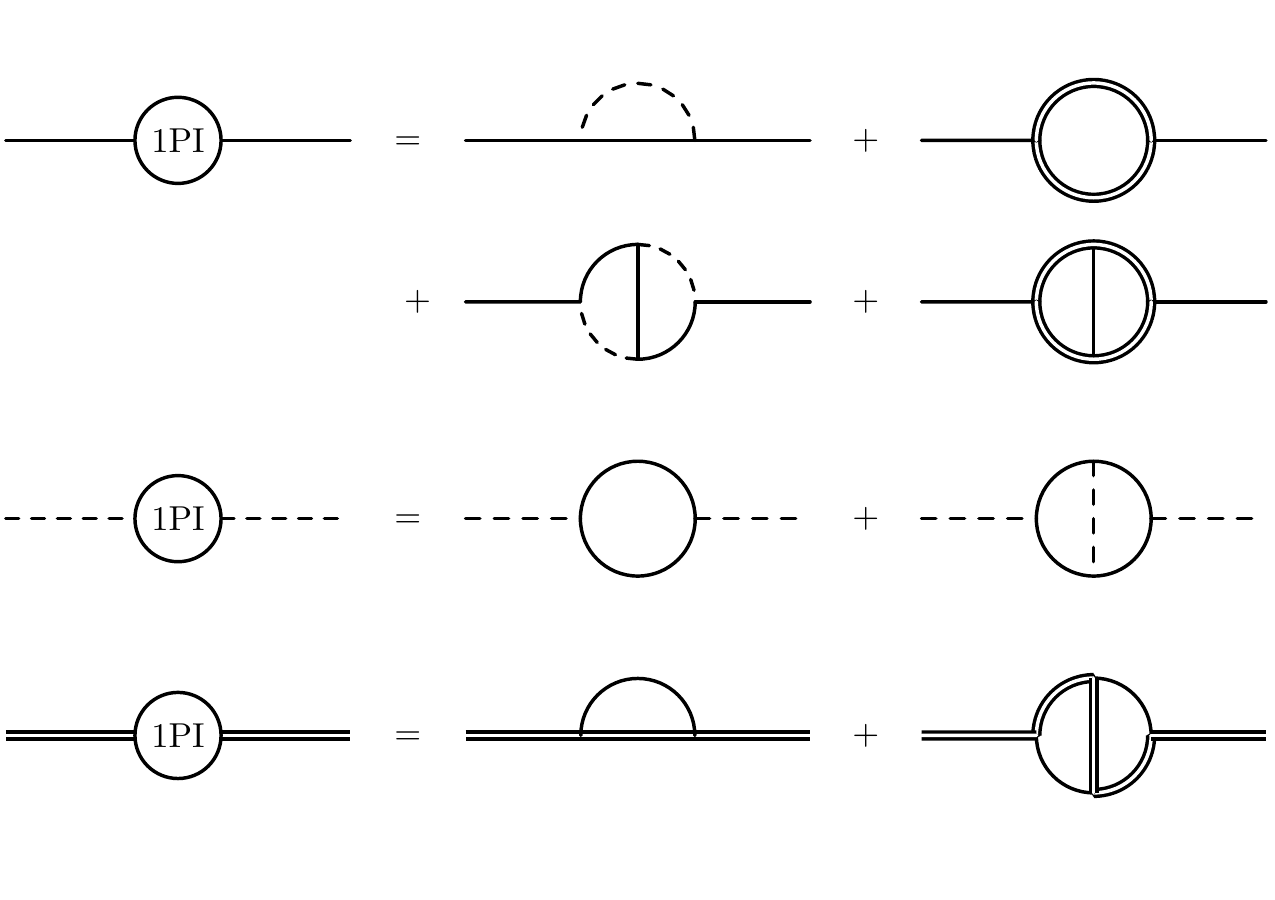}
\caption{ The two-loop truncated Schwinger-Dyson equations with bare cubic coupling. The solid, dashed, and double lines represent the boson propagator $\Delta_n$, the auxiliary field propagator $\epsilon_n$, and the fermion propagator $g_r$, respectively. }
\label{fig:sd2loop}
\end{center}
\end{figure}

Let us look at the two-loop truncated Schwinger-Dyson equations, which are shown diagrammatically in Figure~\ref{fig:sd2loop}.  As it turns out, the two-loop diagrams contribute an additional term\footnote{ The log comes from cutting off the harmonic sum at $\sqrt{\B / \sigma_0}$. }
\ie
{5 \sigma_0 \over \pi} \left( {1 \over 1 + \sqrt3} - {1 \over 1 + \sqrt5} \right) \left( {\B\sigma_0 \over 2} \right)^{3/2} \ln {\B \over \sigma_0} \sim \cO(\B^{-2} \ln \B)
\fe
to the RHS of the Schwinger-Dyson equation for $\sigma_0$ (\ref{sigma0SD}). This is, up to a logarithmic factor, the same as the next-to-leading order term in the one-loop contribution. However, as seen earlier, the scaling solution to the one-loop S-D equations requires complete cancelation at next-to-leading order, and $\sigma_0\sim \beta^{-7/5}$ would be determined only at next-to-next-to-leading order. This is now spoiled by the two-loop contribution. This means that either the ansatz (\ref{soft}) is incorrect, or that $x_n$ is at least of order $\cO(1)$ which invalidates the approximate SUSY Ward identities (\ref{approxward}). So it appears that the two-loop truncated S-D equation of the ${\cal N}=2$ Wess-Zumino quantum mechanics does not give rise to a simple scaling solution in the low temperature limit.

\begin{figure}[htb]
\begin{center}
\begin{tabular}{cc}
\vspace{.15in} & \\
\includegraphics[width=0.2\textwidth]{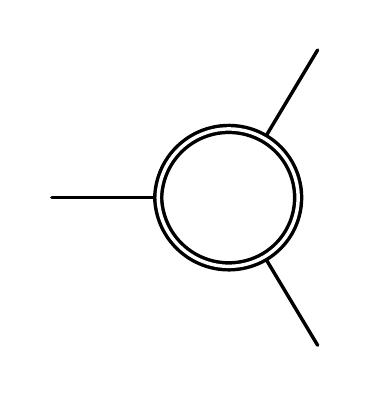} \hspace{.2in} & \hspace{.2in}
\includegraphics[width=0.2\textwidth]{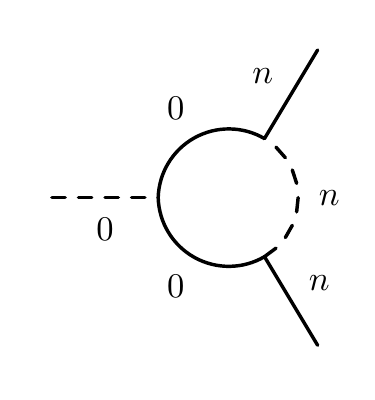} \\
(a) \hspace{.2in} & \hspace{.2in} (b) \\
\vspace{.05in}
\end{tabular}
\caption{ Diagrams that appear in the pertubative expansion of the exact three-point vertex. }
\label{fig:3point}
\end{center}
\end{figure}

Inclusion of three- and higher-loop diagrams would give corrections that even overwhelm the one-loop leading order terms in the low temperature limit, and hence completely invalidates the ansatz (\ref{soft}).  Figure~\ref{fig:3point} shows two diagrams appearing in the pertubative expansion of the exact three-point vertex that are larger than their tree-level counterparts.  Diagram~(a) does not exist at tree-level, and Diagram~(b) is of order
\ie
\B^{-3/2} (\Delta_0)^2 \epsilon_n \sim \B^{-3/2} (\B^{7/5})^2 \B^{-6/5} \sim \B^{-1/2} \B^{3/5},
\fe
which is $\B^{3/5}$ larger than tree-level ($\B^{-1/2}$).  It is possible that a scaling behavior of the type we found still holds in the full theory at large $N$,\footnote{We will comment on the relevance of the large $N$, i.e. planar, limit in the next section.} but if so, an alternative truncation of the Schwinger-Dyson equation would be needed to reveal the correct low temperature scaling beyond the one-loop result.

\section{$\cN = 4$ Wess-Zumino Matrix Quantum Mechanics}
\label{sec:N=4}

The $\cN = 4$ Wess-Zumino matrix quantum mechanics, whose one-loop truncated Schwinger-Dyson equations take an identical form as those of the $\cN = 2$ theory, on the other hand does not receive two-loop contributions, nor planar three-loop contributions.
Moreover, the perturbative expansion of the exact three-point vertex only contains diagrams that are at most of the same order as the leading order at one-loop.
It is therefore conceivable that one-loop truncated S-D equations serve as a better approximation of the ${\cal N}=4$ model. An $\cN=4$ Wess-Zumino matrix quantum mechanics can also be obtained as a truncation (or deformation) of the BFSS matrix model.  In the $\cN = 4$ language, the BFSS matrix model consists of one vector multiplet and three chiral multiplets,\footnote{ Since 1D $\cN = 4$ is the dimensional reduction of 4D $\cN = 1$, we name the supermultiplets using the 4D language. } with an $SU(3)$ invariant cubic potential.

\subsection{Action, Schwinger-Dyson Equations, and Free Energy}

Consider an $\cN = 4$ Wess-Zumino matrix quantum mechanics with a flavor symmetry group $G$. Suppose the chiral multiplets are in an $N_f$-dimensional representation of $G$, governed by a cubic superpotential of the same form as in explained at the beginning of Section~\ref{subsec:N=2action} (except that the real ${\cal N}=2$ matter multiplets are now replaced by the holomorphic ${\cal N}=4$ chiral multiplets).  For the $\cN = 4$ WZ model coming from the matter sector of BFSS, $G=SU(3)$ and $N_f = 3$, $c = 2$. Our conventions for the 1D $\cN = 4$ superspace are defined in Appendix~\ref{app:N=4}.  We have $N_f$ chiral superfields
\ie
\Phi^a = \phi^a(y) + \sqrt2 \psi^a_\A(y) \theta^\A + i f^a(y) \theta^2  \label{Phi4}
\fe
in the adjoint representation of $SU(N)$ and in the fundamental representation of the flavor symmetry $G$. The superpotential is given by
\ie
W = - {i \over 6 \sqrt c} \kappa \epsilon^{abc} \Tr \left( \Phi^a [\Phi^b, \Phi^c] \right),
\fe
and the action is
\ie
S &= \int d\tau ~ \Tr \bigg\{ \dot{\overline \phi^a} \dot \phi^a + \bar\psi^{a\A} \dot \psi^a_\A + \bar f^a f^a \\
& \hspace{1.5in} + \left({\kappa \over 2\sqrt c} \epsilon^{abc} f^a [\phi^b, \phi^c] + {i\kappa \over 2\sqrt c} \epsilon^{\A\B} \epsilon^{abc} \phi^a [\psi^b_\A, \psi^c_\B] + \text{h.c.} \right) \bigg\}.
\label{SN=4}
\fe

Let us write the finite temperature propagators in momentum (frequency) space as
\ie
\la \bar\phi^a_n \phi^b_m \ra &\equiv \Delta_n \delta^{ab} \delta_{n, -m}, \\
\la \bar\psi^{a\A}_{r} \psi^b_{\B, s} \ra &\equiv -i  g_r \delta^{ab} \delta^\A_\B \delta_{r, -s}, \\
\la \bar f^a_n f^b_m \ra &\equiv \epsilon_n \delta^{ab} \delta_{n,-m}  \label{prop4}
\fe
The self-energies are defined as in (\ref{prop2self}).  The $\cN = 4$ SUSY Ward identities, given in (\ref{N=4susyward}), take an identical form as the $\cN=2$ identities (\ref{ward}).  We have normalized the coupling constant $\kappa$ in (\ref{SN=4}) such that the one-loop truncated Schwinger-Dyson equations are identical to the $\cN = 2$ ones (\ref{SDself}), whose solution is given in Section~\ref{subsec:bosonzm}.

The free energy can also be computed to be
\ie
\B F = \text{const} - {5 \over 2} ({\pi \over 3})^{6/5} N^2 N_f \B^{-6/5} + {\cal O}(\B^{-9/5}).
\fe
which is twice that of the $\cN = 2$ free energy (\ref{BF}) because the fields are now complex. To restore the dependence on $\kappa^2$, one replaces $\B$ by the dimensionless combination $\B (\kappa^2N)^{1/3}$.

\subsection{Higher Loops}
\label{subsec:higherloops}

\begin{figure}[htb]
\begin{center}
\begin{tabular}{cc}
\vspace{.15in} & \\
\includegraphics[width=0.45\textwidth]{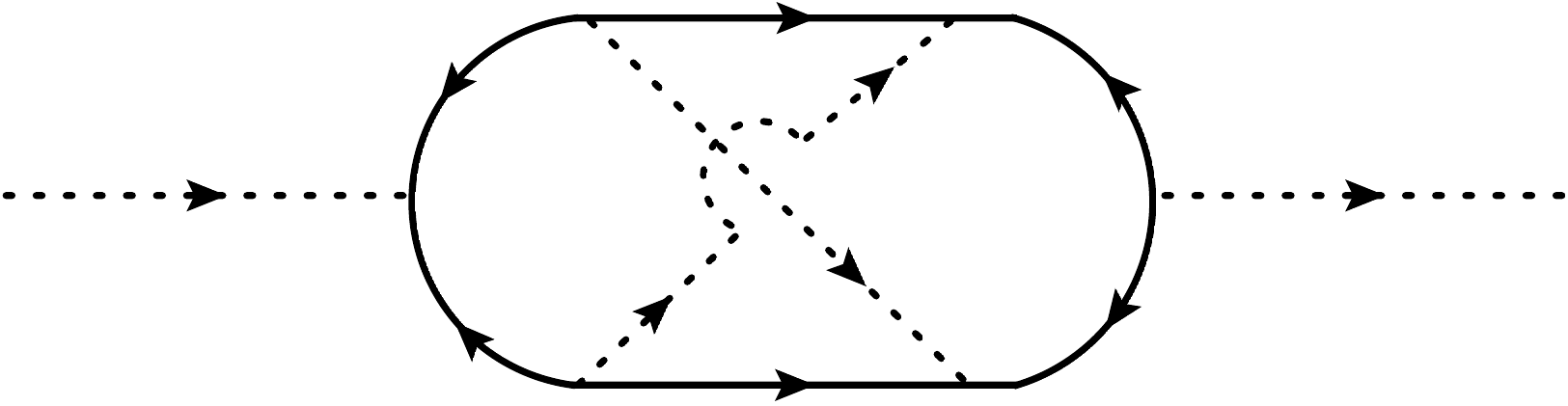} \quad & \quad
\includegraphics[width=0.45\textwidth]{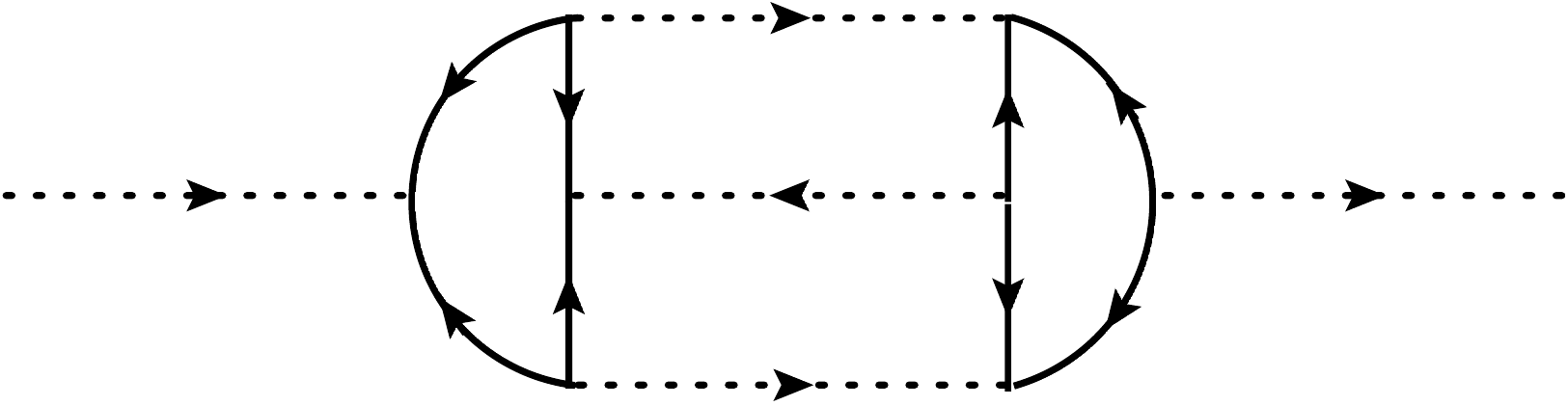} \\
& \\
(a) \quad & \quad (b) \\ \vspace{.05in}
\end{tabular}
\caption{Corrections to the auxiliary self-energy $\eta_n$ from (a) a non-planar three-loop diagram and (b) a planar four-loop diagram.}
\label{fig:34loop}
\end{center}
\end{figure}

A key difference between the $\cN = 2$ and $\cN = 4$ Wess-Zumino quantum mechanics is the holomorphicity of the superpotential. 
In the Feynman diagrams of the ${\cal N}=4$ model, propagators carry arrows since the fields are complex, and vertices have either all three propagators pointing inwards or all pointing outwards; in other words, the Feynman diagrams are bipartite graphs.
It is not hard to see that no two-loop or {\it planar} three-loop diagrams can appear on the RHS of the Schwinger-Dyson equations for the self-energies.  Thus without taking the planar limit, the first nontrivial corrections come from non-planar three-loop diagrams, and in the planar limit, the first nontrivial corrections come from four-loop diagrams.  Examples of these non-planar three-loop and planar four-loop diagrams are shown in Figure~\ref{fig:34loop}.  

\begin{figure}[htb]
\begin{center}
\includegraphics[width=0.7\textwidth]{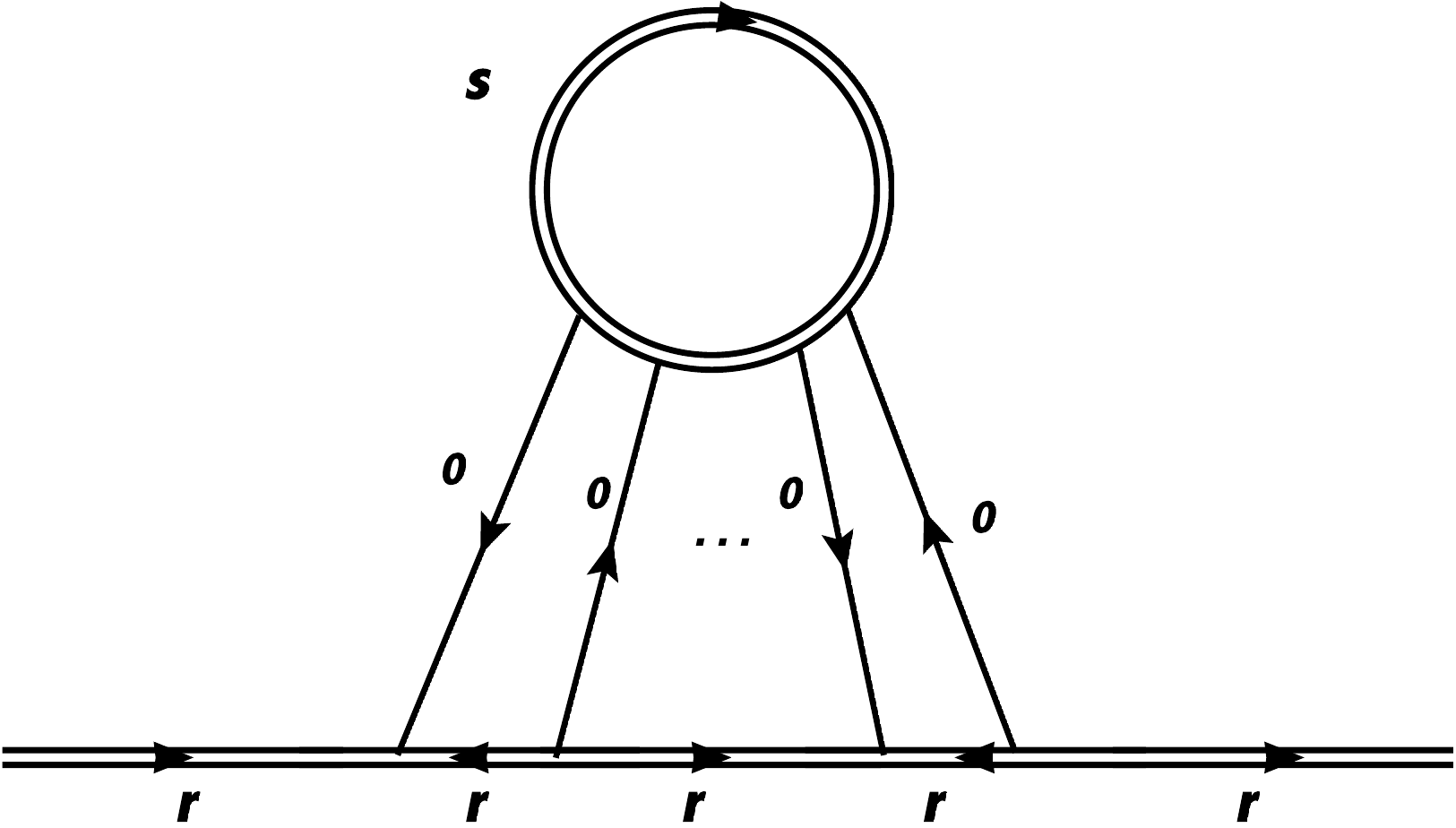}
\caption{A leading order correction to the fermion self-energy $h_r$ from a planar $L$-loop diagram ($L$ has to be even).  There are $L$ boson zero mode propagators (solid line).}
\label{L}
\end{center}
\end{figure}

Although we have not explicitly evaluated the four-loop contributions and its effect on the scaling solution in the low temperature limit, a naive counting of powers of propagators of zero modes versus nonzero modes based on the one-loop scaling ansatz suggests that the higher loop contributions are of the same order as the one-loop contribution. While such contributions are expected to correct the coefficient of the self-energies and the free energy, it is conceivable that the scaling behavior found in the solution to the one-loop truncated equations continue to hold when all-loop contributions are included, at the planar level.
The large $N$ limit is potentially important here, in order for the loop expansion in the Schwinger-Dyson equation to converge.  At $L$-loop order, there are roughly $L!~\cO(C^L)$ diagrams that contribute to the Schwinger-Dyson equations at leading order in $1/\B$, but only $\cO(C'^L)$ {\it planar} diagrams that do so, where $C$ and $C'$ are some constants \cite{Shenker:1990uf}.  For example, Figure~\ref{L} shows an $L$-loop correction to the fermion self-energy $h_r$.  Assuming the scalings in (\ref{soft}), the scaling of this diagram is
\ie
\B^{-L} g_r^{L-1} \Delta_0^{L} g_s^L
\sim \B^{-L} \B^{(-L +1) / 5} \B^{7L / 5} \B^{-L / 5} \sim \B^{1/5},
\fe
which indeed is the leading order in the Schwinger-Dyson equation for $h_r$.  There are two such planar diagrams, coming from the two (alternating) assignments of the R-symmetry index $\alpha = 1, 2$ at the vertices in the fermion loop.  On the other hand, there are $2 (L/2)! (L/2-1)!$ non-planar diagrams, where the extra $(L/2)! (L/2-1)!$ comes from inequivalent permutations among the boson propagators.  While summing over all non-planar diagrams gives an asymptotic series, restricting to planar diagrams may give rise to a convergent sum.

We conjecture that the $\B^{-6/5}$-scaling of the free energy is preserved to all loop orders for the $\cN=4$ Wess-Zumino quantum mechanics in the planar limit, with only the coefficient to be corrected. Further, it is possible that such scaling behavior holds when perturbative $1/N$ corrections are included as well, and is only violated by non-perturbative effects in $1/N$. One reason to anticipate this is the existence of exponentially long lived metastable states in BFSS matrix quantum mechanics, which dominate the thermodyamics as predicted by the gravity dual. As explained in the introduction section, the scaling behavior cannot possibly hold for the exact free energy in the general Wess-Zumino quantum mechanics at finite $N$: depending on the precise matter content and the form of the superpotential, the model could be either gapped, in which case the free energy is exponentially suppressed in the very low temperature limit (for fixed $N$), or infinite, when there is a moduli space (flat directions of the superpotential). Nonetheless, in the infinite $N$ limit, we expect that there is at least a phase of the theory in which the free energy exhibits the nontrivial power scaling in the temperature, as we found from the one-loop Schwinger-Dyson equation. In ${\cal N}=4$ theories where the superpotential is such that the spectrum is gapped at finite $N$, we expect a near-continuum is developed in the large $N$ limit, and if we first take $N\to \infty$, and then take the temperature to be small in 't Hooft units, the free energy should exhibit the scaling behavior. In ${\cal N}=4$ theories where the superpotential has flat directions, there are alternative, and singular, solutions to the Schwinger-Dyson equation, where the self-energies of the bosonic fields corresponding to the flat directions are set to zero, while the self-energies of the fields along the non-flat directions diverge. Such singular solutions may be regularized by introducing an IR regulator that cuts off the volume divergence in the target space. In such theories, at infinite $N$, we expect our scaling solution to describe an ``unbroken" phase (the flavor group is unbroken), as opposed to the ``broken" phase describing the flat directions of the potential. The different phases correspond to different solutions to the Schwinger-Dyson equations.

\section{Towards the Low Temperature Expansion of BFSS}
\label{sec:BFSS}

In the previous sections we have studied the low temperature expansion of the $\cN=2$ and $\cN=4$ Wess-Zumino quantum mechanics. To extend our results for the Wess-Zumino quantum mechanics to BFSS matrix theory, we need to couple the vector multiplet to it and perform the appropriate supersymmetric gauge fixing, which also introduces the ghost multiplet. In this section we discuss some preliminary results toward understanding the planar free energy of BFSS.

%
%

\subsection{$\cN = 2$ Gauge-Fixed BFSS}


The BFSS action can be decomposed into one $\cN=2$ vector multiplet $\Gamma_\A$ plus seven matter multiplets $\Phi^a, \, a=1,\cdots,7$ in the fundamental representation of the $G_2$ flavor symmetry, 
\ie
S_{vec}+S_{matter}=& 
{1\over g_{YM}^2 }\int d\tau d^2\theta \Tr \Big\{ 
- {1\over 4} \left( { i \over 2} \epsilon_{\alpha\beta} \right)  
\nabla_\alpha \mathcal{F}^i \nabla_\beta \mathcal{F}^i \Big\}\\
& \hspace{.5in} +
 \int d\tau d^2\theta \, 
\Tr \left(  -{i\over 4}\epsilon_{\alpha\beta} \nabla_\alpha \Phi^a \nabla_\beta \Phi^a -{ig_{YM}\over 3} \epsilon^{abc} \Phi^a[\Phi^b,\Phi^c]\right),
\label{N=2vecmat}
\fe
where ${\cal F}^i$ is the field-strength superfield for the real connection superfield $\Gamma_\A$.  See Appendix \ref{app:N=2} for our notations.

The $\cN=2$ gauge-fixing condition is
\ie
\epsilon_{\A\B}  D_\A \Gamma_\B = 0,\label{N=2gf}
\fe
If we expand the real connection superfield $\Gamma_\A$ as
\ie
\Gamma_\A=\chi_\A + iA_0 \theta_\A
+X^i \gamma^i_{\A\B}\theta_\B
+d \epsilon_{\A\B} \theta_\B +2\epsilon_{\A\B}\lambda_\B \theta^2.
\fe
(\ref{N=2gf}) becomes
\ie\label{ntcomp}
\dot A_0=0, \quad d= 0, \quad \lambda_\A ={i\over 2}\dot\chi_\A.
\fe
The $\cN=2$ gauge-fixing introduces a ghost multiplet 
\ie
C=\A + \B_\A\theta_\A +i \gamma\theta^2
\fe
with $\A$ and $\gamma$ complex fermonic fields and $\B_\A$ a complex bosonic field.  The ghost action is
\ie
S_{ghost} = \int d\tau d\theta^2 ~ \Tr \bigg\{ {i \over 2} \epsilon_{\A\B} D_\A \bar{C} \nabla_\B C \bigg\}.
\label{N=2ghost}
\fe
(\ref{N=2vecmat}) together with (\ref{N=2ghost}) gives the complete BFSS action in the $\cN=2$ gauge (\ref{N=2gf}).

\subsection{The ${\cal N}=2$ Ghost Determinant}

In the usual Wess-Zumino gauge-fixing, which breaks all supersymmetries, the eigenvalues of the zero mode of $A_0$ have the $2\pi$ periodicity coming from the large gauge transformation. Or equivalently, the eigenvalues of the holonomy matrix $U = \exp \left(\oint d\tau A_0\right)$ are gauge invariant. However, there is no such periodicity for the $\cN=2$ gauge-fixing. To see this, let us note the coupling between $X^i$, $\B_\A$, and the gauge field $A_0$ in the BFSS action,
\ie
S_{vec} &\ni(\partial_\tau X^i + i [A_0,X^i] )^2,\\
S_{ghost} &\ni\bar\B_\A \left( \dot\B_\A +  {i\over 2}  [A_0,\B_\A] \right) + {1\over2} \gamma^i_{\A\B} \bar\B_\A [X^i ,\B_\B],
\fe
where $\gamma^1=\sigma_1,\gamma^2 = \sigma_3$. Note the curious $1/2$ factor in the coupling between the $\B$-ghost and the gauge field. Suppose we shift one of the eigenvalues of $A_0$ by $4\pi$, then this can be undone by shifting the momentum modes of $X^i$ and $\B_\A$ by 2 and 1, respectively. However, the coupling between $X^i$ and $\B_\A$ is not invariant under this shift, so the $1/2$ factor in the gauge coupling of $\B_\A$ completely destroys the periodicity of the eigenvalues of $A_0$.

In fact, the ${\cal N}=2$ gauge condition (\ref{N=2gf}), (\ref{ntcomp}) leaves no residual large gauge transformations (except for the constant gauge rotation). First consider the zero temperature case. Recall the infinitesimal ${\cal N}=2$ gauge transformations form Appendix A.2.1, generated by $\Lambda = \Omega + i\omega_\A \theta_\A + W\theta^2$. The ones that preserve the gauge condition (\ref{N=2gf}) obey
\ie
0 &= {d\over d\tau}\left(\dot\Omega + i[A_0,\Omega] + {i\over 2}\{\chi_\A, \omega_\A\}\right),
\\
0 &= -W - {1\over 2}\epsilon_{\A\B}\{\chi_\A, \omega_\B\},
\\
0 &= \dot\omega_\A + {i \over 2} [A_0, \omega_\A] + {1 \over 2}\epsilon_{\A\B}[\chi_\B, W] - [\lambda_\A, \Omega] + {i\over 2}{d\over d\tau}[\chi_\A,\Omega] - {1 \over 2}\gamma^i_{\A\B} [X^i, \omega_\B].
\fe
The last equation can also be written as
\ie
0 &= \dot\omega_\A + {i \over 2} [A_0, \omega_\A] + {1 \over 2}\epsilon_{\A\B}[\chi_\B, W] + {i\over 2}[\chi_\A, \dot\Omega] - {1 \over 2}\gamma^i_{\A\B} [X^i, \omega_\B].
\fe
For generic $\chi_\A$, these conditions are satisfied only if $\omega_\A=0$, $W=0$, and $\dot\Omega=0$. Therefore, the residual unfixed gauge transformations are constant bosonic gauge rotations. This is in contrast with the non-supersymmetric gauge fixing (Wess-Zumino gauge together with $\dot A_0=0$) where $\Omega$ can be linear in time $\tau$, provided that it also commutes with $A_0$. In the finite temperature case, the finite gauge transformations must be such that the bosonic components of $e^\Lambda$ is periodic on the Euclidean time circle while the fermionic components are anti-periodic. While in the case of non-supersymmetric gauge fixing, there are residual large gauge transformations of the form $e^{i\zeta \tau}$ with $[A_0,\zeta]=0$ and $e^{i\beta\zeta}=1$, in the case of ${\cal N}=2$ gauge fixing the only residual gauge transformations are $e^\Omega$ with constant bosonic $\Omega$. Therefore, in the latter case, the eigenvalues of $A_0$ are {\it not} periodically identified, and in the path integral we must integrate over all range of $A_0$.

If we set $\chi_\A$ to zero, the ${\cal N}=2$ ghost determinant as a functional of (constant) $A_0$ and $X^i$ is
\ie
{\det{}'(\partial_\tau + i[A_0,\cdot])\over \det\left( (\partial_\tau + {i\over 2}[A_0,\cdot])\delta_{\A\B} + {1\over 2} \gamma^i_{\A\B} [X^i,\cdot]\right)}.
\fe
As already mentioned, if we further restrict to $X^i=0$, this ghost determinant is invariant under the shift of eigenvalues of $\beta A_0$ by $4\pi$, but not by $2\pi$. The path integral measure in $A_0$ can be written in terms of the integration over its eigenvalues $\A_j$ ($j=1,2,\cdots,N$) as
\ie
\int \prod_{i=1}^N d\A_i \prod_{i<j} {\sin^2(\A_{ij}/ 2) \over \cos^4(\A_{ij}/4)} .
\fe
In the large $N$ limit, in terms of the eigenvalue distribution function $\rho(\A)$ (whose integral is normalized to 1), the effective potential for $\A$ due to the ghost determinant is (when $\chi_\A$ and $X^i$ are restricted to zero) $N^2$ times
\ie
& {1\over 2}\int d\A d\A' \rho(\A)\rho(\A') \left[ -\ln \sin^2\left({\A-\A'\over 2}\right) + 2\ln\cos^2\left({\A-\A'\over 4}\right) \right]
\\
& \hspace{.5in} = \sum_{n=1}^\infty {\rho_n^2\over n} - 2 \sum_{n=1}^\infty {(-1)^n \rho_{n/2}^2\over n} = \sum_{n=0}^\infty {\rho_{n+{1\over 2}}^2\over n+{1\over 2}},
\fe
where 
\ie
\rho_n \equiv \int d\A \rho(\A) \cos(n\A).
\fe
Note that the coupling of matter multiplets to the vector multiplet still respects the shift of $\beta A_0$ by $2\pi$, and consequently, the effective action due to integrating out matter multiplets should depend only on $\rho_n$ for integer $n$ and not $\rho_{n+{1\over 2}}$. To minimize the effective action, one would set $\rho_{n+{1\over 2}}$ to zero for all integer $n$, and effectively the eigenvalues no longer repel one another. This is rather surprising. Let us note cautiously, however, that we have essentially ignored the scalar $X^i$ and fermion $\chi_\A$ in the vector multiplet, in the above discussion. Integrating them out could have important consequences on the effective action of $A_0$.

\subsection{The Free Energy}

To obtain the free energy for BFSS, one needs to incorporate the vector sector in a consistent manner.  For example, one could try to derive the effective action for the vector multiplet by integrating out the matter multiplets, using the solutions to the one-loop truncated Schwinger-Dyson equations of the Wess-Zumino quantum mechanics, around a certain background of $A_0$. It turns out that the Schwinger-Dyson equations are not particularly useful in solving the vector sector with the effective potential $V(A_0)$ generated by the matter sector. The difficulty lies in that, the soft collinear approximation doesn't hold for the vector and ghost multiplet. Nonetheless, we can infer some properties of $V(A_0)$ from our results. 

Assuming that the matter sector has the same scaling behavior as in the ${\cal N} = 2$ Wess-Zumino matrix quantum mechanics, integrating out the matter sector to leading order in $\B^{-1}$ generates an effective mass term for the vector multiplet
\ie
m^2_V \int d\tau ~ \Tr \left( A_0^2 + (X^i)^2 + \chi_\A \dot\chi_\A \right), \quad m_V^2 \approx {1 \over \B\sigma_0} \sim \B^{2/5}.
\label{vmass}
\fe

In the more familiar case of non-supersymmetric gauge fixing, while typically the effective potential for $A_0$ drives the eigenvalues of $A_0$ toward the origin, there is a competing repelling effect between the eigenvalues due to the ghost determinant (or the measure in the integration over the eigenvalues). Naively, the effective mass $m_V$ would be too small to overcome the repelling of the eigenvalues. However, as seen in the previous subsection, in the ${\cal N}=2$ gauge fixed path integral, the eigenvalues of $A_0$ in fact do not repel, as far as the coupling to matter multiplets is concerned, as the latter only knows about the eigenvalues of $A_0$ up to shifts by $2\pi/\beta$. It then seems plausible that the free energy of the vector multiplet coupled to matter multiplets is dominated by the contribution from the matter multiplets alone, with the vector multiplet turned off. If this is the case, the free energy of BFSS matrix quantum mechanics would have the same low temperature scaling as the truncated Wess-Zumino quantum mechanics, and we would have $\beta F \sim - C T^{6/5}$. We do not know the source of the discrepancy between this $6/5$ scaling exponent and the $9/5$ as predicted from the gravity dual. Clearly, a better understanding of the dynamics of the gauge multiplet, through perhaps a better set of truncated Schwinger-Dyson equations that do capture the correct low temperature behavior, is needed.

Let us also note that, unlike the typical large $N$ gauged matrix models with adjoint matter, where one expects a Gross-Witten-Wadia phase transition \cite{Gross:1980he, Wadia:1980cp} going from high temperatures to low temperatures, such a transition should be absent in the planar limit  of BFSS matrix quantum mechanics, because there is no Hawking-Page transition of the D0 black hole in the gravity dual, for temperatures of order 1 in units of the 't Hooft coupling.

Previously, the authors of \cite{Kabat:1999hp,Kabat:2000zv,Kabat:2001ve} studied the Schwinger-Dyson equations of this $\cN = 2$ gauge-fixed BFSS model (presented in Appendix~\ref{app:N=2gap}), and produced numerical results for the free energy.  A $\B F \sim \B^{-1.7}$ scaling was found there, which appeared to be close to the $\B^{-1.8}$ predicted by the dual black hole.  However, this scaling was obtained by fitting the numerical results within the inverse temperature range $1 < \B < 4$. If one modifies the fitting range to, say, $0.25 < \B < 4$, the result already changes drastically.  In addition, the numerics break down at $\B \sim 6$, essentially due to the ghost multiplet.  In fact, the Schwinger-Dyson equations for just the pure gauge system (\ref{N=2muSD}), (\ref{N=2ghostSD}) do not admit a real solution due to the negative signs in the equations for $\Pi^G_n$ and $\Sigma^G_r$.  When $\B > 6$, the terms from coupling to matter become smaller than the pure gauge terms, and it appears that no real solution exists in this regime. Further, it is unclear to us why \cite{Kabat:2000zv,Kabat:2001ve} treated the gauge field as a holonomy matrix model after gauge-fixing {\it and} including the ghosts in the S-D equations: such a treatment would seem to double-count the ghost determinant. Though, this double-counting turns out to have little effect on the numerical result for the free energy computed from the S-D equations in the temperature range considered in \cite{Kabat:2000zv,Kabat:2001ve}.

\subsection{$\cN = 4$ Gauge-Fixing}

As mentioned before, the $\cN = 4$ Wess-Zumino matrix quantum mechanics with $SU(3)$ flavor can be regarded as a deformation of BFSS.  In ${\cal N} = 4$ language, the BFSS action can be written as
\ie
\mathcal{L}_{BFSS} = \Tr&\Bigg[
\left(-{1\over 4g_{YM}^2} \epsilon^{\A\B} W_\A W_\B \Big|_F+h.c.\right)
+\bar\Phi^{a} e^{-2 V} \Phi^a \Big|_D
\\
&
\hspace{1in} +\left(-{i\over 6\sqrt{2}} g_{YM} \epsilon^{abc} \Phi^a [\Phi^b,\Phi^c]\Big|_F+h.c.\right)
\Bigg], \label{bfss}
\fe
where $V$ is a real superfield representing the vector multiplet and $W$ is the field strength.  The advantage of writing the BFSS action in ${\cal N} = 4$ language is that we can impose a manifestly $\cN = 4$ supersymmetric gauge fixing condition
\ie
\epsilon^{\dot\A\dot\B}\bar {\cal D}_{\dot\A}\bar {\cal D}_{\dot\B} V=0.\label{N=4gf}
\fe
where $V$ is the real superfield representing the vector multiplet. $\bar{\cal D}_{\dot \A} $ is defined as
\ie
\bar{\cal D}_{\dot \A} = - {\partial\over \partial \bar\theta^{\dot \A}} -\theta^\gamma \delta_{\gamma\dot \A} \partial_\tau
\fe

Expanding the real superfield $V(\tau,\theta,\bar\theta)$ as
\ie
V(\tau,\theta,\bar\theta) &= R(\tau) + \theta \chi (\tau)+\bar\theta \bar\chi(\tau) +\theta\theta M (\tau)+ \bar\theta \bar\theta \bar M(\tau)\\
& \hspace{.5in} +\theta^\A \bar\theta^{\dot \B}\left(i
\delta_{\A\dot \B} A_0 (\tau)+ \sigma^{i}_{\A\dot \B} X^i (\tau)
\right)
 +\theta\theta\bar\theta \bar\lambda(\tau)
+\bar\theta\bar\theta \theta \lambda(\tau)
+{1\over 2} \theta\theta\bar\theta\bar\theta D(\tau),
\fe
then (\ref{N=4gf}) implies
\ie
&M=\bar M=0,\\
&\lambda_\A = -{1\over 2} \delta_{\A\dot \B} \dot{\bar \chi}^{\dot \B}, \quad \bar\lambda_{\dot \A} = {1\over 2} \delta_{\B\dot \A} \dot\chi^\B\\
&D=  \dot A_0-{1\over2}\ddot R.
\fe
Note that since the lowest component field $R(\tau)$ is not zero in the $\cN=4$ supersymmetric gauge-fixing condition (\ref{N=4gf}), there are infinitely many terms involving $R(\tau)$ in the action.

For now, we wish to decouple the vector multiplet and only study the matter (chiral) multiplets described by the $\cN = 4$ Wess-Zumino quantum mechanics.  This can be done by deforming the BFSS action with a superpotential $\delta W = {-i \over 6 \sqrt{2}} \kappa \epsilon^{abc} \Tr( \Phi^a [\Phi^b ,\Phi^c])$ and taking $\kappa / g_{YM} \to \infty$.  Note that this deformation is in the same supermultiplet as $\Tr(\Phi^a \bar\Phi^{a})$, which is presumably dual to a stringy mode in the bulk.  Hence the gravity dual of the deformed theory has a very stringy description.

\subsection{Phases of BFSS}

There can be different solutions to the one-loop truncated Schwinger-Dyson equations. In Section \ref{sec:N=4}, we have only considered the solution to the ${\cal N}=4$ matter sector WZ model that respects the $SU(3)$ flavor symmetry. To obtain a solution where the $SU(3)$ flavor symmetry is spontaneously broken, we can take the zero mode self-energy $\sigma_0$ of, say, $\Phi^1$ to be very small, which corresponds to $\Phi^1$ being massless and can acquire a large vacuum expectation value. This in turn generates large masses for $\Phi^2$ and $\Phi^3$.  We can then integrate out $\Phi^2$ and $\Phi^3$ and obtain some effective action for $\Phi^1$. In the effective action there are scattering states described by wave packets at large $\Phi^1$, that becomes just that of the free theory of one chiral superfield $\Phi^1$. The free energy is then given by $N^2$ times the volume divergence from the scattering states, which is certainly different from the $\B$-scaling we found in Section~\ref{sec:N=4}. As already discussed, we expect these different solutions to the planar Schwinger-Dyson equation to describe different phases of the theory at infinite $N$; the tunneling from one phase to another, which resembles the Hawking radiation of D0 black holes in the gravity dual of BFSS matrix theory, is expected to be exponentially suppressed in the large $N$ limit.

When the matter WZ model is coupled to the gauge multiplets, the $\sim N^2$ flat directions are reduced to $\sim N$ flat directions, due to gauge symmetry. From the perspective of the planar Schwinger-Dyson equation, there is no longer the singular solution that breaks the $SU(3)$ flavor symmetry, as the loops of gauge multiplets contribute to the self-energy of all matter chiral multiplets. Given that the planar free energy is supposed to capture metastable microstates of the D0 black hole, an intriguing question is whether one can study quantitatively the tunneling amplitudes which are non-perturbative in $1/N$, in the framework of the Schwinger-Dyson equations. This is left for the future.


\section{The High Temperature Limit}

\label{sec:highT}
Compared to the low temperature analysis of BFSS in the previous sections, the Schwinger-Dyson equations in the high temperature limit are much simpler to deal with, because the nonzero modes are weakly coupled (the zero modes remain strongly coupled). Prior work in this aspect includes \cite{Kawahara:2007ib}, where the high temperature behavior of BFSS was studied by first dimensionally reducing to IKKT matrix model, and then utilizing Monte Carlo method to extract the two point functions. Here, we shall employ the one-loop truncated Schwinger-Dyson equations to perform the integrating out procedure.

While the low temperature limit of BFSS matrix model describes a black hole in the bulk, its high temperature limit corresponds to the stringy regime in the dual theory. Therefore the high temperature analysis should tells us about the stringy black hole in the bulk. Note that while the horizon size of such stringy black holes are small in units of the string length, they are large with respect to the Planck length determined by the local string coupling (which goes to zero at large radii).

For the following analysis, we will work with the $\mathcal{N}=2$ gauge fixing, solve the one-loop truncated Schwinger-Dyson equations in the $\B\ll1$ limit pertubatively, while keeping the next to leading corrections to the self-energies of both zero and nonzero modes (the appropriate expansion parameter in the high temperature turns out to be $\B^{3/2}$), and obtain the free energy up to $\mathcal{O}(\B^3)$. Note that to obtain the next to leading correction to the free energy, we only need to know the $\B^{3/2}$ corrections for the zero mode self-energies. We shall also calculate the ``size" of the bound state $\la R^2\ra_\B$, and compare with the Monte Carlo result from \cite{Kawahara:2007ib}. 

\subsection{Schwinger-Dyson Equations}
At leading order, the Schwinger-Dyson equations reduce to those involving only the zero modes, and those which determine the nonzero mode self-energies from the zero mode self-energies. The equations for zero modes at leading order are
\ie
\Pi^M_0&={4\over\B\Pi_0^V}+{2\over\B\mu}+{12\over\B\Pi^M_0(\Xi^M_0+1)},
\\
\Pi^V_0&={2\over\B\Pi_0^V}+{14\over\B\Pi^M_0}+{2\over \B\mu},
\\
\Xi^M_0&={6\over\B({\Pi^M_0})^2},
\\
{\mu \over 2} &= {2\over \beta \Pi^V_0} +{7\over \beta\Pi^M_0},
\fe
whose solution is given by 
\ie
\Pi_0^V=\mu={b\over \B^{1/2}},~~\Pi_0^M={a\over\B^{1/2}},~~\Xi^M_0=d,
\label{highTzm}
\fe
with
\ie
b=c=4.78511, \quad a={14\over b-4/b}=3.54504, \quad d={6\over a^2}=0.477429.
\label{highTN[zm]}
\fe
Then one can readily obtain the leading self-energies for the nonzero modes

\ie
\begin{array}{llll}
{\rm Boson:} & ~~ \Pi^M_n\simeq (\frac{6 b}{7}-\frac{38}{7 b})\B^{-1/2} , & ~~ \Pi^V_n\simeq(b-{8\over b})\B^{-1/2}, & ~~ \Xi^M_n\simeq{3\over \pi^2 n^2 a}\B^{3/2},
\\
{\rm Fermion:} & ~~ \Sigma_r^V\simeq{2\pi r\over \B}(b-{13\over 2 b})\B^{-1/2}, & ~~ \Sigma^M_r\simeq{\B\over 2\pi r}{6\over 7}(b-{5\over 3b})\B^{-1/2}, & ~~
\\
{\rm Ghost:} & ~~ \Pi^G_n\simeq -{2\over b}\B^{-1/2}, & ~~ \Sigma^G_r \simeq {\B\over 2\pi r}{1\over 2b}\B^{-1/2}, & ~~ \Xi^G_n\simeq {1\over 16\pi^2 n^2}\B^3.
\end{array}
\label{highTnzm}
\fe


Proceeding to next order in $\B^{3/2}$, we find for zero modes, 
\ie
\Pi^{(1)M}_0=-2.52\B, \quad \Pi^{(1)V}_0=-0.32\B, \quad \Xi_0^{(1)M}=0.68\B^{3/2}, \quad \mu^{(1)}=5.53\B,
\fe
while for nonzero modes, we have
\ie
\B^{-1}\Pi^{(1)M}_n&=5.24183\, +\frac{0.0274248}{n^2}, \quad \B^{-1}\Pi^{(1)V}_n=4.86502\, +\frac{0.131842}{n^2}.
\fe
The detailed calculation is presented in Appendix~\ref{app:highT}.

\subsection{The Free Energy}
Now that we have the self-energies, we can immediately obtain the free energy in the high temperature limit. To leading order, we have
\ie
\B F\simeq{\rm const}+6\,{\rm log}\B.
\fe
This agrees with the numerical result in \cite{Kabat:2001ve}.

Taking into account the nonzero modes, and also the subleading correction to the zero mode self-energies, we compute the free energy to the subheading order $\mathcal{O}(\B^{3/2})$ in Appendix~\ref{app:highT},
\ie
\B F={\rm const}+6\,{\rm log}\B-3.89872\B^{3/2}+\mathcal{O}(\B^{3}).
\fe

\subsection{Size of the Wave Function}
From the knowledge about the self-energies of the scalar fields, we may calculate the observable that measures the size of the wave function,\footnote{ Note that our normalization for the fields differs from \cite{Kawahara:2007ib} by a factor of $\sqrt{N}$. }
\ie
\la R^2\ra_\B\equiv {1\over N^2}\la \Tr X^i X^i\ra_\B+{1\over N^2}\la \Tr\phi^a \phi^a\ra_\B,
\fe
where
\ie
{1\over N^2}\la \Tr X^i X^i\ra_\B&=2\B^{-1}\sum_n \sigma_n^2,
\\
{1\over N^2}\la \Tr\phi^a \phi^a\ra_\B&=7\B^{-1}\sum_n\Delta_n^2,
\fe
which to the leading order in high temperature expansion become 
\ie
{1\over N^2} \la \Tr X^i X^i\ra_\B&\simeq{2\over \B\Pi^V_0}={2\over b}\B^{-1/2},
\\
{1\over N^2} \la \Tr\phi^a \phi^a\ra_\B&\simeq{7\over \B\Pi^M_0}={7\over a}\B^{-1/2}=({b\over 2}-{2\over b})\B^{-1/2}.
\fe
Hence we have,
\ie
\la R^2\ra_\B\simeq{b\over 2}\B^{-1/2}(g_{YM}^2N)^{{1\over2}}=2.392g_{YM}N^{1/2}T^{1/2},
\fe
where we have restored the `t Hooft coupling $g_{YM}^2N$, which was set to 1 previously, by dimensional analysis\footnote{Recall that $X^i$ has mass dimension 1 and $g_{YM}^2$ has mass dimension 3.}.
In the dual supergravity geometry, the IIA D0 black hole has radius $r\sim N^{1/5} T^{2/5}$ in the $g_{YM}=1$ unit, which agrees with the spread of the wave function $\sqrt{\la R^2\ra_\B}$ in the 't Hooft region $T\sim N^{1/3}$.

In Appendix~\ref{app:highT} we continue to compute $\la R^2\ra_\B$ to the subleading order $\mathcal{O}(\B)$,
\ie
\la R^2\ra_\B\equiv\omega_1\B^{-1/2}+\omega_2\B+\mathcal{O}(\B^{5/2}),
\fe
where 
\ie
\omega_1&={b\over 2}=2.392, \quad \omega_2={\mu^{(1)}\over 2\B}-{7\over 12}=2.180.
\fe
and $\mu^{(1)}$ is the next-to-leading correction to the $A_0$ self-energy.  Note that $\omega_{1,2}$ will receive corrections from higher-loop diagrams.

In \cite{Kawahara:2007ib}, the authors computed $\la R^2\ra_\B$ numerically. To match the notation there, we have $\omega_1=\chi_1$, and $\omega_2={4 (\chi_3-\chi_4) \over 3} + {3 \over 4}$, where $\chi_1=2.298$, $\chi_3=0.719$ and $\chi_4=-0.082$ according to Monte Carlo method, leading to $\omega_1=2.298$ and $\omega_2=1.818$ in \cite{Kawahara:2007ib}.

\section{Further Discussions}
\label{sec:discuss}

\subsection{More Supersymmetric Gauges}

As discussed in \cite{Kabat:1999hp, Kabat:2000zv, Kabat:2001ve}, manifest off-shell SUSY is essential for solving the Schwinger-Dyson equations. The importance of having an off-shell description can be seen from the fact that our final solutions for the nonzero modes (\ref{finalsol}) respect the SUSY Ward identity (\ref{ward}), which only makes sense for an off-shell SUSY formulation, to leading order in large $\B$. 

A related fact is that having kept the auxiliary field $f$ explicit, all the couplings in the $\cN = 4$ Wess-Zumino quantum mechanics are \textit{cubic}. That is, the quartic interaction $\phi^4$ is represented by the cubic coupling $f\phi^2$. It appears that having cubic couplings only is the key for the soft collinear approximation (\ref{soft}) to work.

If we want to apply the soft collinear ansatz to Schwinger-Dyson equations for the BFSS matrix model and thus solve them analytically, it seems crucial to trade every quartic interaction with cubic couplings by ``integrating in" auxiliary fields. Recall that the BFSS action can be decomposed into a $\cN = 4$ vector multiplet $V$ and three $\cN = 4$ chiral superfields $\Phi^a$ with superpotential $W=-i {1\over 6\sqrt{2}} g_{YM} \epsilon^{abc} \Tr\left(\Phi^a[\Phi^b,\Phi^c]\right)$, see (\ref{bfss}). The quartic interactions among $\phi$'s are already traded with cubic interactions involving the auxiliary fields $f$. However, there are still quartic interaction between the vector and chiral multiplets, e.g. from $D_\mu \phi^a D^\mu \phi^a$. It would be desirable to trade these quartic couplings with cubic couplings as we did for the self-interaction in the chiral multiplet.

Manifest ${\cal N}=8$ supersymmetric formulation using the harmonic superspace \cite{Galperin:2001uw} provides a natural solution. In the ${\cal N}=8$ language, the BFSS matrix model consists of one vector multiplet and one hypermultiplet. The action for the (gauged) hypermultiplet takes a particularly simple form,
\ie
S_{hyper} = - \int du d\zeta^{(-4)} \tilde q^+(D^{++} + V^{++})q^+.
\fe
$q^+$ is the (analytic) superfield representing the hypermultiplet and the $+$ superscript labels a $U(1)$ charge which is not important here. $D^{++}$ is a derivative acting on the $u$-space and $dud\zeta^{(-4)}$ is the measure on the analytic superspace (the analog of chiral superspace in the $\cN = 4$ case). $\tilde ~$ is a special conjugation whose details are again not important. At last, $V^{++}$ is the superfield for the vector multiplet.

The key observation here is that, in the harmonic superspace formulation, the couplings between the hypermultiplet $q^+$ and the vector multiplet $V^{++}$ are cubic. Note, however, that vector multiplet self-interactions involve quartic and higher couplings.

The complication about the harmonic superspace formulation is that each superfield contains infinitely many component fields when expanded with respect to $u$. Therefore, to preserve manifest ${\cal N}=8$ SUSY, we necessarily have to deal with infinitely many coupled Schwinger-Dyson equations. An ${\cal N}=8$ supersymmetric gauge fixing condition
\ie
D^{++}V^{++}=0
\fe
will kill all but a finite number of component fields in $V^{++}$.  However, the hypermultiplet superfield $q^+$ still contains infinitely many component fields.

It may also be possible to formulate Schwinger-Dyson equations for BFSS matrix theory with all ${\cal N}=16$ supersymmetries manifest, by working with a gauge fixed version of the Batalin-Vilkovisky action written in terms of pure spinor superfields \cite{Berkovits:2001rb}. These are left to future investigation.

\subsection{Schwinger-Dyson Equations for Supersymmetric Field Theories in Other Dimensions}

As another potential application of the technology we developed, one may employ the loop truncated Schwinger-Dyson equations to supersymmetric field theories in more than one dimension, and compute correlation functions and free energy along an entire RG flow trajectory. We illustrate this with the example of
the two-dimensional ${\cal N}=(2,2)$ Wess-Zumino model with $\Phi^3$ superpotential. 

It is well known that the ${\cal N}=(2,2)$ Landau-Ginzburg model, with the superspace action
\ie
S=\int d^2 x d^4 \theta ~\bar{\Phi}\Phi-\int d^2 x d^2 \theta~ {g\Phi^{k+2}\over k+2}-\int d^2 x d^2 \bar\theta~ {g\bar{\Phi}^{k+2}\over {k+2}},
\label{lg}
\fe
flows to the $k$-th ${\cal N}=2$ minimal model in the infrared \cite{Kastor:1988ef,Vafa198951}.
The equivalence between the IR fixed point of the Landau-Ginzburg model and the ${\cal N}=2$ minimal model has been checked in many ways: the central charge of the SCFT at the fixed point was derived in \cite{Vafa198951}, the elliptic genus of the two theories were shown to agree  \cite{Witten:1993jg}, the integrable RG flow of the 2d $N=2$ LG model was worked out in \cite{Fendley:1993pi}, and the exact $S$-matrices and the central charge were determined along the entire RG trajectory \cite{Fendley:1992dm}. 
However, apart from numerical results obtained via lattice simulation \cite{Kawai:2010yj,Kamata:2011fr}, the understanding of this RG flow is based on either symmetry arguments (exact scaling dimension of $\Phi$ in the IR limit  from the R-charge \cite{Vafa198951}) or very special integrable structure of the theory at the IR fixed point. The computation of correlation functions along the entire RG flow is not straightforward in such approaches. 
 
The loop truncated Schwinger-Dyson equations for the ${\cal N} =2$ LG model, on the other hand, appear to provide a useful approximation and expansion scheme along the entire flow. In particular, the one-loop Schwinger-Dyson equation already gives the correct scaling dimension of the fields in the IR limit.
For the moment, let us consider the zero temperature limit of the one-loop truncated Schwinger-Dyson equations for the self-energies of the matter multiplet. Unlike in the case of supersymmetric quantum mechanics, here the solution to the S-D equations is non-singular in the zero temperature limit. Clearly, SUSY Ward identity (\ref{N=4susyward}) is obeyed for the zero temperature solution, and we can express the fermion and auxiliary field self-energies in terms of that of the scalar field. The one-loop S-D equation then takes the simple form
\ie\label{cdlg}
{\sigma(p)\over p^2}=g^2\int {d^2 q\over (2\pi)^2}~ {1\over \left(q^2+\sigma(q)\right)\left((p-q)^2+\sigma(p-q)\right)}
\fe
In the $p\rightarrow 0$ limit, the equation reduces to
\ie
{\sigma(p)\over p^2}\approx g^2\int {d^2 q\over (2\pi)^2}~ {1\over \sigma(q)\sigma(p-q)}
\fe
Hence we find the solution for the self-energy in the IR limit,
\ie
\sigma(p)\approx {\Gamma({1\over 3})\over\pi^{1/3}\Gamma({2\over 3})}~g^{2\over3}p^{4\over 3}.
\fe
Indeed, we have rediscovered the ${1\over3}$ scaling dimension of $\Phi$ in the IR. 
While it may seem that this simple computation of the IR scaling dimension is essentially equivalent to the statement that the cubic superpotential is marginal in the IR, the Schwinger-Dyson equation is not specialized to the IR fixed point. One may solve (\ref{cdlg}) at all momenta $p$, which gives an approximation for the self-energy of $\Phi$ along the entire RG flow. It appears that, by taking into account higher loop corrections (the two-loop correction is absent and the first correction comes in at three-loop), we will be able to obtain an expansion for exact correlators along the RG flow.

In addition, one can calculate the free energy by solving the Schwinger-Dyson equations at finite temperature. In this case, we discretize the momentum $\vec p\rightarrow (p,{2\pi n\over\B})$. The S-D equations become an infinite set of coupled integrable equations. In the low temperature limit, scale invariance determines the free energy up to an overall constant, $F = -\pi cVT^2/6$. Here $c$ is the central charge at the IR fixed point. It would be interesting to compute the value of $c$ by solving the loop truncated S-D equations, and compare the result to the $k=1$ ${\cal N}=2$ minimal model value $c=1$ ($c={3k\over k+2}$ for the $k$-th minimal model). 
It would also be interesting to see how the two-point function for $\Phi$ flows using the Schwinger-Dyson equations. We leave these for future development.


\bigskip

\section*{Acknowledgments}

We would like to thank Tatsuo Azeyanagi, Masanori Hanada, Hong Liu, Shiraz Minwalla, Steve Shenker and Washington Taylor for useful discussions. XY would like to thank the hospitality of Tata Institute for Fundamental Research, Mumbai, India, during the course of this work. 
This work is supported in part by the Fundamental Laws Initiative 
Fund at Harvard University, and by NSF Award PHY-0847457.


\appendix

\section{Hawking Decay Rate of D0-Branes}
\label{app:decay}

As argued in the introduction, a D0 black hole decays by emission of D0-branes.  The fact that these emitted D0-branes are unbounded by a D0 black hole background (\ref{IIAbkgd}) is due to the cancellation between the gravitational potential and the Ramond-Ramond potential in the DBI action for D0-branes.  The decay rate of the black hole is governed by Hawking's formula~\cite{Hawking:1974sw}
\ie
\Gamma = \sum_{n = 1}^\infty \int {d^9 \vec k \over (2\pi)^9} {\sigma_\text{abs} v \over e^{\B(\omega - \mu n)} - 1},
\fe
where $\vec k$ is the spatial momentum, $\omega$ the energy, $v = d\omega/dk$ the speed (given by the dispersion relation at infinity), and $n$ the D0-brane charge; $\sigma_\text{abs}$ is the absorption cross section and $\mu$ is the chemical potential for the D0-brane charge.  The phase space and thermal factors make the spectrum of the Hawking emission strongly peaked at $\omega \sim 1/\B$.  The absorption cross section in the regime of interest can be obtained through a classical computation of the capture cross section, or a semi-classical (WKB) computation of the greybody factor and then summing over all partial waves.  Here we present the latter.

\subsection{Greybody Factor}
\label{app:grey}

Consider a scalar field propagating in the D0 black hole background (\ref{IIAbkgd}).  Because supergravity can only be trusted for $r_0 \ll N^{1/3} l_P$, and this background is only a stable solution $r_0 \gg N^{1/9} l_P$, we shall take $N^{1/9} l_P \ll r_0 \ll N^{1/3} l_P$.\footnote{ When $0<r_0 < N^{1/9}$, the solution (\ref{IIAbkgd}) is unstable against the inhomogeneous 11D black hole background \cite{Polchinski:1999br}. }  To incorporate the effects of the graviton, the dilaton, and the Ramond-Ramond 1-form background altogether, it is convenient to lift (\ref{IIAbkgd}) to M-theory:
\ie
ds^2_M &= f (dx_{10} + C_1)^2 + f^{-1/2} ds_{IIA}^2 \\
&= f dx_{10}^2 - (1 + A) dx_{10} dt + {(1-A)^2 \over 4} f^{-1} dt^2 + A^{-1} dr^2 + r^2 d\Omega_8^2, \\
f &= {\tilde c_0 N l_P^9 \over R_{10}^2 r^7}, \quad \tilde c_0 \equiv {15 \over 2}, \quad A = 1 - {r_0^7 \over r^7}
\fe
where $x_{10}$ is periodic under $x_{10} \sim x_{10} + 2\pi R_{10}$, $R_{10} = g_s l_s$, and $l_P \equiv (2\pi g_s)^{1/3} l_s$.  Note that we have written all quantities in the M-theory scales $l_P$ and $R_{10}$.  To simplify the notation, let us set the Planck length $l_P$ to 1, and rescale
\ie
x_{10} \to x_{10} R_{10}, \quad t \to t / R_{10}, \quad \tilde c_0 N \to N.
\label{rescale}
\fe
The Hawking temperature (\ref{hawkingT}) is then
\ie
T_H \equiv {1 \over \B} = {7 \over 4\pi} \sqrt{r_0^5 \over N}.
\fe

In this background, the Klein-Gordon equation for the $n$-th Kaluza-Klein mode $\phi$ is
\ie
\left( -A r^{-8} \partial_r r^8 A \partial_r - 2n \omega \left(1 + {R^7 \over r^7} \right) + {A L^2 \over r^2} \right) \phi = 0,
\fe
where $L^2 \equiv \ell (\ell + 1)$ is the total angular momentum and
\ie
R^7 \equiv {N\omega \over 2n} \left(1 - {n r_0^7 \over 2N \omega} \right)^2.  \label{R7}
\fe
Under a redefinition $\tilde\phi \equiv \sqrt{r^8 A} \phi$, the Klein-Gordon equation turns into a standard wave-in-an-effective potential problem
\ie
- \partial_r^2\tilde\phi+ U \tilde\phi = 0,  \label{effKG}
\fe
where
\ie
U \equiv A^{-1/2} r^{-4} \partial_r^2 (r^4 A^{1/2}) - 2n \omega A^{-2} \left(   1 + {R^7 \over r^7} \right) + {A^{-1} L^2 \over r^2}.  \label{effpot}
\fe
The greybody factor is then just the transmission amplitude $\cT$ of an incoming plane wave in this potential.  Due to the complicated form of $U$, we cannot solve the wave equation (\ref{effKG}) exactly; instead, we will solve it for large and small $r$ using the WKB method, and then match the solutions in an overlapping region.  We show the existence of such a region presently.

The potential $U$ (\ref{effpot}) involves two natural scales, $r_0$ and $R$.  When $\omega \ll nr_0^7 / N$,
\ie
R^7 \simeq {n r_0^{14} \over 8N \omega} = {r_0^7 \over 4} {n r_0^7 \over 2N \omega} \gg r_0^7,
\fe
and when $\omega \gg nr_0^7 / 2N$,
\ie
R^7 \simeq {N\omega \over 2n}  = {r_0^7 \over 4} {2N \omega \over n r_0^7} \gg r_0^7.
\fe
In either case, $R \gg r_0$, so if we define the far region to be $r \gg r_0$ and the near region to be $r \ll R$, then we have an overlapping region $r_0 \ll r \ll R$.  The only part of the spectrum where this fails is when $\omega \sim {n r_0^7 \over 2N}$.  However, for $\omega \sim {n r_0^7 \over 2N} \gg 1/\B$, the thermal factor suppresses the emission of D0-branes exponentially, so the contribution from $\omega \sim {n r_0^7 \over N}$ is negligible.  Similarly, the phase space factor suppresses the contribution from $\omega \lesssim{ 1 \over n R^2}$, so we may neglect this region and assume $n \omega R^2 \gg 1$.  We will examine these two regions more carefully at the end of the section.

In the far region $r \gg r_0$, we can set $A \simeq 1$.  In terms of $\rho \equiv r / R$, the wave equation (\ref{effKG}) becomes
\ie
- \partial_\rho^2\tilde\phi  +  V \tilde\phi = 0,
\fe
where
\ie
V = {12 + L^2 \over \rho^2} - 2n \omega R^2 \left(1 + {1 \over \rho^7} \right).  \label{farpot}
\fe
To check if WKB is valid, let us compute
\ie
V' = -2 \left( 12 + L^2 \over \rho^3 \right) + {14 n \omega R^2 \over \rho^8}.
\fe
For $L^2 \sim 1$, each of $V$ and $\partial_\rho V$ is dominated by the second term since  $n \omega R^2 \gg 1$, and so the WKB criterion $|\partial_\rho V|^2 / |V|^3$ is small for all $\rho$.  Hence WKB is valid, and the far solution is
\ie
\phi_\text{far}^\text{WKB} \propto \rho^{-4} (-V)^{-1/4} e^{\pm i \int d\rho \sqrt{-V}}.  \label{farsol}
\fe

In the near region $r_0 < r \ll R$, let us define $\rho' \equiv r / r_0$.  Then the wave equation (\ref{effKG}) becomes
\ie
-\partial_{\rho'}^2 \tilde\phi+ W \tilde\phi = 0,
\fe
where
\ie
W = - { 1 + 4(24+L^2)\rho'^7 + (8n\omega R^7/r_0^5)\rho'^9 - 4(12+L^2 )\rho'^{14} \over 4\rho'^2 (1-\rho'^7)^2 }.
\fe
If $L^2 \sim 1$, then the $\rho'^9$ term dominates over 1 and the $\rho'^7$ term because both $R/r_0$ and $n\omega R^2$ are large, and because the range of $\rho'$ is bounded below by 1; the $\rho'^9$ term also dominates over the $\rho'^{14}$ term because $\rho' \ll R / r_0$.  Thus the potential can be approximated by
\ie
W \simeq - {2n\omega R^7 \over r_0^5} {\rho'^7 \over (1 - \rho'^7)^2}.  \label{nearpot}
\fe
The WKB criterion
\ie
{|\partial_{\rho'}W|^2 \over |W|^3} \simeq {r_0^5 \over 2n \omega R^7} {49 (1 + \rho'^7)^2 \over \rho'^9}
\fe
is peaked near $\rho' \sim 1$, the value thereof is small since both $R/r_0$ and $n\omega R^2$ are large.  Hence WKB is valid, and the near solution is
\ie
\phi_\text{near}^\text{WKB} \propto \rho'^{-4} (-W)^{-1/4} e^{\pm i \int d\rho' \sqrt{-W}}.  \label{nearsol}
\fe

Now we match the two solutions in the overlapping region $r_0 \ll r \ll R$, which corresponds to $\rho \ll 1$ and $\rho' \gg 1$.  But the two sides match trivially, since the two solutions (\ref{farsol}) and (\ref{nearsol}) become the same in this region.  The WKB method can give a nontrivial reflection amplitude only if there exist turning points in the potential.  Under our assumptions, the potential (\ref{effpot}) is well below 0 for the entire range of $r$, and therefore the transmission amplitude $\cT$ is just 1.  However, the potential (\ref{effpot}) has turning points when $L^2$ is large.  From the approximate potential in the far region (\ref{farpot}), this happens when
\ie
12 + L^2 > \left( 2 \over 5 \right)^{1/7} L^2_\text{max}, \quad L^2_\text{max} \equiv 2n \omega R^2.  \label{L>Lmax}
\fe
It is not hard to see that the exact potential (\ref{effpot}) gives the same bound up to multiplication by an $\cO(1)$ constant and subleading order corrections.  If $L^2$ exceeds the threshold, then the potential barrier will provide an exponential suppression of the transmission amplitude.

\subsection{Partial Wave Summation}

The absorption cross section is related to the greybody factor by~\cite{Gubser:1997qr}
\ie
\sigma_\text{abs} = \sum_{\ell = 0}^\infty \sigma^\ell_\text{abs},  \label{pwsuminfty}
\fe
where
\ie
\sigma^\ell_\text{abs} = 2^7 \pi^{7/2} \Gamma(7/2) (\ell + 7/2)
\begin{pmatrix}
\ell+6 \\
\ell
\end{pmatrix}
{\cT \over k^8}
\fe
is the absorption cross section for each partial wave.  Let $\ell_\text{max}$ be defined such that $L^2_\text{max} = \ell_\text{max} (\ell_\text{max} + 1)$, then because $\cT \simeq 1$ for $\ell < \ell_\text{max}$ and $\cT$ is exponentially suppressed for $\ell > \ell_\text{max}$, we can approximate the sum (\ref{pwsuminfty}) by a hard cutoff
\ie
\sigma_\text{abs} \simeq \sum_{\ell = 0}^{\ell_\text{max}} \sigma^\ell_\text{abs} \simeq { 2^7 \pi^{7/2} \Gamma(7/2) \over 8 \times 6! } {\ell_\text{max}^8 \over k^8},
\fe
where
\ie
\ell_\text{max} \simeq \sqrt{ L^2_\text{max} } \simeq \sqrt{2n \omega} R.
\fe
Note that even though we have kept the prefactor, corrections to the hard cutoff approximation will multiply this prefactor by an $\cO(1)$ constant, so one should not take its value too seriously.

\subsection{Decay Rate}

We are now ready to compute the decay rate.  The dispersion relation at infinity is $\omega = k^2 / 2n$, so $v = k / n = \sqrt{2\omega / n}$ and $d^9\vec k = ((2n)^{9/2} \pi^{9/2}/\Gamma(9/2)) ~ \omega^{7/2} d\omega$.  The chemical potential is\footnote{ It is the difference between the gauge potential at the horizon and that at the ``far-horizon'' infinity, which is 0 by our gauge choice.  Alternatively, without going to far-horizon, it can be fixed by the first law where one uses the regularized ADM mass for the black hole. }
\ie
\mu dt = C_1|_{r = r_0} = - {r_0^7 \over 2N} dt.
\fe
Putting these together, the decay rate is
\ie
\Gamma \simeq {2^4 \over 8! \pi } \sum_{n = 1}^\infty n^4 e^{-n\B r_0^7 / 2N} \int_0^\infty d\omega ~ {\omega^4 R^8 \over e^{\B\omega} - 1}.
\fe
Due to the exponential suppression from the thermal factor, the dominant contribution to $\Gamma$ comes from $\omega \ll n r_0^7 / 2N$, where (\ref{R7}) can be approximated by
\ie
R^7 \simeq {n r_0^{14} \over 8N \omega}.
\fe
Then
\ie
\Gamma &\simeq { 2^{4/7} \over 8! \pi } { r_0^{16} \over N^{8/7} }\sum_{n = 1}^\infty n^{36/7} e^{-n\B r_0^7 / 2N} \int_0^\infty d\omega ~ {\omega^{20/7} \over e^{\B\omega} - 1} \\
&\simeq { 2^{4/7} \Gamma(27/7) \zeta(27/7) \over 8! \pi } { r_0^{16} \over N^{8/7} \B^{27/7} } ~ e^{-\B r_0^7 / 2N}, \\
&\simeq { 2^{4/7} \Gamma(27/7) \zeta(27/7) \over 8! \pi } \left( 7 \over 4\pi \right)^{27/7} { r_0^{359/14} \over N^{43/14} } ~ e^{-(2\pi / 7) \sqrt{ r_0^9 / N} }.
\label{Gammafinal}
\fe
where the sum over $n$ is replaced by the $n = 1$ term because $\B r_0^7 / N \sim \sqrt{r_0^9 / N} \gg 1$.  To conform with the units in the rest of the paper, one can undo the rescalings (\ref{rescale}) and restore $l_P$.

Now let us examine the regions $\omega \sim n r_0^7 / 2N$ and $\omega \lesssim 1 / n R^2$ where the computation of $\cT$ in Appendix~\ref{app:grey} fails.  For $\omega \sim n r_0^7 / 2N$, we lose the overlapping region because the two terms in the brackets in (\ref{R7}) may cancel to give a small $R$.  However, looking at the range of $L^2$ (\ref{L>Lmax}) for which $\cT$ becomes exponentially suppressed, we see that a smaller value of $R$ only lowers the hard cutoff and makes the contribution to $\Gamma$ smaller.  Since we already get an exponential suppression from the thermal factor, we can safely neglect the contribution from this region.  For $\omega \lesssim 1 / n R^2$, the cutoff is at most $\cO(1)$, so the suppression from the phase space factor again allows us to neglect its contribution.


\section{1D $\cN=2$ SUSY}
\label{app:N=2}

In this appendix we will discuss the structure of 1D $\cN=2$ SUSY, the vector and the matter multiplet. It is previously discussed in \cite{Kabat:1999hp}. We will work in the Euclidean signature.

\subsection{Superspace}

The (Euclidean) 1D $\cN = 2$ superspace consists of one bosonic coordinate $\tau$ and two real Grassmann coordinates $\theta_\A$ in the vector representation of R-symmetry group $SO(2)_R$.  We choose the $SO(2)_R$ Dirac matrices to be real and symmetric, $\gamma^1 = \sigma_1$ and $\gamma^2 = \sigma_3$, where $\sigma^i$ are the standard Pauli matrices.

There are two invariant symbols, $\delta_{\A\B}$ and $\epsilon_{\A\B}$, which represent the inner and exterior products for 2D vectors. We normalize them by $\delta_{11}=\delta_{22}=1,\delta_{12}=\delta_{21}=0$, and $\epsilon_{12} =-\epsilon_{21} =1,\epsilon_{11}=\epsilon_{22}=0$. If we were to distinguish upper with lower indices, there is a potential confusion whether one uses $\delta_{\A\B}$ or $\epsilon_{\A\B}$ to raise and lower the index. To avoid this ambiguity, we will write every index downstairs. 

Let us define
\ie
\theta^2 \equiv {i \over 2} \epsilon_{\A\B} \theta_\A \theta_\B \quad \Leftrightarrow \quad \theta_\A\theta_\B = -i \epsilon_{\A\B}\theta^2,
\fe
so that $\theta^2$ is real.\footnote{By real we mean the it is real when we Wick rotate to the Lorentzian signature. This will always be assumed implicitly in the following.}
We define our fermionic integral measure such that
\ie
\int d^2\theta ~ \theta^2 = 1.
\fe
Given a $2 \times 2$ matrix $B_{\A\B}$, we can expand it as
\ie
B_{\A\B}={1\over 2} B_{\rho\rho} \delta_{\A\B} +{1\over 2} \gamma^i_{\rho\sigma}B_{\rho\sigma} \gamma^i _{\A\B} + {1\over 2} \epsilon_{\rho\sigma}B_{\rho\sigma}\epsilon_{\A\B}.
\fe
Some useful identities are
\ie
& \epsilon_{\A\gamma} \epsilon_{\B\gamma} = \delta_{\A\B}, \quad \epsilon_{\A\gamma} \epsilon_{\B\delta} =\delta_{\A\B} \delta_{\gamma_\delta} - \delta_{\A\delta} \delta_{\B\gamma}, \\ & \epsilon_{\A\gamma} \gamma^i_{\gamma\B} = \epsilon^{ij} \gamma^j_{\A\B}, \quad \gamma^i_{\A\B} \gamma^j_{\A\B} = 2\delta^{ij}, \\
& \gamma^i_{\A\B} \gamma^i_{\rho\sigma} = \delta_{\A\rho} \delta_{\B\sigma} + \delta_{\A\sigma} \delta_{\B\rho} - \delta_{\A\B} \delta_{\rho\sigma}, \\
& \epsilon_{\gamma\delta} \gamma^i_{\A\gamma} \gamma^j_{\B\delta} = -\epsilon^{ij} \delta_{\A\B} - \delta^{ij} \epsilon_{\A\B}. 
\fe

The supercharges and super-derivatives are
\begin{align}
Q_\A & = { \partial \over \partial \theta_\A} - \theta_\A \partial_\tau, \\
D_\A & = { \partial \over \partial \theta_\A} + \theta_\A \partial_\tau,
\end{align}
They obey the algebra
\begin{align}
&\{ Q_\A ,Q_\B \} = -2 \delta_{\A\B} \partial_\tau,\\
&\{ Q_\A ,D_\B \} =0, \\
&\{ D_\A ,D_\B \} = 2 \delta_{\A\B} \partial_\tau.
\end{align}

\subsection{Vector Multiplet}

\subsubsection{Connections}

The vector multiplet is constructed as follows. We first introduce a real connection on superspace
\begin{align}
&\nabla_\alpha=D_\alpha +\Gamma_\alpha,\\
&\nabla_\tau=\partial_\tau + i\Gamma_\tau.
\end{align}
Note that $\Gamma_\alpha$ is an anticommuting real superfield while $\Gamma_\tau$ is a commuting real superfield.

For reason that will become clear in a minute, $\Gamma_\tau$ is determined by $\Gamma_\alpha$ by a constraint equation, so we only have to expand the real connection $\Gamma_\alpha$ as
\begin{align}
\Gamma_\alpha=\chi_\alpha +i A_0 \theta_\alpha
+X^i \gamma^i_{\alpha\beta}\theta_\beta
+d \epsilon_{\alpha\beta} \theta_\beta +2\epsilon_{\alpha\beta}\lambda_\beta \theta^2.
\end{align}
The super gauge transformations are
\begin{align}
&\delta_\Lambda \Gamma_\alpha= i \nabla_\alpha \Lambda = i D_\alpha \Lambda + i [ \Gamma_\alpha, \Lambda ],\label{gauge1}\\
&\delta_\Lambda \Gamma_\tau =  \nabla_\tau \Lambda
= \partial_\tau \Lambda + i [  \Gamma_\tau , \Lambda   ]\label{gauge2},
\end{align}
with $\Lambda$ a real superfield. If we expand
\begin{align}
\Lambda = \Omega + i\omega_\alpha \theta_\alpha +W\theta^2,
\end{align}
then the super gauge transformation in the component form is
\begin{align}
\delta_\Lambda \chi_\alpha &= \omega_\alpha + i [\chi_\alpha , \Omega],\\
\delta_\Lambda A_0 
&=\dot \Omega +i [A_0 , \Omega ] 
+{i\over 2} \{ \chi_\alpha, \omega_\alpha \} ,\\
 \delta_\Lambda X^i &= i [X^i , \Omega]
 -{1\over 2} \gamma^i_{\alpha\beta} \{ \chi_\alpha , \omega_\beta \},\\
  \delta_\Lambda d &= -W +i [ d,\Omega]  - {1\over 2} 
 \epsilon_{\alpha\beta} \{ \chi_\alpha ,\omega_\beta\},\\
 \delta_\Lambda \lambda_\alpha &= 
 -{i\over 2} \dot \omega_\alpha -{i\over 2} \epsilon_{\alpha\beta } [\chi_\beta,W] 
 +i [\lambda_\alpha, \Omega]
 +{1\over 2} [A_0 ,\omega_\alpha] 
 + {i\over 2} \gamma^i _{\alpha\beta} [X^i ,\omega_\beta] 
 - {i\over 2} \epsilon_{\alpha\beta} [d,\omega_\beta].
 \end{align}

\subsubsection{Supersymmetry Transformations}

The SUSY transformations are generated by $Q_\alpha$ in the following way
\begin{align}
\delta_\varepsilon \Gamma_\alpha &= \varepsilon_\beta Q_\beta \Gamma_\alpha\notag\\
\Rightarrow \quad
\delta_\varepsilon \chi_\alpha &=
i \varepsilon_\alpha A_0 + \varepsilon_\beta \gamma^i_{\beta\alpha}X^i + \varepsilon_\beta \epsilon_{\alpha\beta} d,\\
 \delta_\varepsilon A_0 &= 
  \varepsilon_\alpha \lambda_\alpha - {i\over 2} \varepsilon_\alpha \dot\chi_\alpha,\\
 \delta_\varepsilon X^i &= 
 -i \gamma^i_{\alpha\beta} \varepsilon_\alpha \lambda_\beta
 +{1\over 2}\gamma^i_{\alpha\beta}\varepsilon_\alpha \dot\chi_\beta , \\
 \delta_\varepsilon d &=
 -i \epsilon_{\alpha\beta} \varepsilon_\alpha \lambda_\beta 
 -{1\over 2} \epsilon_{\alpha\beta}\varepsilon_\alpha \dot\chi_\beta,\\
 \delta_\varepsilon \lambda_\alpha &=
 {1\over 2}\varepsilon_\alpha \dot A_0
+{i\over 2}\varepsilon_\beta \dot X^i \gamma^i_{\alpha\beta}
 -{i\over 2}\varepsilon_\beta \dot d \epsilon_{\alpha\beta}.
\end{align}

\subsubsection{Field Strengths}

The field strengths $\mathcal{F}_{\alpha\beta}$ and $\mathcal{F}_\alpha$ are
\begin{align}
\{ \nabla_\alpha , \nabla_\beta \} &= 2\delta_{\alpha\beta} \nabla_\tau + \mathcal{F}_{\alpha\beta},\label{F1}\\
[ \nabla_\tau , \nabla_\alpha ] &=-i \mathcal{F}_\alpha. \label{F2}
\end{align}
Explicitly, they are given by
\begin{align}
\mathcal{F}_{\alpha\beta} &= 
D_\alpha \Gamma_\beta + D_\beta \Gamma_\alpha
+\{ \Gamma_\alpha , \Gamma_\beta \} 
-2i \delta_{\alpha\beta} \Gamma_\tau,\\
\mathcal{F}_\alpha &=
i\partial_\tau \Gamma_\alpha +D_\alpha \Gamma_\tau -
 [ \Gamma_\tau ,\Gamma_\alpha ].
\end{align}
Under super gauge transformations, the field strengths transform covariantly:
\begin{align}
\delta_\Lambda \mathcal{F}_{\alpha\beta}
&=i [ \mathcal{F}_{\alpha\beta} ,\Lambda],\\
\delta_\Lambda \mathcal{F}_{\alpha}
&=i [ \mathcal{F}_{\alpha} ,\Lambda].
\end{align}
Note that the lowest component ($\theta=0$) of the field strength is neutral under the super gauge transformations which depend on $\theta$.

Next we impose the constraint $\delta_{\alpha\beta}\mathcal{F}_{\alpha\beta}=0$, from which we can express $\Gamma_\tau$ in terms of others
\begin{align}
\Gamma_\tau = -{i\over 2} \left( D_\alpha \Gamma_\alpha + \Gamma_\alpha\Gamma_\alpha \right). \label{constraint}
\end{align}
Note the super gauge transformations (\ref{gauge1}) and (\ref{gauge2}) are compatible with the constraint (\ref{constraint}). Now $\mathcal{F}_{\alpha\beta}$ is a symmetric traceless matrix, so it can be viewed as a vector in $SO(2)_R$ by defining ${\cal F}^i$ as
\begin{align}
& \mathcal{F}^i \equiv {1\over 4} \gamma^i_{\alpha\beta}\mathcal{F}_{\alpha\beta}
= {1\over 2} \gamma^i_{\alpha\beta} \left( D_{(\alpha}\Gamma_{\beta)} +\Gamma_{(\alpha}\Gamma_{\beta)}\right)
\end{align}

\subsubsection{Action}

The 1D $\cN=2$ super Yang-Mills action is given by
\ie
S_{YM} =& {1\over g_{YM}^2 }\int d\tau d^2\theta \Tr \Big\{ 
- {1\over 4} \left( { i \over 2} \epsilon_{\alpha\beta} \right)  
\nabla_\alpha \mathcal{F}^i \nabla_\beta \mathcal{F}^i \Big\}
\fe
In the Wess-Zumino gauge, the component fields $\chi_\alpha$ and $d$ in $\Gamma_\alpha$ are gauged away by appropriately choosing the super gauge parameter $\omega_\alpha$ and $W$. While the part of the super gauge symmetry is fixed by the Wess-Zumino gauge, we still have the conventional bosonic gauge transformation generated by $\Omega$. In the Wess-Zumino gauge, the expansion of the real connection superfield reduces to
\begin{align}
\Gamma_\alpha = iA_0\theta_\alpha
+X^i \gamma^i_{\alpha\beta}\theta_\beta
 +2\epsilon_{\alpha\beta}\lambda_\beta \theta^2.
\end{align}
The field strength $\mathcal{F}^i$ in the component form is then
\begin{align}
\mathcal{F}^i = 
X^i
 -i \gamma^i_{\alpha\beta} \lambda_\alpha \theta_\beta
 + \epsilon^{ij} \left( i\dot X^j - [A_0, X^j ] \right) \theta^2.
 \end{align}
The Yang-Mills action in the WZ gauge is
\begin{align}
S_{YM} 
&= {1\over g_{YM}^2 }\int d\tau \, \Tr \Big\{ \, 
{1\over 2} \lambda_\alpha D_0 \lambda_\alpha - {1\over 2} \lambda_\alpha \gamma^i_{\alpha\beta} [ X^i , \lambda_\beta ] \\
&\hspace{1.5in} + {1\over 2} D_0 X^i D_0X^i - {1\over 4} [X^i,X^j][X^i,X^j] \Big\},\label{SYM}
\end{align}
where $D_0 = \partial_\tau + i[A_0 , \cdot~]$. Note again that (\ref{SYM}) is only invariant under the conventional (bosonic) gauge transformation, but not under the super gauge transformation.

Next we want to write down the action without fixing a gauge. Define the ``covariant" component fields to be
\begin{align}
X^i|_{cov} \equiv& \mathcal{F}^i |_{\theta=0}
= {1\over 2} \gamma^i_{\alpha\beta} \left( D_{(\alpha}\Gamma_{\beta)} +\Gamma_{(\alpha}\Gamma_{\beta)}\right)|_{\theta=0}\notag\\
=& X^i +{1\over 2} \gamma^i_{\alpha\beta} \chi_\alpha \chi_\beta,\label{cov1}\\
A_0|_{cov}\equiv& \Gamma_\tau|_{\theta=0} =-i {1\over 2} \left( D_\alpha \Gamma_\alpha + \Gamma_\alpha\Gamma_\alpha \right) |_{\theta=0}        \notag   \\
=& A_0 -i{1\over 2}\chi_\alpha \chi_\alpha   \label{cov2}       \\
\lambda_\alpha |_{cov} \equiv & \mathcal{F}_\alpha|_{\theta=0}= 
i\partial_\tau \Gamma_\alpha +D_\alpha\Gamma_\tau -[\Gamma_\tau,\Gamma_\alpha] |_{\theta=0}\notag\\
=&\lambda_\alpha +{i\over2}\dot \chi_\alpha -{1\over 2}[A_0 ,\chi_\alpha]-{i\over 2}\gamma^i_{\alpha\beta} [X^i ,\chi_\beta ]+{i\over 2}\epsilon_{\alpha\beta}[d,\chi_\beta]-{i\over 2} [\chi_\alpha,\chi_\beta\chi_\beta]  \label{cov3}
\end{align}

To obtain the vector multiplet action without fixing a gauge, one has to find a extension of  (\ref{SYM}) such that it is invariant under the full super gauge transformation. Suppose we replace every $X^i$, $A_0$, and $\lambda_\alpha$ in (\ref{SYM}) by $X^i|_{cov}$, $A_0|_{cov}$, and $\lambda_\alpha|_{cov}$ defined above, then the action is still invariant under the conventional gauge transformation because the gauge indices are properly contracted and the derivatives always appear in the form of gauge covariant derivatives $D_0$. In addition, since the covariant fields are the lowest components of $\mathcal{F}^i$, $\mathcal{F}_\alpha$, and $\Gamma_\tau$, they are neutral under gauge transformations that depend on $\theta$. As a result, the new action is invariant under the full super gauge transformation as desired. Thus the 1D $\cN=2$ super Yang-Mills action is given by (\ref{SYM}) with every component field replaced by the covariant component fields defined in (\ref{cov1}-\ref{cov3})

\subsection{Matter Multiplet}

Consider a real scalar superfield $\Phi$ in the adjoint representation of the gauge group:
\begin{align}
\Phi = \phi + i\psi_\alpha \theta_\alpha +if\theta^2.
\end{align}
where the $i$ in front of $f$ is such that the kinetic term for $f$ has the right sign. 

The SUSY transformations are 
\begin{align}
\delta_\varepsilon \Phi &=\varepsilon_\alpha Q_\alpha \Phi \\
\Rightarrow \quad 
\delta_\varepsilon \phi &= -i\varepsilon_\alpha\psi_\alpha,\\ 
\delta_\varepsilon \psi_\alpha &= 
 -i\varepsilon_\beta f \epsilon_{\alpha\beta}
+i\varepsilon_\alpha \dot \phi,\\
\delta_\varepsilon f &= 
\varepsilon_\alpha \epsilon_{\alpha\beta}\dot\psi_\beta.
\end{align}

The super gauge transformation is
\begin{align}
\delta_\Lambda \Phi& = i [\Phi,\Lambda]\end{align}
Written in terms of the component fields, we have
\begin{align}
\delta_\Lambda\phi &=  i[\phi ,\Omega] ,\\
\delta_\Lambda \psi_\alpha &= i [\psi_\alpha,\Omega] +i [\phi ,\omega_\alpha],\\
\delta_\Lambda f &= \epsilon_{\alpha\beta}  \{\psi_\alpha ,\omega_\beta \}
 + i[f,\Omega] 
 +i [\phi ,W ].
\end{align}

The action we will consider is
\begin{align}
S_{M} =\int d\tau d^2\theta \, 
\Tr \left(  -{i\over 4}\epsilon_{\alpha\beta} \nabla_\alpha \Phi^a \nabla_\beta \Phi^a -{ig_{YM}\over 3} \epsilon^{abc} \Phi^a[\Phi^b,\Phi^c]\right),
\end{align}
where $a$ is in some representation of the flavor symmetry $G$ and $\epsilon^{abc}$ is a totally antisymmetric $G$-invariant tensor. If we write the BFSS action in the $\cN=2$ language, then it consists of one vector multiplet and the matter multiplet in the ${\bf 7}$ of the the flavor symmetry $G_2$.

\subsection{Schwinger-Dyson Equations for BFSS in the $\cN=2$ Language}\label{app:N=2gap}

Let us write down the one-loop truncated Schwinger-Dyson equations for BFSS with manifest $\cN = 2$ supersymmetry.  The full action is $S_\text{vec} + S_\text{matter} + S_\text{ghost}$ given in (\ref{N=2vecmat}) and (\ref{N=2ghost}).  Let us denote the propagators for the vector, matter, and ghost multiplets by
\ie
\begin{array}{llll}
{\rm vector:} & ~~ \sigma_n^2, & ~~  a_r, & 
\\
{\rm matter:}  & ~~  \Delta_n^2,  & ~~  g_r,  & ~~ \epsilon_n^2,
\\
{\rm ghost:}  & ~~  s_n, & ~~  t_r,  & ~~ u_n,
\end{array}
\fe
and write them in terms of the self-energies as
\ie
\begin{array}{*4{>{\displaystyle}l}}
\sigma_n^2 = {1\over ({2\pi n\over \beta})^2 + \Pi^V_n }, & \quad  a_r = {i\over ({2\pi r\over \beta})^3 + \Sigma^V_r}, &
\\
\Delta_n^2 = {1\over  ({2\pi n\over \beta})^2 + \Pi^M_n }, & \quad  g_r = -{i\over {2\pi r\over\beta} + \Sigma^M_r},
 & \quad  \epsilon_n^2 = {1\over 1+ \Xi_n^M},
\\
s_n = - {1\over ({2\pi n\over\beta})^2 + \Pi^G_n}, & \quad  t_r = - {i\over {2\pi r\over\beta} + \Sigma^G_r},
 & \quad  u_n = {1\over 1+ \Xi^G_n}.
\end{array}
\fe
There is also a propagator for the zero mode of the gauge field, $\langle A_{00} A_{00}\rangle=\rho_0^2=1/\mu$, where $A_{00}$ is normalized by $A_0 = \beta^{-{1\over 2}} A_{00}$.

The one-loop truncated Schwinger-Dyson equations are
\ie
\Pi^V_n &= {2\over \beta} \sum_m {1\over ({2\pi m\over \beta})^2 + \Pi^V_m} - {3\over \beta}\sum_r {1\over ({2\pi r\over \beta})^2 + {\beta\over 2\pi r}\Sigma^V_r} + {14\over \beta} \sum_m {1\over ({2\pi m\over \beta})^2 + \Pi^M_m} 
\\
&~~~ + {14\over \beta} \sum_r {1\over ({2\pi r\over \beta} + \Sigma_r^M)({2\pi (n-r)\over \beta} + \Sigma_{n-r}^M)}
- {1\over\beta} \sum_r {1\over ({2\pi r\over \beta} + \Sigma_r^G)({2\pi (n-r)\over \beta} + \Sigma_{n-r}^G)}
\\
&~~~ + {2\over \beta\mu}
- {8\over \beta\mu} {1\over 1+ ({\beta\over 2\pi n})^2 \Pi^V_n},
\\
\Sigma^V_r &= {3\over \beta} {2\pi r\over \beta} \sum_m {1\over ({2\pi m\over \beta})^2+\Pi^V_m}
- {4\over\beta} {2\pi r\over \beta} \sum_s {1\over ({2\pi s\over\beta})^2 + {\beta\over 2\pi s} \Sigma^V_s}
+{14\over \beta} {2\pi r\over\beta} \sum_m {1\over ({2\pi m\over \beta})^2 + \Pi^M_m}
\\
&~~~ + {5\over 2\beta\mu} {2\pi r\over \beta}  - {8\over\beta\mu} {2\pi r\over\beta} {1\over 1+({\beta\over 2\pi r})^3 \Sigma^V_r} + {14\over\beta} \sum_n {1\over (1 + ({\beta\over 2\pi n})^2 \Pi^M_n) ({2\pi (r-n)\over \beta} + \Sigma^M_{r-n}) } 
\\
& ~~~ + {14\over\beta} \sum_n {1\over (1 + \Xi^M_n) ({2\pi (r-n)\over \beta} + \Sigma^M_{r-n}) } 
- {1\over \beta} \sum_n {1\over (1+\Xi^G_n)({2\pi (r-n)\over \beta} + \Sigma^G_{r-n})}
\\
&~~~ - {1\over \beta} \sum_n {1\over (1+({\beta\over 2\pi n})^2 \Pi^G_n) ({2\pi (r-n)\over \beta}+\Sigma^G_{r-n})},
\label{N=2vectorSD}
\fe
\ie
\Pi^M_n &=  {4\over\beta} \sum_m {1\over ({2\pi m\over \beta})^2 + \Pi^V_m} - {4\over \beta} \sum_r {1\over ({2\pi r\over \beta})^2 + {\beta\over 2\pi r} \Sigma_r^V} + {12\over \beta} \sum_m {1\over (({2\pi m\over \beta})^2 + \Pi^M_m) (1+\Xi^M_{n-m})} 
\\
&~~~ + {12\over \beta} \sum_r {1\over ({2\pi r\over \beta} + \Sigma^M_r)({2\pi (n-r)\over \beta} + \Sigma^M_{n-r})} + {4\over \beta} ({2\pi n\over \beta})^2 \sum_r {1\over (({2\pi r\over \beta})^3 + \Sigma^V_r)({2\pi (n-r)\over \beta} + \Sigma^M_{n-r})} 
\\
&~~~  + {2\over \beta\mu} - {8\over\beta\mu} {1 \over 1+ ({\beta\over 2\pi n})^2 \Pi^M_n },
\\
\Sigma^M_r &= {12\over \beta} \sum_n {1\over (({2\pi n\over \beta})^2 + \Pi^M_n)({2\pi (r-n)\over \beta} + \Sigma^M_{r-n})} + {2\over \beta} \sum_n {1\over (1+ ({\beta\over 2\pi n})^2 \Pi^M_n)( ({2\pi(r-n)\over \beta})^3 + \Sigma^V_{r-n})}
\\
&~~~ + {2\over\beta} \sum_n {1\over (1+\Xi^M_n)(({2\pi(r-n)\over\beta})^3 + \Sigma^V_{r-n})}
+ {4\over \beta} \sum_n {1\over (({2\pi n\over \beta})^2 + \Pi^V_n)({2\pi(r-n)\over \beta} + \Sigma^M_{r-n})}
\\
&~~~ - {2\over\beta\mu}{1\over {2\pi r\over \beta} + \Sigma^M_r},
\\
\Xi^M_n &= {6\over \beta} \sum_m {1\over (({2\pi m\over \beta})^2 + \Pi^M_m)(({2\pi (n-m)\over \beta})^2 + \Pi^M_{n-m})} + {4\over \beta} \sum_r {1\over (({2\pi r\over \beta})^3 + \Sigma^V_r)({2\pi(n-r)\over\beta} + \Sigma^M_{n-r})},
\label{N=2matterSD}
\fe
\ie
\Pi^G_n &= {1\over \beta} ({2\pi n\over\beta})^2 \sum_r {1\over (({2\pi r\over \beta})^3 + \Sigma^V_r)({2\pi(n-r)\over\beta} + \Sigma^G_{n-r})} - {2\over \beta\mu} {1\over 1+ ({\beta\over 2\pi n})^2 \Pi^G_n},
\\
\Sigma^G_r &= {1\over\beta} \sum_n {1\over (({2\pi n\over\beta})^2 + \Pi^V_n)({2\pi(r-n)\over \beta} + \Sigma^G_{r-n})} + {1\over 2\beta} \sum_n {1\over (1+\Xi^G_n)(({2\pi(r-n)\over\beta})^3+\Sigma^V_{r-n})}
\\
&~~~ + {1\over 2\beta} \sum_n {1\over (1+ ({\beta\over 2\pi n})^2 \Pi^G_n)(({2\pi(r-n)\over\beta})^3+\Sigma^V_{r-n})}
- {1\over 2\beta\mu} {1\over {2\pi r\over\beta} + \Sigma^G_r},
\\
\Xi^G_n &= {1\over\beta} \sum_r {1\over (({2\pi r\over\beta})^3 + \Sigma^V_r)( {2\pi(n-r)\over\beta} + \Sigma^G_{n-r})},
\label{N=2ghostSD}
\fe
\ie
{\mu \over 2} &= {2\over \beta} \sum_n {1\over ({2\pi n\over\beta})^2 + \Pi^V_n} - {5\over 2\beta} \sum_r {1\over ({2\pi r\over\beta})^2 + {\beta\over 2\pi r} \Sigma^V_r} + {7\over \beta} \sum_n {1\over ({2\pi n\over\beta})^2 + \Pi^M_n}
\\
&~~~ - {4\over\beta} \sum_n {1\over ({2\pi n\over\beta} + {\beta\over 2\pi n}\Pi^V_n)^2}
+ {4\over\beta} \sum_r {1 \over ({2\pi r\over\beta} + ({\beta\over 2\pi r})^2 \Sigma^V_r)^2}
-{14\over\beta} \sum_n {1\over ({2\pi n\over\beta} + {\beta\over 2\pi n}\Pi^M_n)^2}
\\
&~~~ + {7\over\beta} \sum_r {1\over ({2\pi r\over\beta} + \Sigma^M_r)^2} + {1\over \beta}\sum_n {1\over ({2\pi n\over\beta} + {\beta\over 2\pi n}\Pi^G_n)^2} - {1\over 2\beta} \sum_r {1\over ({2\pi r\over \beta} + \Sigma^G_r)^2}.
\label{N=2muSD}
\fe

\subsection{Free energy of BFSS in the $\cN=2$ gauge-fixing condition}\label{N=2free}
The free energy can be approximated using the mean-field method \cite{Kabat:1999hp}. It consists of four parts,
\ie
\B F_0=&-{1\over 2}{\rm log}\,\mu-\sum_l {\rm log}\,\sigma^2_l+\sum_r {\rm log}\,a_r-{7 \over 2}\sum_l {\rm log}\,\Delta^2_l+7\sum_r {\rm log}\,g_r
\\&-{7\over 2}\sum_l {\rm log}\,\epsilon^2_l+\sum_{l\neq 0} {\rm log}\,s_l-2\sum_r {\rm log}\,t_r+\sum_l {\rm log}\,u_l,
\fe
\ie
\B F_2=& \sum_l \left(({2\pi l\over \B})^2\sigma^2_l-1\right)+\sum_r \left(i({2\pi r\over \B})^3 a_r+1\right)+{7\over 2}\sum_l \left(({2\pi l\over \B})^2\Delta^2_l-1\right)
\\&
+7\sum_r \left(-i{2\pi r\over \B}g_r+1\right)+{7\over 2}\sum_l \left( \epsilon_l^2-1\right)+\sum_{l\neq 0} \left(({2\pi l\over \B})^2 s_l+1\right)
\\&
+2\sum_r \left(i{2\pi r\over \B}t_r-1\right)-\sum_l \left(u_l-1\right),
\fe
\ie
\B F_4=&-{2\over\B}\sum_{r,s} {2\pi r\over \B} {2\pi s\over \B} a_r a_s+{3i\over \B}\sum_{l,r}  {2\pi r\over \B} a_r\sigma_l^2+{1\over \B}\sum_{l,m}\sigma^2_l\sigma^2_m+{14\over \B}\sum_{l,m}\Delta^2_l\sigma^2_m
\\&
+{14i\over \B}\sum_{l,r}\Delta^2_l{2\pi r\over \B}a_r+{7\over\B}\sum_l {\Delta^2_l\over \mu}+{5i\over 2\B}\sum_r{2\pi r\over \B}{a_r\over \mu}+{2\over \B}\sum_l {\sigma^2_l\over\m},
\fe
\ie
\B F_3=&-{4\over\B}\sum_l ({2\pi l\over \B})^2{\sigma_l^4\over\m}-{4\over\B}\sum_r ({2\pi r\over\B})^4 {a_r^2\over\m}+{14\over \B}\sum_{l+r+s=0}({2\pi r\over \B})^2\Delta^2_l a_r g_s
\\&
+{14\over \B}\sum_{l+r+s=0}\epsilon^2_l a_r g_s-{14\over \B}\sum_{l+r+s=0}\sigma^2_l g_r g_s-{14\over \B}\sum_l({2\pi l\over \B})^2{\Delta_l^4\over\m}
\\&
-{7\over\B}\sum_r {g_r^2\over \m}+{21\over\B}\sum_{l+m+n=0} \Delta_l^2\Delta_m^2\epsilon_n^2-{42\over\B}\sum_{l+r+s=0} \Delta_l^2 g_r g_s
\\&+{1\over\B}\sum_{l+r+s=0} \left(\sigma_l^2 t_r t_s-u_l t_r a_s+({2\pi l\over \B})^2s_l t_r a_s\right)
\\&
+{1\over 2\B}\sum_r {t^2_r \over\m}+{1\over \B}\sum_l ({2\pi l\over\B})^2{s^2_l\over\m}.
\fe
Here $\B F_0$ is the free energy of the trial action, $\B F_2$ is that from the difference between the quadratic part of the BFSS action and the trial action, and $\B F_3$ and $\B F_4$ from the cubic and quartic couplings.

\section{1D $\cN = 4$ SUSY}
\label{app:N=4}

The 1D $\cN=4$ superspace is the same as the 4D $\cN=1$ superspace, so we will mostly adopt the 4D notations here. We will work in the Euclidean signature.

\subsection{Superspace and Convention}

Let us start with the 4D $\cN=1$ superspace. We will denote $(\bf 2,1)$ and $(\bf 1,2)$ of 4D rotation group $SO(4)=SU(2)_\ell\times SU(2)_r$ by the lower $\A$ and $\dot \A$, respectively. The $SO(4)$ invariant symbols $\epsilon_{\A\B}$, $\epsilon_{\dot \A\dot \B}$ and their inverses are normalized by 
\ie
\epsilon^{12} =\epsilon^{\dot 1\dot 2}=+1,
~~\epsilon_{12}=\epsilon_{\dot 1\dot 2}=-1.
\fe
We also have two other $SO(4)$ invariant symbols $\sigma^\mu_{\A\dot \A}$ and $\bar\sigma^{\mu\dot \A \A}$ with $\mu=0,1,2,3$ defined as
\ie
&\sigma^\mu_{\A\dot \A} = (iI, \vec{\sigma}),\\
&\bar\sigma^{\mu \dot \A \A} = \epsilon^{\A\B}\epsilon^{\dot \A\dot \B} \sigma^\mu _{\B\dot \B} = (iI,-\vec{\sigma}).
\fe
Under dimension reduction, the time direction is singled out and the $SO(4)$ rotation group is broken to the $SO(3)_R\cong SU(2)_R$ $R$-symmetry of the 1D quantum mechanics. In addition, the 4D $\cN=1$ SUSY theory has an $U(1)_R$ symmetry, so the full $R$-symmetry of the 1D $\cN=4$ quantum mechanics is $U(1)_R\times SU(2)_R$. Note that the $SU(2)_R$ is embedded in the 4D rotation group $SU(2)_\ell\times SU(2)_r$ in such a way that the zeroth component of $\sigma^\mu$ and $\bar\sigma^\mu$, i.e. $i\delta_{\A\dot \B}$, is its invariant symbol.

The supercharges $Q_\A, \bar Q_{\dot \A}$ and the supercovariant derivatives ${\cal D}_\A, \bar{\cal D}_{\dot \A}$ are defined as
\ie
&Q_\A = {\partial\over \partial\theta^\A} - \delta_{\A\dot \B} \bar \theta^{\dot \B} \partial_\tau,\\
&\bar Q_{\dot \A} = -{\partial\over \partial\bar\theta^{\dot \A}} +\delta_{\B\dot \A} \theta^{\B} \partial_\tau,\\
&{\cal D}_\A = {\partial\over \partial\theta^\A} +\delta_{\A\dot \B} \bar \theta^{\dot \B} \partial_\tau,\\
&\bar{\cal D}_{\dot \A} =- {\partial\over \partial\bar\theta^{\dot \A}} -\delta_{\B\dot \A}  \theta^{ \B} \partial_\tau.
\fe
The only nontrivial anticommutators are
\ie
\{Q_\A , \bar Q_{\dot \A}\} = 2 \delta_{\A\dot \A} \partial_\tau,~~
\{{\cal D}_\A,\bar{\cal D}_{\dot \A} \} = -2 \delta_{\A\dot \A} \partial_\tau.
\fe

As opposed to the $\cN=2$ case, we will distinguish the upper index with the lower one. The indices are raised and lowered by the invariant symbols $\epsilon_{\A\B}$, $\epsilon^{ \A\B}$ and their inverses,
\ie
\psi^\A = \epsilon^{\A\B}\psi_\B,~~\psi_{ \A} = \epsilon_{ \A \B} \psi^{ \B},
\fe
and similarly for the dotted indices.
We introduce the following notation for the spinor bilinear,
\ie
&\chi \psi = \chi^\A \psi_\A =- \epsilon^{\A\B} \chi_\A\psi_\B= \psi \chi,\\
&\overline{ \psi} \overline{\chi} = \overline{\psi}_{\dot \A} \overline{\chi}^{\dot \A} = \epsilon^{\dot \A\dot \B} \overline{\psi}_{\dot \A} \overline{\chi}_{\dot \B}=\overline{\chi}\overline{\psi}.
\fe
In particular,
\ie
&\theta_\A \theta_\B = {1\over 2} \theta\theta \epsilon_{\A\B} ,~~~
\theta^\A \theta^\B =- {1\over 2} \theta\theta \epsilon^{\A\B} ,\\
&\bar\theta_{\dot \A}\bar \theta_{\dot \B} =- {1\over 2}\bar \theta\bar\theta \epsilon_{\dot \A\dot \B} ,~~~
\bar\theta^{\dot \A}\bar \theta^{\dot \B} ={1\over 2}\bar \theta\bar\theta \epsilon^{\dot \A\dot \B}.
\fe

\subsection{Chiral Multiplet}

The chiral superfield $\Phi(\tau,\theta,\bar\theta)$ satisfies
\ie
\bar{\cal D}_{\dot a} \Phi(\tau,\theta,\bar\theta)=0.
\fe
We can solve this constraint by first introducing 
\ie
y =\tau +\delta_{\A\dot \A} \theta^\A\bar\theta^{\dot \A}.
\fe
By using $\bar{\cal D}_{\dot \A} \theta_\A =0$ and $\bar{\cal D}_{\dot \A}  y=0$ we find that any superfield $\Phi(y,\theta)$ that is a function of $y$ and $\theta$ is chiral.

We can expand $\Phi(y,\theta)$ in terms of component fields,
\ie
\Phi^a(y,\theta) = \phi^a(y) + \sqrt{2} \psi_\A^a(y)\theta^\A + i f^a(y) \theta^2,
\fe
with $a$ being the index of some flavor symmetry group $G$.
Note that we are working in the Euclidean signature so there is an $i$ for the auxiliary field $f(y)$ to make its quadratic term in the action positive.

The SUSY transformations are
\ie
& [ Q_\A , \phi^a ] = i\sqrt{2} \psi_\A^a,~~~
\{ Q_\A , \psi_\B^a \} = \sqrt{2} \epsilon_{\A\B} f^a
,~~~[Q_\A,f^a]=0,\\
& [ Q_\A , \bar\phi^a ] =0,~~~
\{ Q_\A , \bar\psi^{a}_{\dot\B} \} =-i \sqrt{2} \delta_{\A\dot\B} \dot {\overline{\phi^a}}
,~~~[Q_\A,\bar f^a]=\sqrt{2}\delta_{\A\dot \B} \dot{\overline\psi}^{a\dot\B}.\\
& [ \bar Q_{\dot\A} , \phi^a ] = 0,~~~
\{ \bar Q_{\dot\A} , \psi_\B^a \} =-i \sqrt{2} \delta_{\B\dot\A} \dot\phi^a
,~~~[\bar Q_{\dot\A},f^a]=-\sqrt{2} \delta_{\B\dot\A}\dot\psi^{a\B},\\
& [ \bar Q_{\dot\A} , \bar\phi^a ] = i\sqrt{2} \bar\psi^{a}_{\dot \A},~~~
\{ \bar Q_{\dot\A} , \bar\psi^{a}_{\dot\B} \} =- \sqrt{2}\epsilon_{\dot\A\dot\B} \bar f^a
,~~~[\bar Q_{\dot\A},\bar f^a]=0.
\fe

The Lagrangian with a general superpotential $W(\Phi)$ is given by\footnote{Note that the complex conjugate is defined by first Wick rotating to Lorentzian signature, taking the complex conjugate, and then Wick rotating back to the Euclidean signature.}
\ie
\mathcal{L} = \int d^2\theta d^2\bar\theta ~ \bar\Phi\Phi+\left(
\int d^2\theta~W(\Phi) + h.c.
\right).
\fe

\subsection{Vector Multiplet}

Consider a real superfield $V(\tau,\theta,\bar\theta)$ in the adjoint representation of some gauge group with the following component field expansion
\ie
V (\tau,\theta,\bar\theta) &= R(\tau) + \theta \chi +\bar\theta \bar\chi (\tau)+\theta\theta M (\tau)+ \bar\theta \bar\theta \bar M(\tau)\\
& \hspace{.5in} +\theta^a \bar\theta^{\dot b}\left(i
\delta_{a\dot b} A_0(\tau) + \sigma^{i}_{a\dot b} X^i (\tau)
\right)
 +\theta\theta\bar\theta \bar\lambda(\tau)
+\bar\theta\bar\theta \theta \lambda(\tau)
+{1\over 2} \theta\theta\bar\theta\bar\theta D(\tau).
\fe

The gauge transformation is
\ie
e^{-2V} \rightarrow e^{-2i \bar\Lambda} e^{-2V} e^{2i\Lambda}
\fe
where $\Lambda$ and $\bar\Lambda$ are chiral and anti-chiral superfield, respectively.

The field-strength superfield is defined by
\ie
W_\A = -{1\over 8} \bar{\cal D}_{\dot \A} \bar {\cal D}^{\dot \A} e^{2V}{\cal D}_{\A} e^{-2V}
\fe
The kinetic action for the vector multiplet is then given by
\ie
\mathcal{L} = \Tr\left( \int d^2\theta ~{1\over 4g_{YM}^2} W^\A W_\A +h.c.\right).
\fe

\subsection{$\cN=4$ SUSY Ward Identities}

The $\cN=4$ SUSY Ward identities for the chiral multiplet can be similarly derived as in the $\cN=2$ case. We define the self-energies $\sigma(p )$, $\eta(p)$, and $h(p)$ for the boson $\phi^a$, the auxiliary field $f^a$, and the fermion $\psi^a_\A$, $\bar \psi^{a \A}$ to be 
\ie
&\la \bar\phi^a(p) \phi^b(-p) \ra = {\delta^{ab}\over p^2+\sigma(p)},~~~\\
&\la \bar f^a (p ) f^b(-p) \ra = { \delta^{ab} \over 1+ \eta(p)},~~~\\
&\la \bar \psi^{a}_{\dot \A} (p)\psi^b_\B (-p) \ra= -i { \delta^{ab}\delta_{\dot \A\B}\over p + h( p)},
\fe
respectively. Note that $\sigma(-p)=\sigma(p)$, $\eta(-p)=\eta(p)$, and $h(-p)=-h(p)$.

The SUSY Ward identity is an exact relation between the exact propagators for fields in the same SUSY multiplet. Suppose the SUSY is unbroken in the model, we have
\ie
0 = \la \{ Q_\A ,\phi^a(\tau) \bar\psi^{b}_{\dot \B}(\tau')\}\ra
= i\sqrt{2} \la \psi^a_\A(\tau) \bar\psi^{b}_{\dot\B}(\tau')\ra
-i\sqrt{2}\delta_{\A\dot \B} \la \phi^a(\tau) \dot{\overline{\phi^b}}(\tau')\ra.
\fe
Going to the momentum space, this implies
\ie
\la \bar\psi^{b}_{\dot \B}(p)\psi^a_\A(-p)\ra
=-ip\delta_{\A\dot\B} \la \bar\phi^a(p)\phi^b(-p)\ra
\fe
that is
\ie
\sigma(p)= ph(p).
\fe
This is an exact relation between the self-energies for the boson $\phi^a$ and the fermion $\psi^a_\A$.

Similarly, we can consider
\ie
0 = \la \{ Q_\A , \psi^a_\B (\tau)\bar f^b(\tau')\}\ra
=\sqrt{2}\epsilon_{\A\B} \la f^a(\tau)  \bar f^b(\tau')\ra
-\sqrt{2} \delta_{\A\dot\gamma}\la\psi^a_\B(\tau) \dot{\overline \psi}^{b\dot\gamma}(\tau')\ra.
\fe
In the momentum space it is
\ie
\epsilon_{\A\B}\la \bar f^b(p)f^a(-p) \ra =-  ip \delta_{\A\dot\gamma}\epsilon^{\dot\gamma \dot\rho}\la {\overline \psi}^{b}_{\dot\rho}(p) \psi^a_\B(-p)\ra.
\fe
Using $\epsilon^{\dot \gamma\dot\rho} \delta_{\A\dot\gamma}\delta_{\B\dot\rho} = - \epsilon_{\A\B}$, we have
\ie
\eta(p) ={h(p)\over p}.
\fe
In summary, we have obtained the exact relation
\ie 
h(p) = {\sigma(p)\over p} = p \eta(p),\label{N=4susyward}
\fe
which takes exactly the same form as in the $\cN=2$ case (\ref{ward}).

\subsection{Vector Multiplet Zero Mode Action under $\cN=4$ Gauge-Fixing}

In this subsection we will truncate the vector multiplet action to the zero mode sector in the $\cN=4$ gauge-fixing condition (\ref{N=4gf}). We only need to focus on the component fields $A_0, X^i,$ and $R$ since $D=\dot A_0 -{1\over 2}\ddot R$ is set to zero in the zero mode truncation. We will write $A_0$ and $X^i$ collectively as $A_\mu$ with $\mu =0,1,2,3$.

The field-strength superfield is
\ie
W_\A= -{1\over 8} \bar{\cal D}_{\dot \A} \bar {\cal D}^{\dot \A} e^{2V} {\cal D}_\A e^{-2 V}\label{W}.
\fe
To get the terms with only zero modes, we can replace every $\cal D$ in (\ref{W}) by $\partial/\partial \theta$ and similarly for $\cal{\bar D}$. Using
\ie
& \hspace{-.5in} \exp\left({2R + 2\theta^\A\bar\theta^{\dot \B} \sigma^\mu_{\A\dot \B} A_\mu}\right) \\
&= e^{2R} + \int_0^1 ds \, e^{2sR }\, 2\theta^\A\bar\theta^{\dot \B} \sigma^\mu_{\A\dot \B} A_\mu e^{2(1-s)R}\\
& \hspace{.5in} + \int^1_0 ds \int^s_0 du\, e^{2uR}\, 2\theta^\A\bar\theta^{\dot \B} \sigma^\mu_{\A\dot \B} A_\mu e^{2(s-u)R}\, 2\theta^\gamma\bar\theta^{\dot \delta} \sigma^\nu_{\gamma\dot \delta} A_\nu}e^{2(1-s)R\\
&=e^{2R} + 2\int_0^1 ds \, e^{2sR }\, \theta^\A\bar\theta^{\dot \B} \sigma^\mu_{\A\dot \B} A_\mu e^{2(1-s)R}\\
& \hspace{.5in} -2\theta^2\bar\theta^2\int^1_0 ds\int^s_0du \, e^{2uR} A_\mu e^{2(s-u)R} A^\mu e^{2(1-s)R},
\fe
we then expand
\ie
W_\A& =
 -2\theta_\A \int^1_0 ds\int^s_0du \, e^{2(1-u)R} A_\mu e^{-2(s-u)R} A^\mu e^{-2(1-s)R}\\
&
\hspace{1in} -\theta^\gamma \epsilon^{\dot \delta\dot \epsilon} \sigma^\mu_{\gamma\dot \delta} \sigma^\nu_{\A\dot \epsilon} \int^1_0 ds e^{2sR}A_\mu e^{2(1-s)R}\int^1_0 due^{-2sR} A_\nu e^{-2(1-s)R}.
\fe
We can simplify the above equation by using the identity
\ie
\epsilon^{\dot \delta\dot \epsilon}\sigma^\mu_{\gamma\dot \delta}\sigma^\nu_{\A\dot \epsilon}=\epsilon_{\A\B}\sigma^\mu_{\gamma\dot \delta}\sigma^{\nu \dot \delta \B}
=-\epsilon_{\A\gamma} \delta^{\mu\nu} -2i\epsilon_{\A\B} (S^{\mu\nu}_L)_\gamma^{~\B},
\fe  
where
\ie
(S^{\mu\nu}_L)_\A^{~\B} = {i\over 4} (\sigma^\mu\bar\sigma^\nu -\sigma^\nu \bar\sigma^\mu )_\A^{~\B}.
\fe
Now
\ie
W_\A &= \theta_\A\left(-2 \int^1_0 ds\int^s_0du \, e^{2(1-u)R} A_\mu e^{-2(s-u)R} A^\mu e^{-2(1-s)R}\right.\\
& \hspace{.5in} \left.
+  \int^1_0 ds e^{2sR}A_\mu e^{2(1-s)R}\int^1_0 due^{-2sR} A^\mu e^{-2(1-s)R}\right)\\
& \hspace{.5in} -2i (S^{\mu\nu}_L)_\A^{~\delta}\theta_\delta\int^1_0 ds e^{2sR}A_\mu e^{2(1-s)R}\int^1_0 due^{-2sR} A_\nu e^{-2(1-s)R},
\fe
where we have used $\epsilon_{\A\B}\epsilon^{\gamma\delta} = -\delta _\A^{~\gamma} \delta_\B^{~\delta}+\delta_\A^{~\delta} \delta_\B^{~\gamma}$ and $(S^{\mu\nu}_L)_\A^{~\A}=0$. Note that in the case of $R=0$, only the term with $S^{\mu\nu}_L$ survives and is contracted with $[A_\mu,A_\nu]$, as expected in the usual Wess-Zumino gauge.

The zero mode terms in the Lagrangian can then be written as
\ie
\mathcal{L}=& {1\over4g_{YM}^2} \epsilon^{\A\B} W_\A W_\B\Big|_{F} +h.c.\\
\ni&
-{2\over g_{YM}^2} \left(-2 \int^1_0 ds\int^s_0du \, e^{2(1-u)R} A_\mu e^{-2(s-u)R} A^\mu e^{-2(1-s)R}\right.\\
&\left.
+  \int^1_0 ds e^{2sR}A_\mu e^{2(1-s)R}\int^1_0 due^{-2uR} A^\mu e^{-2(1-u)R}\right)^2\\
&-{2\over g_{YM}^2}\left[  (\delta^{\mu\rho}\delta^{\nu\sigma}-\delta^{\mu\sigma}\delta^{\nu\rho})+\epsilon^{\mu\nu\rho\sigma}\right]{\cal A}^+_\mu{\cal A}^-_\nu{\cal A}^+_\rho{\cal A}^-_\sigma,
\fe
where
\ie
{\cal A}_\mu ^{\pm} \equiv\int^1_0 ds e^{\pm2sR}A_\mu e^{\pm2(1-s)R}.
\fe
We have used
\ie
(S^{\mu\nu}_L)_\A^{~\gamma} (S^{\rho\sigma}_L)_\gamma^{~\A} =  {1\over 2} (\delta^{\mu\rho}\delta^{\nu\sigma}-\delta^{\mu\sigma}\delta^{\nu\rho})+{1\over2} \epsilon^{\mu\nu\rho\sigma},
\fe
with $\epsilon^{\mu\nu\rho\sigma}$ normalized as $\epsilon^{0123}=1$.

\section{Computation of $x_n$}\label{appx}

In the following we will frequently encounter divergent sums over integers or half-integers.  Since these divergences come from expanding propagators in series of say $(2\pi n \over \B)^2$, the sums should be understood as regularized sums cut off at $n \approx \sqrt{\B / \sigma_0}$.

Using (\ref{hr}), (\ref{ansatz}) and (\ref{sn}), the equation for $x_n$ is
\ie
(1 + x_n)^2 = \left[ {2 \over \B\sigma_0}{1 \over {2\pi n \over \B} + s_n (1 + x_n)^{-1} }  +  A_n \right]
\left[{2 \over \B\sigma_0}{1 \over {2\pi n \over \B} + s_n (1 + x_n) } + B_n   \right]^{-1},
\fe
where
\ie
A_n &= {1 \over \pi n} \left( 
\sum_{k \neq 0, n} {n-k \over k} {1 \over s_k s_{n-k}}{1 + x_{n-k} \over 1 + x_k} + \sum_r {1 \over h_r h_{n-r}} 
+ {\B \over 2\pi |n| \eta_0} \sqrt{ {\B\sigma_0\over 2} } \right) \left[ 1+ \cO(\B^{-6/5}) \right] \\
B_n &= {1 \over \pi} \sum_{k \neq 0, n} {1\over k s_k s_{n-k}} {1 \over (1 + x_k) (1 + x_{n-k})} \left[ 1 + \cO(\B^{-6/5}) \right].  \label{AB}
\fe


Since $A_n$ contains $\eta_0$, to solve for $x_n$ we also need to know $\eta_0$ to the relevant order,
\ie
\eta_0 = {1 \over \B\sigma_0^2} + \mathcal{O}(\B^{3/5}).
\fe
The Schwinger-Dyson equation for $x_n$ can thus be written as
\ie
{2\over\B\sigma_0} ({2\pi n\over \B}) {(1+x_n)^2-1\over s_n^2}= A_n -(1+x_n)^2 B_n 
+\cO(\B^{-{11/ 5}}),
\fe
or
\ie
& ({\B\over 2\pi n})^2 {2\over \B} \left[ n \sum_{k \neq 0,n} {1\over k s_k s_{n-k}} \left( {1+x_{n-k} \over 1+x_k}
  - {(1+x_n)^2\over (1+x_k)(1+x_{n-k})} \right)
  \right.\\
  & \hspace{1in} \left.
  - \sum_{k \neq 0,n} {1\over s_k s_{n-k}} {1+x_{n-k} \over 1+x_k}  + \sum_r {1\over h_r h_{n-r}} 
+{\B \over 2\pi |n| \eta_0} \sqrt{ {\B\sigma_0\over 2}  }
\right]
\\
& \hspace{.5in} = (1+x_n)^2-1+ \cO(\B^{-3/5}).
\fe
Note that the correction in replacing $s_n$ by $s_n^{(0)}$ is of order $\cO(\B^{-3/5})$ because $A_n-(1+x_n)^2B_n \sim \B^{-1}$.
Write
\ie
s_k = \text{sign}(k) \sqrt{2\over \B\sigma_0} + s_k^{(1)},~~~ h_r = \text{sign}(r) \sqrt{2\over\B\sigma_0} + h_r^{(1)}\label{s0},
\fe
where $s_k^{(1)}, h_r^{(1)}$ are of order $\B^{-2/5}$.

$s_n^{(1)}$ can be determined by expanding the Schwinger-Dyson equation for $s_n$ as written earlier. Let us also write for now the Schwinger-Dyson equation for $h_r$ to the relevant order,
\ie
h_r &= {2\over \B\sigma_0} {1\over h_r } + {1\over \pi} \sum_{n \neq 0} {1\over k s_k(1+x_k) h_{r-k}} + \cO(\B^{-1}).
\fe
We have
\ie
& s_k^{(1)} = {\B\sigma_0\over 4\pi}\sum_{\ell\neq0,k} {\text{sign}(\ell) \text{sign}(k-\ell)\over\ell}+\mathcal{O}(\B^{-1}),\\
& h_r^{(1)} =  {\B\sigma_0\over 4\pi} \sum_{\ell \neq 0} {\text{sign}(\ell) \text{sign}(r-\ell)\over \ell}+\mathcal{O}(\B^{-1}).\label{s1}
\fe

The equation for $x_n$ can be written as
\ie
& (1+x_n)^2-1 \\
& = ({\B\over 2\pi n})^2 {2\over \B} \left[ {\B\sigma_0\over 2} n \sum_{k \neq 0,n} {\text{sign}(k) \text{sign}(n-k)\over k} \left( {1+x_{n-k} \over 1+x_k}- {(1+x_n)^2\over (1+x_k)(1+x_{n-k})} \right)
\right.
\\
&\left.
\hspace{.5in} -{\B\sigma_0\over2}  \sum_{k\neq 0,n} \text{sign}(k)\text{sign}(n-k)\left( {1+x_{n-k} \over1+x_k } -1 \right)
\right.
\\
& \left. + 2({\B\sigma_0\over 2})^{3\over 2}\left( \sum_{k \neq 0,n} \text{sign}(n-k) s_k^{(1)} - \sum_r \text{sign}(n-r) h_r^{(1)} \right) 
+{\B \over 2\pi |n| \eta_0} \sqrt{ {\B\sigma_0\over 2}  }
\right] + \cO(\B^{-3/5}).
\fe
Note that we have set $x_n$ to zero in the third line since it is already of order $\B^{0}$. Using
\ie
& \sum_{k \neq 0,n} \text{sign}(n-k) s_k^{(1)} - \sum_r \text{sign}(n-r) h_r^{(1)} 
\\
& = {\B\sigma_0\over 4\pi} \sum_{\ell \neq 0} {1\over |\ell|} \left[ \sum_{k \neq 0,n,\ell} {\text{sign}(n-k) \text{sign}(k-\ell)} - \sum_r \text{sign}(n-r) \text{sign}(r-\ell) \right] + \mathcal{O}(\B^{-1}),
\fe
we have
\ie
& {2\pi^2n^2} \left[ (1+x_n)^2-1\right] =  {\B^2\sigma_0\over 2} n \sum_{k \neq 0,n} {\text{sign}(k) \text{sign}(n-k)\over k} \left(  {1+x_{n-k} \over 1+x_k}- {(1+x_n)^2\over (1+x_k)(1+x_{n-k})} \right) 
\\
&+({\B\sigma_0\over 2})^{5\over 2}{\B\over \pi}\sum_{\ell \neq 0} {1\over |\ell|} \left[ \sum_{k \neq 0,n,\ell} {\text{sign}(n-k) \text{sign}(k-\ell)} - \sum_r \text{sign}(n-r) \text{sign}(r-\ell) \right] \\
&+{\B ^3\sigma_0^2\over 2\pi |n| } \sqrt{ {\B\sigma_0\over 2}  }
+ \cO(\B^{-3/5}) ,
\fe
where we have used $\eta_0 = 1/\B^2\sigma_0+\cO(\B^{3/5})$. Note that on the RHS, even though the $\B^2\sigma_0$ term is naively of order $\B^{3/5}$, but the two pieces in the bracket cancel to leading order, and so the term is actually of order $\B^0$.  Without further cancellation, the whole RHS is of order 1.  On the other hand, the LHS is of order $\B^{-3/5}$.  Therefore we have an inconsistency unless the RHS vanishes to leading order.

Write
\ie
x_n =  ({\B\sigma_0\over 2})^{3\over 2} {a_n\over \pi} +\cO(\B^{-6/5}),
\fe
where $a_n\to 0$ as $n\to \infty$. We assume that $a_n$ is of order 1. The equation for $x_n$ becomes
\ie
& 0 = 2 {n} \sum_{k \neq 0,n} {\text{sign}(k) \text{sign}(n-k)\over k} \left(a_{n-k} -  a_n \right) +{3\over |n| } + \mathcal{O}(\B^{-3/5}).
\fe
Here we used
\ie
& \sum_{\ell \neq 0} {1\over |\ell|} \left[ \sum_{k \neq 0,n,\ell} {\text{sign}(n-k) \text{sign}(k-\ell)} - \sum_r \text{sign}(n-r) \text{sign}(r-\ell) \right]
\\
&= \sum_{\ell \neq 0} {1\over |\ell|} \left(\text{sign}(n)\text{sign}(\ell) + \delta_{n\ell} \right)
= {1\over |n|}.
\fe

$a_n$ now obeys the equation

\ie
4 \left[ \mathrm{sign}(n)HN(|n|-1)+{1\over 2n} \right] a_n = 3{\text{sign}(n)\over n^2} + 2\sum_{k \neq 0,n} {\mathrm{sign}(k) a_k\over |n-k|}.
\fe
where
\ie
HN(m)\equiv\sum_{k=1}^{m}{1\over k}.
\fe
A solution is given by
\ie
a_n={3\over 2|n|},
\fe
that is,
\ie
x_n = ({\B\sigma_0\over 2})^{3\over2} {3\over 2\pi |n|} +\cO(\B^{-6/5}).\label{x}
\fe
This confirms our intuition that SUSY Ward identity should hold for the nonzero modes to leading order in the low temperature expansion (\ref{approxward}) ($x_n$ essentially measures the violation of the the SUSY Ward identity).
Together with (\ref{ansatz}), (\ref{s0}), and (\ref{s1}), we have solved $\sigma_n$, $\eta_n$, and $h_r$ to the next-to-leading order in large $\B$ expansion in terms of $\sigma_0$.

\section{Computation of $s_n^{(1)}, h_r^{(1)}, s_n^{(2)}$ and $h_r^{(2)}$}  \label{app:shcorr}

In this appendix we will compute $s_n$ and $h_r$ to the second order, $\cO(\B^{-1})$. These terms are relevant for solving the Schwinger-Dyson equation for $\sigma_0$. Let us collect the equations for $s_n$ and $h_r$ to the relevant order.
\ie
& s_n = ({2\over\B\sigma_0})^2 {1\over s_n^3} \left( 1 - {4\pi n\over\B} {1\over s_n} \right)
+ {A_n B_n\over s_n} + {2\over \B\sigma_0} {1\over s_n^2}\left[ {A_n\over 1+x_n} + B_n (1+x_n) \right] + \cO(\B^{-8/5})
\\
&~~~= ({2\over\B\sigma_0})^2 {1\over s_n^3} \left( 1 - {4\pi n\over\B} {1\over s_n} \right)
+ {A_n B_n\over s_n} + {2\over \B\sigma_0} {1\over s_n^2}\left( A_n + B_n \right) + \cO(\B^{-8/5}) ,
\\
& h_r = {2\over \B\sigma_0} {1\over {2\pi r\over\B}+h_r} + {1\over \pi} \sum_{k \neq 0} {1\over k s_k (1+x_k) h_{r-k}}
+ \cO(\B^{-8/5})
\\
&~~~=  {2\over \B\sigma_0} {1\over h_r} - {4\pi r\over \B^2\sigma_0 h_r^2}+ {1\over \pi} \sum_{k \neq 0} {1\over k s_k h_{r-k}}
- {\B\sigma_0\over 2\pi} \sum_{k \neq 0} {\text{sign}(k) \text{sign}(r-k)\over k} x_k+ \cO(\B^{-8/5}),
\\
& A_n = {1\over \pi n} \left( n \sum_{k \neq 0,n} {1\over k s_k s_{n-k}} - \sum_{k \neq 0,n}{1\over s_k s_{n-k}} + \sum_r {1\over h_r h_{n-r}} \right) 
\\
&~~~~+ {\B\sigma_0\over 2\pi }\sum_{k\neq0,n}{n-k\over kn}\mathrm{sign}(k)\mathrm{sign}(n-k)(x_{n-k}-x_k)+
{2\text{sign}(n)\over \pi^2 n^2 } ( {\B\sigma_0\over 2} )^{{5\over2}}
+\cO(\B^{-8/5}),
\\
& B_n = {1\over \pi} \sum_{k \neq 0,n} {1\over k s_k s_{n-k}} -{\B\sigma_0 \over 2\pi} \sum_{k \neq 0,n} {\text{sign}(k) \text{sign}(n-k)\over k} ( x_k + x_{n-k} ) + \cO(\B^{-8/5}) .
\fe
Define
\ie
g = \sqrt{\B\sigma_0\over 2},~~~ C_k = \sum_{\ell \neq 0,k} {\text{sign}(\ell) \text{sign}(k-\ell)\over \ell}.
\fe
We can then write
\ie
& s_n = \text{sign}(n) g^{-1} + {g^2\over 2\pi} C_n + s^{(2)}_n,
\\
& h_r =\text{sign}(r) g^{-1} + {g^2\over 2\pi} C_r + h^{(2)}_n,
\fe
with $s_n^{(2)}$ and $h^{(2)}_n$ of $\B^{-1}$ order. $A_n$ and $B_n$ can be expanded as
\ie
& A_n = {g^2\over \pi} C_n - {g^5\over 2\pi^2} \sum_{k \neq 0,n} {\text{sign}(n-k) C_k
+ \text{sign}(k) C_{n-k}\over k}  + {g^5\over \pi^2 n} \sum_{k \neq 0,2n} (-1)^k \text{sign}(n-{k\over 2}) C_{k\over 2}
\\
& ~~~~+{g^2\over \pi }\sum_{k\neq0,n}{n-k\over kn}\mathrm{sign}(k)\mathrm{sign}(n-k)(x_{n-k}-x_k)
+{2\text{sign}(n)\over \pi^2 n^2 }g^5
+ \cO(g^8),
\\
& B_n = {g^2\over \pi} C_n - {g^5\over 2\pi^2} \sum_{k \neq 0,n} {\text{sign}(n-k) C_k
+ \text{sign}(k) C_{n-k}\over k}\\
&~~~~  - {g^2\over \pi}  \sum_{k \neq 0,n} {\text{sign}(k) \text{sign}(n-k)\over k} ( x_k + x_{n-k} )
+ \cO(g^8).
\fe
The equations for $s_n$ and $h_r$ are equivalent to
\ie
& (g s_n)^4 = 1 - {4\pi n\over \B s_n} + g^4 A_n B_n s_n^2 + g^2 s_n (A_n + B_n) + \cO(g^9)
\\
&~~~~~~~~= 1 - {4\pi n\over \B s_n} + {g^6\over \pi^2} C_n^2 + {2g^4 s_n C_n \over \pi}
- {g^7s_n\over \pi^2} \sum_{k \neq 0,n} {\text{sign}(n-k) C_k
+ \text{sign}(k) C_{n-k}\over k}  
\\
&~~~~~~~~ + {g^7s_n\over \pi^2 n} \sum_{k \neq 0,2n} (-1)^k \text{sign}(n-{k\over 2}) C_{k\over 2}
- {g^4 s_n\over \pi} \sum_{k \neq 0,n} {\text{sign}(k) \text{sign}(n-k)\over k} ( x_k + x_{n-k} )
\\
&~~~~~~~~  +{g^4s_n\over \pi }\sum_{k\neq0,n}{n-k\over kn}\mathrm{sign}(k)\mathrm{sign}(n-k)(x_{n-k}-x_k)
+{2\text{sign}(n)\over \pi^2 n^2 }g^7s_n
+\cO(g^9)
\\
&~~~~~~~~ = 1 - {4\pi g|n|\over \B} + {2g^3 \text{sign}(n) C_n \over \pi} + {2g^6 C_n^2\over \pi^2} 
\\
&~~~~~~~~ - {g^6\over \pi^2}\text{sign}(n) \sum_{k \neq 0,n} {\text{sign}(n-k) C_k + \text{sign}(k) C_{n-k}\over k}
\\
&~~~~~~~~ + {g^6\over \pi^2 |n|} \sum_{k \neq 0,2n} (-1)^k \text{sign}(n-{k\over 2}) C_{k\over 2} - {2g^3\over \pi}\text{sign}(n) \sum_{k \neq 0,n} {\text{sign}(k) \text{sign}(n-k)\over k} x_k \\
&~~~~~~~~ +{2\over \pi^2 n^2 }g^6 + \cO(g^9),
 \fe
 
 \ie
& (g h_r)^2 = 1 - {2\pi r\over \B h_r} + {g^2 h_r\over \pi} \sum_{k \neq 0} {1\over k s_k h_{r-k}} - {g^3\over \pi} \text{sign}(r) \sum_{k \neq 0} {\text{sign}(k) \text{sign}(r-k)\over k} x_k 
+ \cO(g^9)
\\
&~~~~~~~~= 1 - {2\pi g |r|\over \B} + {g^3 \over \pi}\text{sign}(r) C_r + {g^6 C_r^2\over 2\pi^2}  - {g^6\over 2\pi^2} \text{sign}(r) \sum_{k \neq 0} {\text{sign}(r-k) C_k
+ \text{sign}(k) C_{r-k}\over k}   
\\
&~~~~~~~~- {g^3\over \pi} \text{sign}(r) \sum_{k \neq 0} {\text{sign}(k) \text{sign}(r-k)\over k} x_k + \cO(g^9) .
\fe
$s^{(2)}_n$ and $h^{(2)}_r$ can then be solved to be
\ie
gs_n^{(2)} =&
-{g\pi n \over \B} +\text{sign}(n){g^6\over8\pi^2} C_n^2 
-{g^6\over 4\pi^2} \sum_{k\neq 0,n}
{ \text{sign}(n-k) C_k +\text{sign}(k) C_{n-k} \over k  }\\
&+{g^6\over 4\pi^2 n} \sum_{k\neq 0,2n} (-)^k \text{sign}(n-{k\over2}) C_{k/2}
-{g^3\over 2\pi} \sum_{k\neq0,n} 
{  \text{sign}(k) \text{sign}(n-k)  x_k\over k }+{g^6\over2\pi^2n^2}+\cO(g^9),\\
gh_r^{(2)} =&
-{g\pi r \over \B} +\text{sign}( r){g^6\over8\pi^2} C_r^2 
-{g^6\over 4\pi^2} \sum_{k\neq 0}
{ \text{sign}(r-k) C_k +\text{sign}(k) C_{r-k} \over k  }\\
&
-{g^3\over 2\pi} \sum_{k\neq0} 
{  \text{sign}(k) \text{sign}(r-k)  x_k\over k }+\cO(g^9)
,
\fe
It follows that
\ie
& \sum_{n \neq 0} (g s_n)^{-2} - \sum_r (g h_r)^{-2} 
= \sum_{\ell \neq 0} (-1)^\ell - {g^3\over \pi} \sum_{\ell \neq 0} (-1)^\ell \text{sign}(\ell) C_{\ell\over 2} + {g^6\over 2\pi^2}
\sum_{\ell \neq 0} (-1)^\ell C^2_{\ell\over 2} \\
& - {g^6\over 2\pi^2} \sum_{k\neq0} (-1)^k  C_{k\over 2}\sum_{n \neq 0,{k\over2}}{\text{sign}(n-{k\over 2})\text{sign}(n)\over n}
 +  {2\pi g\over \B} \sum_{\ell \neq 0} (-1)^\ell |{\ell\over 2}|
 \\&+
{g^6\over 2\pi^2} \sum_{\ell \neq 0} (-1)^\ell \text{sign}(\ell) \sum_{k\not =0,{\ell\over 2}} {\text{sign}({\ell\over 2}-k) C_k + \text{sign}(k) C_{{\ell\over 2}-k} \over k} 
\\
&+ {g^3\over \pi} \sum_{n \neq 0} \text{sign}(n) \sum_{k \neq 0,n} {\text{sign}(k) \text{sign}(n-k) x_k  \over k}
-{g^3\over \pi} \sum_r \text{sign}(r)  \sum_{k \neq 0} {\text{sign}(k) \text{sign}(r-k)\over k} x_k\\&
-{1\over 3 }g^6
 + \cO(g^9) 
\\
&=-1-{\pi g\over 2\B}-{5g^6 \over 6},
\fe
where we have used 
\ie
\sum_{n\neq 0}(-)^n C^2_{n /2}
&=
-{\pi^2\over3},
\\
\sum_{\ell \neq 0} (-1)^\ell \text{sign}&(\ell) \sum_{k\not=0,{\ell\over 2}} { \text{sign}(k) C_{{\ell\over 2}-k} \over k}
&= - {\pi^2\over3}.
\fe

\section{Computation of the Free Energy}
\label{app:free}

We will drop the overall $N^2N_f$ factor in the following. Let us compute $\B F_0$ first, the free energy for the trial action $S_0$,
\ie
\B F_0 &= 
{1\over2} \ln {\B^2\sigma_0\over 4\pi^2}  +{1\over 2} \ln (1+\eta_0)\\
&+\sum_{ n=1}^\infty \ln \left[ {2\pi n\over\B} + s_n (1+x_n)
\right]
+ \sum_{ n=1}^\infty \ln \left[ {2\pi n\over\B} + s_n (1+x_n)^{-1}\right]
- 2\sum_{ r>0} \ln \left( {2\pi r\over\B} + h_r\right)\\
%
%
%
%
%
%
& = \text{const} +g\left(\sum_{n\neq0}^\infty s^{(2)}_{|n|} -\sum_{r} h^{(2)}_{|r|}\right)
+g{4\pi\over\B}\left( \sum_{n=1}^\infty n -\sum_{r>0} r\right)
+{5\over 192}(\B\sigma_0)^3 +\mathcal{O}(\B^{-9/5}),
\fe
We leave the second and third term in the last line undone because they will be cancelled by the terms in $-{1\over2} \la S^2_3\ra_0$.

To compute $-{1\over2} \la S^2_3\ra_0 $, it will be convenient to recall our solutions for the propagators,
\ie
\Delta_n& = {\B\over 2\pi n} {1\over {2\pi n\over\B} + s_n(1+x_n)} \\
&= {\B\over 2\pi| n|} g \left[(1+{g^3\over 2\pi}C_{|n|} +gs^{(2)}_{|n|} )(1+x_n) +g{2\pi |n|\over\B} \right]^{-1}+{\cO }(\B^{-1})\\
&\equiv \Delta^{(0)}_n +\Delta^{(1)}_n +\Delta^{(2)}_n+\mathcal{O}(\B^{-1}),\\
\Delta^{(0)}_n &\equiv {\B\over 2\pi |n|} g, \quad
\Delta^{(1)}_n \equiv -{\B\over 2\pi |n|} g \left( {g^3\over 2\pi}C_{|n|} +x_n\right),\\
\Delta^{(2)}_n&\equiv {\B\over 2\pi |n|} g
\left(  -gs^{(2)}_{|n|}+{g^3\over 2\pi }C_{|n|} x_n -g{2\pi |n|\over\B}
+{g^6\over 4\pi^2}C_{|n|}^2+x_n^2
\right).
\fe
\ie
\epsilon_n &= {2\pi n\over \B} {1\over {2\pi n\over\B} + s_n(1+x_n)^{-1}}\\&
= { 2\pi| n|\over \B} g \left[(1+{g^3\over 2\pi}C_{|n|} +gs^{(2)}_{|n|} )(1-x_n+x_n^2) +g{2\pi |n|\over\B} \right]^{-1}+\mathcal{O}(\B^{-3})\\
&\equiv \epsilon^{(0)}_n +\epsilon^{(1)}_n +\epsilon^{(2)}_n+\mathcal{O}(\B^{-3}),\\
\epsilon^{(0)}_n &\equiv { 2\pi |n|\over \B} g,~~~
\epsilon^{(1)}_n \equiv -{ 2\pi |n|\over\B} g \left( {g^3\over 2\pi}C_{|n|} -x_n\right),\\
\epsilon^{(2)}_n&\equiv {2\pi |n|\over\B} g
\left(  -gs^{(2)}_{|n|}-{g^3\over 2\pi }C_{|n|} x_n -g{2\pi |n|\over\B}
+{g^6\over 4\pi^2}C_{|n|}^2
\right).
\fe
\ie
g_r &= {1\over {2\pi r\over\B} + h_r }=g\, \text{sign}(r )
\left[ 1 +{g^3\over 2\pi}C_{|r|} +gh_{|r|}^{(2)} +g{2\pi|r|\over\B}
\right]^{-1} \equiv g_r^{(0)} + g_r^{(1)} +g_r^{(2)}+\mathcal{O}(\B^{-2}),\\
g^{(0)}_r &\equiv g\, \text{sign}(r ),~~~
g^{(1)}_r \equiv -\, \text{sign}(r ){g^4\over 2\pi}C_{|r|},~~~
g^{(2)}_r \equiv  g\, \text{sign}(r )\left( -gh_{|r|}^{(2)}+{g^6\over 4\pi^2}C_{|r|}^2-g{2\pi|r|\over \B} \right).
\fe
Using the above,
\ie
-{1\over2} &\la S^2_3\ra_0 = {1\over 2\B\sigma_0^2 (1+\eta_0)}
+{2\over \B\sigma_0} \sum_{n=1}^\infty \Delta_n \epsilon_n
-{2\over\B\sigma_0} \sum_{r>0} g_r^2+{1\over\B(1+\eta_0)}  \sum_{n=1}^\infty
\Delta_n^2\\
&+{1\over 2\B} \sum_{\ell \neq0 }\sum_{n\neq 0,\ell}
\Delta_{n}\Delta_{\ell-n} \epsilon_{\ell}
+{1\over \B} \sum_{\ell \neq 0}\sum_rg_rg_{\ell-r} \Delta_{\ell} \\
&=  {1\over 2\B\sigma_0^2 (1+\eta_0)}
+{2\over \B\sigma_0} g^2 \sum_{k=1}^\infty(-)^k 
+{1\over(1+\eta_0)}  {g^2\B\over 4\pi^2}\sum_{n=1}^\infty {1\over n^2}
+{2\over \B\sigma_0} \sum_{n=1}^\infty \Delta^{(1)}_n\epsilon^{(0)}_n
\\
&
+{2\over \B\sigma_0} \sum_{n=1}^\infty \Delta^{(0)}_n\epsilon^{(1)}_n - {4\over\B\sigma_0} \sum_{r>0} g^{(0)}_r g^{(1)}_r
+{2\over \B\sigma_0} \sum_{n=1}^\infty \Delta^{(2)}_n\epsilon^{(0)}_n
+{2\over \B\sigma_0} \sum_{n=1}^\infty \Delta^{(1)}_n\epsilon^{(1)}_n
\\
&+{2\over \B\sigma_0} \sum_{n=1}^\infty \Delta^{(0)}_n\epsilon^{(2)}_n
-{4\over\B\sigma_0} \sum_{r>0} g_r^{(2)}g_r^{(0)}
-{2\over\B\sigma_0} \sum_{r>0} (g_r^{(1)})^2
\\
&-{ g^3\over 4\pi}\sum_{\ell\neq 0}\text{sign}(\ell)\sum_{n\neq0,\ell} 
\text{sign}(\ell-n)\text{sign}(n) { \ell\over (\ell-n) n}
+{ g^3\over 2\pi}\sum_{\ell\neq0}{\text{sign}(\ell)\over\ell}  \sum_r \text{sign}(r )\text{sign}(\ell-r)\\
&+{1\over\B}\sum_{\ell\neq0}\sum_{n\neq0,\ell} \Delta^{(1)}_n\Delta^{(0)}_{\ell-n} \epsilon_\ell^{(0)}
+{1\over 2\B}\sum_{\ell\neq0}\sum_{n\neq0,\ell}\Delta^{(0)}_n\Delta^{(0)}_{\ell-n}\epsilon^{(1)}_\ell
+{2\over\B}\sum_{\ell\neq0}\sum_r g^{(1)}_r g^{(0)}_{\ell-r} \Delta^{(0)}_\ell\\&
+{1\over\B} \sum_{\ell\neq0}\sum_r g^{(0)}_rg^{(0)}_{\ell-r}\Delta^{(1)}_\ell+\mathcal{O}(\B^{-9/5}) \\
&=-g\left(\sum_{n\neq0}^\infty s^{(2)}_{|n|} -\sum_{r} h^{(2)}_{|r|}\right)
-g {4\pi\over \B}
\left( \sum_{n=1}^\infty n  -\sum_{r>0} r\right) +{1\over 192}(\B\sigma_0)^3+{\cal O }(\B^{-9/5})\\
\fe
We see that the first two terms above cancel with the second and third term in $\B F_0$.

\section{High Temperature Analysis of One-Loop Truncated $\cN=2$ Gauge-Fixed BFSS}
\label{app:highT}

In this appendix we will present the high temperature expansion to the subleading order of the $\cN=2$ Schwinger-Dyson equations in Appendix~\ref{app:N=2gap}. 

\subsection{Subleading Corrections to the Self-Energies}

For the zero modes, the next-to-leading correction to the self-energies satisfies the following equations (see Appendix~\ref{app:N=2gap}), with $a,b,c,d$ defined in \eqref{highTzm}, 
\ie
\Pi^{(1)M}_0&={4\over \B}\sum_{m\neq 0}({\B\over 2\pi m})^2-{4\over \B}\sum_r({\B\over 2\pi r})^2+{12\over \B}\sum_{m\neq0}({\B\over 2\pi m})^2{1\over 1+\Xi^M_{m}}-{12\over \B}\sum_r({\B\over 2\pi r})^2
\\&~~~- {2 \mu^{(1)}\over \beta\mu^2} -{4\over\B}{\Pi^{(1)V}_0\over (\Pi^V_0)^2}-{12\over\B}\left({\Pi^{(1)M}_0\over(\Pi^M_0)^2(\Xi^M_0+1)}+{\Xi^{(1)M}_0\over \Pi^M_0 (\Xi^M_0+1)^2}\right)
\\&=-{8\B\over 3}- {2 \mu^{(1)}\over b^2} -{4\Pi^{(1)V}_0\over b^2}-12{\Pi^{(1)M}_0\over a^2(d+1)}-{12\over \B^{1/2}}{\Xi^{(1)M}_0\over a (d+1)^2},
\\
\Pi^{(1)V}_0&= {2\over \beta} \sum_{m\neq0} ({\B\over 2\pi m})^2 - {3\over \beta}\sum_r ({\B\over 2\pi r})^2 + {14\over \beta} \sum_{m\neq0} ({\B\over 2\pi m})^2- {13\over \beta} \sum_r ({\B\over 2\pi r})^2
\\
&~~~-{2 \mu^{(1)}\over \beta\mu^2}  -{2\over\B}{\Pi^{(1)V}_0\over (\Pi^V_0)^2}-{14\over\B}{\Pi^{(1)M}_0\over (\Pi^M_0)^2}
\\
&=-{8\B\over 3}-{2 \mu^{(1)}\over b^2}-{2\Pi^{(1)V}_0\over b^2}-{14\Pi^{(1)M}_0\over a^2},
\\
\\
\Xi_0^{(1)M}&=-{12\over\B} {\Pi^{(1)M}_0\over(\Pi^M_0)^3}=-{12\B^{1/2}\Pi^{(1)M}_0\over a^3},
\\
{\mu^{(1)} \over 2} &={4\B\over 3}- {2\Pi^{(1)V}_0\over b^2}-{7\Pi^{(1)M}_0\over a^2},
\fe
where we have used
\ie
&\sum_{m\neq 0}{1\over\pi^2 m^2}={1\over 3},
\\&
\sum_{r}{1\over\pi^2 r^2}=1.
\fe
The solution to these algebraic equations are,
\ie
\Pi^{(1)M}_0=-2.52\B,~~\Pi^{(1)V}_0=-0.32\B,~~\Xi_0^{(1)M}=0.68\B^{3/2},~~\mu^{(1)}=5.53\B.
\fe
Now for nonzero modes,
\ie
\Pi^{(1)M}_n&={4\over \B}\sum_{m\neq 0}({\B\over 2\pi m})^2-{4\over \B}\sum_r({\B\over 2\pi r})^2+{12\over \B}\sum_{m\neq 0}({\B\over 2\pi m})^2{1\over 1+\Xi^M_{n-m}}+{12\over \B}\sum_r{\B\over 2\pi r}{\B\over 2\pi (n-r)}
\\&~~~+ {4\over \beta} ({2\pi n\over \beta})^2 \sum_r  ({\B\over 2\pi r})^3 {\B\over2\pi(n-r)}- {2 \mu^{(1)}\over \beta\mu^2} + {8 \mu^{(1)}\over\beta\mu^2}+ {8\over\beta\mu}({\beta\over 2\pi n})^2 \Pi^M_n 
\\&~~~-{4\over\B}{\Pi^{(1)V}_0\over (\Pi^V_0)^2}-{12\over\B}\left({\Pi^{(1)M}_0\over(\Pi^M_0)^2(\Xi^M_n+1)}+{\Xi^{(1)M}_n\over \Pi^M_0 (\Xi^M_n+1)^2}\right)
\\&
=\B C^M_n+ {6 \mu^{(1)}\over b^2}-{4\Pi^{(1)V}_0\over b^2}-12{\Pi^{(1)M}_0\over a^2},
\\
\Pi^{(1)V}_n&= {2\over \beta} \sum_{m\neq 0} ({\B\over 2\pi m})^2 - {3\over \beta}\sum_r ({\B\over 2\pi r})^2 + {14\over \beta} \sum_{m\neq 0} ({\B\over 2\pi m})^2+ {14\over \beta} \sum_r {\B\over 2\pi r}{\B\over 2\pi (n-r)}
\\
&~~~- {1\over\beta} \sum_r{\B\over 2\pi r}{\B\over 2\pi (n-r)} -{2 \mu^{(1)}\over \beta\mu^2} + {8 \mu^{(1)}\over\beta\mu^2}+ {8\over\beta\mu}({\beta\over 2\pi n})^2 \Pi^V_n 
-{2\over\B}{\Pi^{(1)V}_0\over (\Pi^V_0)^2}-{14\over\B}{\Pi^{(1)M}_0\over (\Pi^M_0)^2}
\\&
=\B C^V_n+{6 \mu^{(1)}\over b^2}-{2\Pi^{(1)V}_0\over b^2}-{14\Pi^{(1)M}_0\over a^2},
\\
\Xi^{(1)M}_n&=\mathcal{O}(\B^3),
\\
{\mu^{(1)} \over 2} &= {2\over \beta}  \sum_{m\neq 0} ({\B\over 2\pi m})^2 - {5\over 2\beta} \sum_r ({\B\over 2\pi r})^2 + {7\over \beta} \sum_{m\neq 0} ({\B\over 2\pi m})^2
 - {4\over\beta} \sum_{m\neq 0} ({\B\over 2\pi m})^2
+ {4\over\beta} \sum_r ({\B\over 2\pi r})^2
\\
&~~~-{14\over\beta}  \sum_{m\neq 0} ({\B\over 2\pi m})^2 + {7\over\beta}\sum_r ({\B\over 2\pi r})^2 + {1\over \beta} \sum_{m\neq 0} ({\B\over 2\pi m})^2 - {1\over 2\beta}\sum_r ({\B\over 2\pi r})^2
\\
&~~~-{2\over\B}{\Pi^{(1)V}_0\over (\Pi^V_0)^2}-{7\over\B}{\Pi^{(1)M}_0\over (\Pi^M_0)^2}
\\
&=2\B(1-{1\over 3})- {2\Pi^{(1)V}_0\over b^2}-{7\Pi^{(1)M}_0\over a^2}={4\B\over 3}- {2\Pi^{(1)V}_0\over b^2}-{7\Pi^{(1)M}_0\over a^2},
\fe
where
\ie
C^M_n
=& {4\over 3}-{2\over \pi^2 n^2}\left(\frac{9}{a^2+6}+\frac{2 (a-6 b)}{a b^2}\right),
\\
C^V_n
=& \frac{7}{12}+{2\over \pi^2 n^2}\left(1-{8\over b^2}\right).
\fe
We have used
\ie
\sum_r{1\over r(n-r)}
=0, \quad
\sum_r{1\over r^3(n-r)}
={\pi^2\over n^2}.
\fe
Therefore, from \eqref{highTN[zm]} and (G.3),
\ie
\B^{-1}\Pi^{(1)M}_n&={4\over 3}-{1\over \pi^2 n^2}\left(\frac{18}{a^2+6}+\frac{4 (a-6 b)}{a b^2}\right)-\frac{12 \Pi^{(1)M}_0}{\B a^2}-\frac{4 \Pi^{(1)V}_0}{\B b^2}+\frac{6 \mu^{(1)}}{\B b^2}
\\
&=5.24183\, +\frac{0.0274248}{n^2},
\\
\B^{-1}\Pi^{(1)V}_n&=\frac{7}{12}+{2\over \pi^2 n^2}\left(1-{8\over b^2}\right)-\frac{14 \Pi^{(1)M}_0}{\B a^2}-\frac{2 \Pi^{(1)V}_0}{\B b^2}+\frac{6 \mu^{(1)}}{\B b^2}
\\&
=4.86502\, +\frac{0.131842}{n^2}.
\fe

\subsection{Free Energy in the High Temperature Limit}

To leading order in $\B$, the free energy receives only contribution from the bosonic sector. Using the fact that $\Pi^M_n$ and $\Pi^V_n$ scales as $\B^{-1/2}$ \eqref{highTnzm}, the same as their corresponding zero modes, we have
\ie
\B F^{(0)}&=f(\mu)-\mu\partial_\mu f-\sum_l {\rm log}\,\sigma^2_l-{7 \over 2}\sum_l {\rm log}\,\Delta^2_l
\\&\simeq{1\over 2} {\rm log}\,\mu+2{\rm log}\,{\rm sinh}\left({\B(\Pi^V_0)^{1/2}\over 2}\right)+7{\rm log}\,{\rm sinh}\left({\B(\Pi^M_0)^{1/2}\over 2}\right)
\\&\simeq \text{const}+6\,{\rm log}\,\B.
\fe
The subleading contributions to the free energy come from the $\mathcal{O}(\B^{3/2})$ corrections to the zero mode self-energies and also the leading nonzero mode self-energies (relative to the kinetic energy),
\ie
\B F_0^{(1)}=&{\Pi^{(1)V}_0\over\Pi^{(0)V}_0}+{7\over 2}{\Pi^{(1)M}_0\over\Pi^{(0)M}_0}+{7\over 2}{\Xi^{(1)M}_0\over 1+\Xi^{(0)M}_0}+{1\over 2}{\mu^{(1)}\over \m^{(0)}}
\\&+\sum_{l\neq 0} ({\B\over 2\pi l})^2\Pi_l^V-\sum_{r} ({\B\over 2\pi r})^3\Sigma_r^V+{7 \over 2}\sum_{l\neq 0}  ({\B\over 2\pi l})^2\Pi_l^M-7\sum_{r} {\B\over 2\pi r}\Sigma_r^M
\\&+{7\over 2}\sum_{l\neq 0} \Xi^M_l-\sum_{l\neq 0} ({\B\over 2\pi l})^2 \Pi^G_l+2\sum_r {\B\over 2\pi r}\Sigma^G_r,
\fe
\ie
\B F_2^{(1)}=&-\sum_{l\neq 0} ({\B\over 2\pi l})^2\Pi_l^V+\sum_{r} ({\B\over 2\pi r})^3\Sigma_r^V-{7 \over 2}\sum_{l\neq 0}  ({\B\over 2\pi l})^2\Pi_l^M+7\sum_{r} {\B\over 2\pi r}\Sigma_r^M
\\&-{7\over 2}\sum_{l\neq 0} \Xi^M_l+\sum_{l\neq 0} ({\B\over 2\pi l})^2 \Pi^G_l-2\sum_r {\B\over 2\pi r}\Sigma^G_r-{7\over 2}{\Xi^M_0\over 1+\Xi^M_0}.
\fe
Note that to this order
\ie
{\Xi^M_0\over 1+\Xi^M_0}\simeq \text{const}+{\Xi^{(1)M}_0\over 1+\Xi^{(0)M}_0}-{\Xi^{(1)M}_0\Xi^{(0)M}_0\over (1+\Xi^{(0)M}_0)^2}.
\fe
Hence
\ie
\B F_0^{(1)}+\B F_2^{(1)}=\B^{3/2}\left[{\Pi^{(1)V}_0\over\B b}+{7\over 2}{ \Pi^{(1)M}_0\over\B a}+{7\over 2}{\Xi^{(1)M}_0 d\over \B^{3/2}( 1+d)^2}+{1\over 2} {\mu^{(1)}\over\B b} \right].
\fe
$\B F_3$ and $\B F_4$ can be similarly computed to be
\ie
\B F_3^{(1)}
\simeq &-{4\over\B\m}\sum_{l\neq0} ({\B\over 2\pi l})^2+{4\over\B\m}\sum_r ({\B\over 2\pi r})^2-{14\over \B\Pi^M_0}\sum_{r} ({\B\over 2\pi r})^2
\\&
-{14\over \B\Pi^V_0}\sum_{r} ({\B\over 2\pi r})^2-{14\over\B\m}\sum_{l\neq0} ({\B\over 2\pi l})^2+{7\over\B\m}\sum_r ({\B\over 2\pi r})^2
\\&
+{21\over\B(\Pi^M_0)^2}\epsilon_0^2+{42\over\B\Pi^M_0}\sum_{l\neq0}  ({\B\over 2\pi l})^2-{42\over \B\Pi^M_0}\sum_{r} ({\B\over 2\pi r})^2
\\&+{1\over \B\Pi^V_0}\sum_{r} ({\B\over 2\pi r})^2
-{1\over 2\B\m}\sum_r ({\B\over 2\pi r})^2+{1\over\B\m}\sum_{l\neq 0} ({\B\over 2\pi l})^2
\\ \simeq&\B^{3/2}\left[-\frac{7 (7 a+36 b)}{24 a b}-{21\Xi^{(1)M}_0 a^2\over \B^{3/2} (a^2+6)^2}-{42\Pi^{(1)M}_0\over \B a(a^2+6)}\right]
\\ \simeq&-2.38766\B^{3/2}.
\fe

Up to an additive constant, 
\ie
\B F_4^{(1)}\simeq&-{3\over \B\Pi^V_0}\sum_{r}  ({\B\over 2\pi r})^2-{2\Pi^{(1)V}_0\over \B(\Pi^V_0)^3}+{2\over \B\Pi^V_0}\sum_{l\neq 0}({\B\over 2\pi l})^2-{14\over \B\Pi^V_0\Pi^M_0}\left({\Pi^{(1)V}_0\over\Pi^V_0}+{\Pi^{(1)M}_0\over\Pi^M_0}\right)
\\&
+{14\over \B\Pi^V_0\Pi^M_0}(\Pi^V_0+ \Pi^M_0)\sum_{l\neq 0}({\B\over 2\pi l})^2
-{14\over \B\Pi^M_0}\sum_{r}({\B\over 2\pi r})^2-{7\over\B\m\Pi^M_0}\left({\Pi^{(1)M}_0\over\Pi^M_0}+{\m^{(1)}\over\m}\right)
\\&
+{7\over\B\m}\sum_{l\neq0}({\B\over 2\pi l})^2-{5\over 2\B\m}\sum_r({\B\over 2\pi r})^2
\\&
-{2\over\B\m\Pi^V_0}\left({\Pi^{(1)V}_0\over\Pi^V_0}+{\m^{(1)}\over\m}\right)+{2\over\B\m}\sum_{l\neq0}({\B\over 2\pi l})^2
\\\simeq&\left(-\frac{12\Pi^{(1)M}_0}{7 \B b^3}-\frac{3 \Pi^{(1)M}_0 b}{28\B} +\frac{6 \Pi^{(1)M}_0}{7\B b}-\frac{\Pi^{(1)V}_0}{\B b}-\frac{\mu^{(1)}}{2 \B b}-\frac{b}{6}+\frac{11}{8 b}\right)\B^{3/2}
\\\simeq&-0.142172\B^{3/2}.
\fe
Therefore, 
\ie
\B F^{(1)}
\simeq &\frac{b}{864} \left(105 b^4-2338 b^2-2220\right)\B^{3/2}
\\\simeq &-3.89872\B^{3/2}.
\fe
Hence
\ie
\B F = {\rm const}+6\,{\rm log}\B-3.89872\B^{3/2}+\mathcal{O}(\B^{3}).
\fe

\bibliographystyle{JHEP}

\bibliography{draft_gap_v2}

\end{document}